\documentclass[preprint2]{aastex}
\shortauthors{JENKINS \& TRIPP}
\shorttitle{THERMAL PRESSURE DISTRIBUTION IN THE ISM}
\begin{document}
\title{The Distribution of Thermal Pressures in the Diffuse, Cold Neutral Medium 
of our Galaxy. II. An Expanded Survey of Interstellar C~I Fine-Structure 
Excitations}
\author{Edward B. Jenkins}
\affil{Princeton University Observatory\\
Princeton, NJ 08544-1001}
\email{ebj@astro.princeton.edu}
\and\author{Todd M. Tripp}
\affil{Department of Astronomy\\
University of Massachusetts\\
710 North Pleasant Street\\
Amherst, MA 01003-9305}
\email{tripp@astro.umass.edu}
\begin{abstract}
We analyzed absorption features arising from interstellar neutral carbon that 
appeared in the UV spectra of 89 stars recorded in the highest resolution echelle 
modes of the {\it Space Telescope Imaging Spectrograph\/} on {\it HST\/} so that 
we could determine the relative populations of collisionally excited fine-structure 
levels in the atom's electronic ground state.   From this information, in 
combination with molecular hydrogen rotation temperatures, we derive the 
distribution of thermal pressures in the diffuse, cold neutral medium.  We find a 
lognormal pressure distribution (weighted by mass) with a mean in $\log (p/k)$ 
equal to 3.58 and an rms dispersion of at least 0.175~dex that plausibly arises 
from turbulence with a characteristic Mach number in the range $1 < M < 4$.  
The extreme tails in the distribution are above the lognormal function however.  
Overall, pressures are well correlated with local starlight intensities and extreme 
kinematics, and they show some anticorrelation with kinetic temperatures.  A 
subsample restricted to low ambient UV intensities reveals a mode in the 
distribution of $\log (p/k)$ that is nearly the same as the complete sample, but 
with a strong negative skewness created by a near absence of a tail at high 
pressures.  Approximately 23\% of this gas is at a pressure that is below that 
allowed for a static cold neutral medium.  Accompanying nearly all of the gas is a 
small fraction ($\sim$0.05\%) that has an extraordinarily large pressure, $\log 
(p/k) > 5.5$, and this condition is more prevalent at high velocities or for regions 
with enhanced starlight densities. This survey suggests that the dispersion of thermal 
pressures in the cold, neutral ISM is predominantly governed by microscopic 
turbulence driven by star-forming regions, with some additional effects from 
macroscopic events (e.g., SN explosions), and these measurements provide 
constraints for future studies of the broader impact of turbulence on the ISM and 
star formation.
\end{abstract}
\section{Introduction}\label{intro}

In his commentary on the classic paper by Chandrasekhar \& M\"{u}nch 
 (1952) on brightness fluctuations in the Milky Way, Scalo (1999) 
presented an insightful discussion of a dichotomy in our perception of the 
structure of the diffuse interstellar medium (ISM) of our Galaxy.  On the one 
hand, we might view the ISM in terms of a collection of isolated, dense clouds 
enveloped in a more tenuous medium, a concept that has directed our thinking 
on the establishment of discrete ``phases'' of the ISM with well established spatial 
domains and vastly different properties that can be justified on some 
fundamental physical grounds
% ****V\'azquez-Semadeni
 (Field et al. 1969; McKee \& Ostriker 1977; Burkert \& Lin 2000; V\'azquez-Semadeni
et al. 2000; Brandenburg et al. 2007).  On the other hand, much of the ISM can be viewed
as a continuous fluid medium containing a texture of seemingly random fluctuations in
density, velocity and temperature.  Adherents to this second picture [e.g., Ballesteros-
Paredes et al.  (1999)] view clouds as an illusion created 
by the most extreme fluctuations in a turbulent medium having an extraordinarily 
high Reynolds number.  While this is undoubtedly true, we must also 
acknowledge the presence of nearly static, sharp boundaries between different 
media, as revealed by dark clouds with well defined edges that have been 
sculpted by ionization, dissociation, and evaporation/condensation fronts.  Both 
pictures have their utility in exploring important issues on the multitude of 
processes that can occur within the ISM.

As we switch our perspective from morphological to dynamical properties of the 
ISM, we find that over macroscopic scales the motions of gases in our Galaxy can 
be governed by the injection and dissipation of mechanical energy from a wide 
range of energy sources that include supernova explosions (McKee \& Ostriker
1977; McCray \& Snow 1979; Mac Low et al. 1989; Kim et al. 2001; de Avillez
\& Breitschwerdt 2005a), disturbances from newly formed H~II regions 
% *** Rodr\'iguez-Gaspar
 (Lasker 1967; Tenorio-Tagle 1979; Rodr\'iguez-Gaspar \& Tenorio-Tagle 1998;
Peters et al. 2008), stellar mass loss  (Abbott 1982; McKee et al. 1984; Owocki
1999), infalling gas clouds from the Galactic halo
% *** Santill\'an 
 (Wakker \& van Woerden 1997; Santill\'an et al. 1999, 2007) , bipolar jets from
 star 
forming regions (Bally 2007),  shocks in spiral arm density waves  
 (Roberts et al. 1975), and the magnetorotational instabililty driven by 
differential galactic rotation (Pinotek \& Ostriker 2004).  These 
processes play a strong role in creating recognizable, discrete structures and flows 
of material in the ISM, but ultimately some of the energy from the resulting 
compressions and vorticity will also be transformed into random turbulent 
motions.  Transient structures of small sizes can arise from the cascade of larger 
turbulent cells into small ones or be created in the interface regions between 
colliding gas flows (Audit \& Hennebelle 2005).  Turbulence can also be 
fed by instabilities in phase transition layers (Inoue et al. 2006) or the 
weak driving forces that arise from the thermal instability of the ISM 
 (Kritsuk \& Norman 2002b; Koyama \& Inutsuka 2006), the latter 
of which can be sustained by abrupt changes in the heating rate from UV 
radiation (Kritsuk \& Norman 2002a).

Over the past several decades, much progress has been made in the study of 
magnetohydrodymical (MHD) turbulence in the ISM.  Our understanding of this 
phenomenon has been facilitated by the rapid emergence of powerful 
3-dimensional computer simulations, and its existence is supported by 
observations of column density distributions, velocity statistics, cloud 
morphologies, deviations in magnetic fields, and various kinds of disturbances in 
the propagation of radio waves in ionized media [for a review, see Elmegreen \& 
Scalo (2004)]. This phenomenon is influential on 
the heating, chemical mixing, radio wave propagation, and cosmic ray scattering 
in the ISM (Scalo \& Elmegreen 2004).  Within denser environments, 
turbulent processes are expected to have a strong influence on the fragmentation 
of density concentrations just before and during the earliest stages of 
gravitational collapse that leads to star formation (Mac Low \& Klessen 2004;
McKee \& Ostriker 2007).

Both the coherent dynamical phenomena and the smaller scale turbulent motions 
have an influence on pressures in the ISM.  These pressures appear in many 
forms: thermal, magnetic, dynamical, and the indirect effects of cosmic rays, and 
their collective magnitude amounts to about $p/k=2.5\times 10^4{\rm 
cm}^{-3}\,$K\footnote{Throughout this paper, we quantify pressures in terms of 
$p/k$ in the units ${\rm cm}^{-3}\,$K instead of simply $p$ in the units of ${\rm 
dyne~cm}^{-2}$ or ${\rm erg~cm}^{-3}$.  Our representation facilitates 
comparisons with actual densities and temperatures in the ISM.} that is 
established by the hydrostatic equilibrium of gaseous material in the gravitational 
potential of the Galactic plane (Boulares \& Cox 1990).  Except for 
very hot media ($T > 10^5\,$K) that have been created by shock heating from 
supernova blast waves (Cox \& Smith 1974; de Avillez \& Breitschwerdt 2005b) or
that reside within wind-blown bubbles around stars  (Castor et al. 1975; Weaver et
al. 1977), the thermal pressures of the ISM represent a small fraction (about
one-tenth for $T\sim 100\,$K) of the total pressure (de Avillez \& Breitschwerdt
2005a). While thermal pressures and their variability may seem unimportant in the 
dynamical development of the general ISM, they nevertheless can be influenced 
by stochastic dynamical effects and thus can provide us with useful information. 

In this paper, we make use of the fact that the two upper fine-structure levels in 
the electronic ground state of the neutral carbon atom, with excitation energies 
$E/k=23.6$ and $62.4\,$K, are easily excited and de-excited by collisions with 
neutral and charged particles at typical densities and temperatures within the 
diffuse, cold gas in the Galactic plane.  The balance between the effects of these 
collisions and spontaneous radiative decays (at wavelengths 609 and 
$371\,\mu$m) establishes fine-structure level population ratios that can serve as 
an indicator of the local density and temperature of the C~I-bearing material.  We 
sense these population ratios by observing the UV multiplets of C~I that appear as 
foreground absorption features in the spectra of hot stars recorded at high 
spectral resolution.

The first widespread study of C~I fine-structure excitations was carried out by 
Jenkins \& Shaya (1979), who analyzed observations 
that came from the UV spectrograph on the {\it Copernicus\/} satellite.  Those 
observations and a more comprehensive survey by Jenkins et al. (1983) were
primitive by today's standards for observing C~I features set by spectrographs on
the {\it Hubble Space Telescope\/} ({\it HST\/}) (Smith et al. 1991; Jenkins et
al. 1998; Jenkins 2002), but they nevertheless established an early framework for
determining the average thermal pressures in the ISM and their variations from
one location to the next.   In addition to these general surveys, special studies
sensed extreme positive deviations in pressure from the C~I features in spectra of
stars within and behind the Vela supernova remnant (Jenkins et al. 1981, 1984,
1998; Jenkins \& Wallerstein 1995; Wallerstein et al. 1995; Nichols \& Slavin
2004), indicating that the blast wave overtook and compressed\footnote{We note
that Nichols \& Slavin  (2004) proposed some possible alternative explanations for
producing an excess of excited C~I.} small clouds in the medium that surrounded
the explosion site (Chevalier 1977).

A new advance in the study of C~I excitation in the general ISM arose from the 
study by Jenkins \& Tripp (2001) (hereafter 
JT01), who used the highest resolution configurations of the echelle spectrograph 
in the {\it Space Telescope Imaging Spectrograph\/} (STIS) (Kimble et al. 1998;
Woodgate et al. 1998) to observe C~I in the spectra of 21 stars.  An important
breakthrough in this work was to make use of the ability of STIS to record many
different C~I multiplets simultaneously, which allowed JT01 to benefit from a
special analysis technique that they developed to unravel the blended absorption
profiles arising from the three fine-structure levels.

The current study of thermal pressures expands on the work of JT01, once again 
using the analysis method employed earlier, but with some technical 
improvements outlined in Appendix~\ref{improvements}.  In \S\ref{observations} 
we describe our new coverage of sightlines that significantly expands on the 
limited selection of stars that were studied by JT01, but we caution that one must 
be aware of a few, mostly unavoidable, selection biases in the sampling.  In 
\S\ref{results} we review the basic principles of the analysis, but leave it to the 
reader to consult JT01 for a more detailed description of the mathematical 
method.  This section also introduces our fundamental approach to interpreting 
the population ratios in terms of the local density and temperature of the 
C~I-bearing gas, a method originally developed by Jenkins \& Shaya  
 (1979).  Table~\ref{sightlines_table} in this section 
lists the 89 target stars whose spectra were analyzed in this study, along with 
some relevant information about the foreground regions probed by the sightlines.  
For each sightline, we show in Table~\ref{obs_quantities_table} some composite 
interstellar conditions that we derived and some reflections on their significance 
in \S\ref{overview}.  In Table~\ref{MRT} we provide more detailed, 
machine-readable information for each velocity bin in the entire survey. 
Section~\ref{properties} describes a number of additional factors that require 
some consideration when the C~I results are interpreted, including ways to 
estimate the total amount of gas that accompanies the C~I 
(\S\ref{ionization_corrections}), the kinetic temperature of the gas (\S\ref{T}), the 
mix of particles that can excite the upper levels (\S\ref{particle_mix}), and the 
local intensity of starlight (\S\ref{starlight}).  The starlight intensity must be known 
in order to correct for optical pumping of the fine-structure levels (as described in 
\S\ref{derivations}), and, as we demonstrate in \S\ref{basic_dist}, such intensities 
seem to be correlated with pressure.

Section~\ref{admixtures} discusses how we interpret our finding that virtually all 
of the measurements do not agree with the expected fine-structure populations 
for collisional excitation for any uniform values of local density or temperature.  
Here, we introduce the idea that small admixtures of gas at extraordinarily high 
pressures accompany virtually all of the gas at normal pressures, and in 
\S\ref{misleading_conclusion} we examine (and ultimately reject) some 
alternative explanations of the observed deviations.   Some discussions on the 
implications and possible origins of the high pressure component appear in 
\S\S\ref{amount_hipress}$-$\ref{origins}, and we show in 
\S\ref{behavior_velocity} and \S\ref{origins} that the gas fractions at high 
pressures are accentuated in material that is moving rapidly or is exposed to a 
high intensity of starlight.

In \S\ref{interpretation} we derive the distribution function for the 
mass-weighted thermal pressures in the dominant low pressure regime for two 
samples: (1) all of the gas and (2) gas that is well removed from intense sources of 
UV radiation.  For the convenience of those who wish to compare our results with 
computer simulations of turbulence based on volume-weighted distributions, we 
convert our mass-weighted sampling to a volume-weighted one in 
\S\ref{vol_weighted}, but with the precarious assumption that the gas responds 
to pressure perturbations with a single polytropic index $\gamma$ (which is left 
as a free parameter).  In \S\ref{pileup}, we address the possibility that the width 
of the pressure distribution understates the true dispersion of pressures in the 
ISM, due to the fact that we view in each velocity bin the superposition of 
absorptions by gases with different pressures and thus only sense an average 
pressure in each case.  In \S\ref{comparisons_turbulence} we derive a range of 
possible characteristic turbulent Mach numbers for the C~I-bearing gas

We explore in \S\ref{time_constants} some time constants for various physical 
processes that are relevant to our work.  For instance, the excitation 
temperatures of the two lowest rotational levels of H$_2$ play a role many of our 
pressure determinations.  In \S\ref{crossing_times} we find that only on scales of 
order 100$-$1000$\,$AU are  lag times of any importance in weakening the 
coupling of H$_2$ rotation temperatures to the local kinetic temperature.  The 
same applies to the equilibrium between heating and cooling of the gas: 
compressions and decompressions of the gas should follow the equilibrium 
relationship except on the smallest scales where the behavior becomes more 
adiabatic in nature.   The time scales for the equilibria of fine-structure 
populations and the balance between C~I and C~II are quite short and hence 
apply on all of the relevant size scales.

In a brief departure from the discussion of turbulence as a source of pressure 
deviations, we consider in \S\ref{coherent_disturbances} the possibility that the 
upper end of our pressure distribution function is consistent with random 
interceptions of supernova remnants in different stages of development.  Finally, 
we summarize our conclusions in \S\ref{summary}.
  
\section{Observations}\label{observations}

Our earlier survey (JT01) covered only 21~stars that were observed with the 
guaranteed observing time granted to the STIS instrument definition team.  In 
order to maximize the observing efficiency, the target stars in this study were 
located within two Galactic longitude intervals, ones where the {\it HST\/} 
continuous viewing zones\footnote{The continuous viewing zones (CVZs) are two 
declination bands centered on $\delta=\pm 61\fdg5$ where observations can be 
performed with high efficiency because the targets are not occulted by the Earth 
as the satellite progresses along its orbit.}  intersected the Galactic plane 
($99\arcdeg < \ell < 138\arcdeg$, $254\arcdeg <  \ell < 313\arcdeg$).  Many 
additional observations, most of which were performed after the study by JT01, 
were more broadly distributed in the sky and were taken before the 5-year hiatus 
of observing brought about by STIS instrument failure in August 2004.  We 
downloaded from MAST (Multimission Archive at the Space Telescope Science 
Institute)  virtually all of the observations performed at wavelengths that covered 
two or more C~I transitions in the E140H and E230H modes.  Once again, we 
made use of the broad wavelength coverage of STIS to examine as many 
multiplets as possible.  These spectra had a resolving power in radial velocity equal 
to $2.6\,{\rm km~s}^{-1}$ (or $1.5\,{\rm km~s}^{-1}$ for the stars observed by 
JT01 because a narrower entrance slit was used) (Proffitt et al. 2010).  Prominent
among these observations were those performed by a SNAP (snapshot) program conducted
by J.~T.~Lauroesch (program nr.~8241) in 1999 and 2000.  In the current new study, we
have also reanalyzed the data presented by JT01 because we have now adopted some new,
more refined analysis procedures (see Appendix~\ref{improvements}).   All of the data
were processed in the manner described by JT01, except that for observations outside
their survey we did not need to implement an intensity rebalancing between MAMA 
half pixels (see their \S4.2 for details), since these half pixel intensities were 
binned together beforehand.

A small fraction of the observations had to be rejected because either (1) there 
was an insufficient amount of C~I present to perform a meaningful analysis with 
the signal-to-noise ratio at hand, or (2) the projected rotational velocity of the 
star was so low that stellar features interfered with the interstellar ones or made 
the continua too difficult to model.   These unsuitable sight lines are identified by 
their target star names in Table~\ref{rejected}.  We also rejected the central stars 
of planetary nebulae, since the interstellar components could be contaminated by 
contributions from gas in the nebular shell.

\placetable{rejected}
\begin{center}
\begin{table}[h!]
\caption{Rejected Sightlines\label{rejected}}
\begin{tabular}{
c   % C I too weak
c   % stellar features
}
%\tablecolumns{2}
%\tablewidth{0pt}
%\tablecaption{Rejected Sightlines\label{rejected}}
\tableline
%\tablehead{
%\colhead{Insufficient} & \colhead{Stellar Line}\\
%\colhead{C~I} & \colhead{Confusion}
%}
Insufficient & Stellar Line\\
C~I & Confusion\\
\tableline
BD+25D2534&CPD$-$64D481\\
BD$-$03D2179&HD$\,$1909\\
HD$\,$1999&HD$\,$3175\\
HD$\,$6456&HD$\,$30122\\
HD$\,$6457&HD$\,$37367\\
HD$\,$23873&HD$\,$43819\\
HD$\,$32039&HD$\,$44743\\
HD$\,$64109&HD$\,$52329\\
HD$\,$79931&HD$\,$62714\\
HD$\,$86360&HD$\,$93237\\
HD$\,$92536&HD$\,$94144\\
HD$\,$164340&HD$\,$106943\\
HD$\,$192273&HD$\,$108610\\
HD$\,$195455&HD$\,$175756\\
HD$\,$196867\\
HD$\,$201908\\
HD$\,$233622\\
\tableline
\end{tabular}
\end{table}
\end{center}
Initially, we had considered using observations recorded at lower resolution with 
the E140M mode of STIS to broaden the selection of targets, but a comparison of 
the results for a few stars that were also observed with the E140H mode indicated 
that unreliable results emerged from the lower resolution data as a result of 
improper treatments of unresolved, saturated profiles.  (The correction scheme 
developed by Jenkins  (1996) could not be used because 
various lines overlap each other, and the optical depths must go through a 
complicated transformation to obtain unique answers for the column densities of 
the three fine-structure levels, as described in \S5.2.1 of JT01.)

Table~\ref{sightlines_table} presents the information on the 89 sightlines included 
in the present study.  They span path lengths that range from about 0.2$\,$kpc to 
6$\,$kpc and have a median length of 1.9$\,$kpc.  We processed all of the data 
that we felt were acceptable, according to the principles outlined in the above 
two paragraphs.  We made 2416 separate measurements, but since we 
oversampled the wavelength resolution of the spectrograph by a factor of 5.3, 
our determinations actually represent only about 460 independent samples in 
radial velocity.

\begin{figure*}
\epsscale{2.2}
\plotone{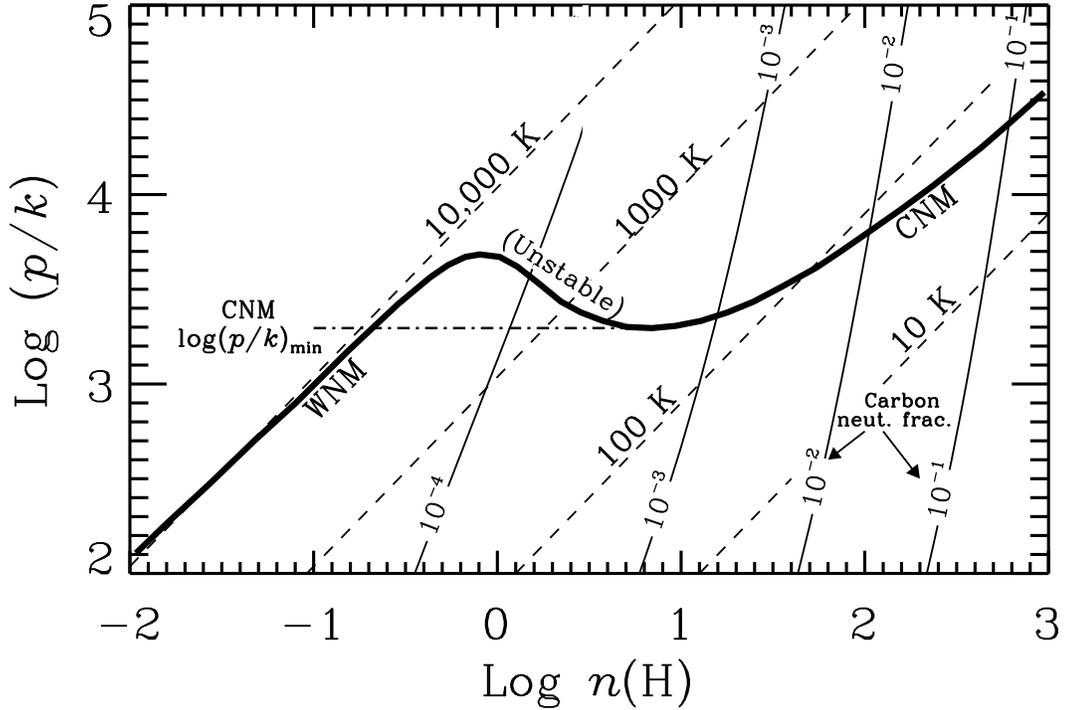}
\caption{A plot of the thermal pressure, $\log(p/k)$, vs. the density of hydrogen 
nuclei, $\log n(H)$, showing the locations where heating equals cooling in the ISM 
near our part of the Galaxy (thick curve), according to the thermal equilibrium 
calculations by Wolfire et al. (2003) (their ``standard model''; see their 
Fig.~8).  Portions of this curve where the slopes are positive are thermally stable 
and form the distinct phases called the warm neutral medium (WNM) and cold 
neutral medium (CNM), as indicated.  The portion of the curve that has a negative 
slope has a balance between heating and cooling, but is thermally unstable 
 (Field 1965).  In the absence of rapidly changing pressures and 
densities due to turbulence, the lowest allowable pressure for the CNM is at the 
horizontal dash dot line labeled ``CNM $\log (p/k)_{\rm min}$.''  Different 
temperatures in this diagram are revealed by the straight, dashed lines, 
constructed using the assumption that He/H=0.09 and, for $T<1000\,$K, $f({\rm 
H}_2)=0.6$ (see \S\protect\ref{particle_mix}).  The thin, gently curved lines show 
constant values for the expected values of ${\rm C~I_{total}/(C~II+C~I_{total}})$, 
as indicated, according to our equation for ionization equilibrium (see 
Eq.~\protect\ref{C_ionization} in \S\protect\ref{starlight} and the accompanying 
text), under the assumption that the starlight intensity is equal to the average level 
$I_0$ given by Mathis et al. (1983).  These curves 
demonstrate that the WNM is virtually invisible in our survey of C~I. 
\label{phase_diag}}
\end{figure*}

We refrained from applying any special selection criteria to make our sampling of 
regions more evenhanded.   For this reason, one must be aware of certain 
selection biases in the composite information presented in \S\ref{results} below.  
The following are some noteworthy considerations about our sample:
\begin{enumerate}
\item All sightlines terminate at the location of a bright, early-type star.  Thus, it is 
inevitable that we will be probing an environment near such a star, or in fact a 
location near a grouping of many such stars, since they tend to be strongly 
clustered in space.  In a number of cases, strong elevations in thermal pressures 
seen in certain radial velocity channels probably arise from either the effects of 
stellar winds or rapidly expanding H~II regions.   As we will show in 
\S\ref{basic_dist},  a large portion of the C~I that we observe resides in regions 
with much higher than normal densities of starlight radiation.  This is probably a 
consequence of the fact that the observations are biased toward a sampling of the 
progenitorial cloud complexes out of which the stars had formed.
\item Regions where the density is low enough (or the local temperature or 
radiation density high enough) to shift the ionization equilibrium of carbon atoms 
more strongly than usual to its ionized form are missed in our sample.   As we 
show in Figure~\ref{phase_diag}, practically all of what is classically known as the 
warm neutral  medium (WNM, with $T\sim 9000\,$K) is invisible to us; our survey 
is restricted to the phase called the cold neutral medium (CNM, with $T\sim 
80\,$K), with possibly some very limited sensitivity to gas in the thermally 
unstable intermediate temperatures.  In addition, there are situations where C~I is 
detected at certain velocities, but the quality of the data is insufficient to measure 
reliably the thermal pressures, as we discuss in some detail in 
\S\ref{uncertainties}.  Hence, such regions are excluded.
\item Regions of moderate size that are very dense will have enough extinction in 
the UV to make stars behind them too faint to observe.   This effect will result in 
our missing clouds that happen to be strongly compressed by turbulence or 
gravity.  It is clear from the results shown in Column~5 of 
Table~\ref{sightlines_table} that a cutoff of our sample corresponds to a $B-V$ 
color excess of about 0.5, which in turn translates approximately to $N({\rm 
H})=3\times 10^{21}{\rm cm}^{-2}$ if we use the standard relation between 
E$(B-V$) and $N({\rm H})$ in the ISM (Bohlin et al. 1978; Rachford et al. 2009). 
Also, portions of some of our strongest C~I absorption profiles were rejected from
consideration because we sensed that they had velocity substructures that were
saturated and not resolved by the instrument. \item Stars in certain programs
were selected by observers because they had interesting properties.  Of special
relevance to our thermal pressure outcomes would be the observations that were
designed to probe regions that were known to be either disturbed (e.g., showing
high velocity gas) or at higher than normal densities (e.g., showing unusually
strong molecular absorptions).  Often, one can sense the characters of such
selections by reading the abstracts of the programs that made the
observations.\footnote{The archive root names listed in Column~9 of
Table~\ref{sightlines_table} can be used as a guide on the MAST {\it HST\/}
search web page to find the Proposal ID and its abstract.}
\end{enumerate}
\placetable{sightlines_table}
\clearpage
%\LongTables
\begin{deluxetable}{
c    % hd nr (1)
c   % l (2)
c   % b (3)
c   % Sp. (4)
c   % E(B-V) (5)
c    % dist. (6)
c   % T_01 (7)
c   %  Ref. (8)
c   % archive exp. nr. (9)
}
\tabletypesize{\small}
\rotate
\tablecolumns{9}
\tablewidth{0pt}
\tablecaption{Properties of the Sightlines\label{sightlines_table}}
\tablehead{
\colhead{Target} & \multicolumn{2}{c}{Galactic Coordinates (deg.)} & 
\colhead{Spectral}&\colhead{}&\colhead{Distance}\tablenotemark{a}&
\colhead{H$_2~T_{01}$\tablenotemark{b}}&\colhead{}&\colhead{Archive 
Exposure}\\
\cline{2-3}
\colhead{Star} & \colhead{$\ell$} & \colhead{$b$}&\colhead{Type}&
\colhead{E($B-V$)\tablenotemark{a}}&\colhead{(kpc)}&\colhead{(K)}&
\colhead{Ref.\tablenotemark{c}}&\colhead{Root Name(s)}\\
\colhead{(1)}& \colhead{(2)}& \colhead{(3)}& \colhead{(4)}&
 \colhead{(5)}& \colhead{(6)}& \colhead{(7)}& \colhead{(8)}& \colhead{(9)}
}
\startdata
CPD$$-$59\arcdeg\,$2603&287.590&$-$0.687&O5$\,$V((f))&0.36&3.5&77&6&
O40P01D6Q\\
&&&&&&&&O4QX03010$-$30\\
HD$\,$108&117.928&1.250&O6pe&0.42&3.8&\nodata&&O5LH01010$-$80\\
HD$\,$1383&119.019&$-$0.893&B1$\,$II&0.37&2.9&\nodata&&O5C07C010\\
HD$\,$3827&120.788&$-$23.226&B0.7$\,$Vn&0.05&1.8&\nodata&&O54309010$-$30\\
&&&&&&&&O54359010$-$30\\
HD$\,$15137&137.462&$-$7.577&O9.5$\,$II$-$IIIn&0.24&3.5&104&7&O5LH02010$-$80\\
HD$\,$23478&160.765&$-$17.418&B3$\,$IV&0.20&0.47&55&7&O6LJ01020\\
HD$\,$24190&160.389&$-$15.184&B2$\,$Vn&0.23&0.82&66&7&O6LJ02020\\
HD$\,$24534 (X Per)&163.083&$-$17.137&O9.5$\,$III&0.31&2.1&57&6&O66P02010\\
&&&&&&&&O64813010$-$20\\
&&&&&&&&O66P01010$-$20\\
HD$\,$27778 (62 Tau)&172.764&$-$17.393&B3$\,$V&0.34&0.23&55&2&O59S01010$-$20\\
HD$\,$32040&196.071&$-$22.605&B9$\,$Vn&0.00&0.16&\nodata&&O56L04010$-$30\\
&&&&&&&&O8MM02010$-$30\\
HD$\,$36408&188.498&$-$8.885&B7$\,$IV&0.11&0.19&\nodata&&O8MM04020$-$30\\
HD$\,$37021 ($\theta^1$~Ori~B)
&209.007&$-$19.384&B3$\,$V&0.42&0.56&\nodata&&O59S02010\\
HD$\,$37061 
($\nu$~Ori)&208.926&$-$19.274&B0.5$\,$V&0.44&0.64&\nodata&&O59S03010\\
HD$\,$37903&206.853&$-$16.538&B1.5$\,$V&0.29&0.83&68&8&O59S04010\\
HD$\,$40893&180.086&4.336&B0$\,$IV&0.31&3.1&78&8&O8NA02010$-$20\\
HD$\,$43818 (11~Gem)&188.489&3.874&B0$\,$II&0.45&1.9&\nodata&&O5C07I010\\
HD$\,$44173&199.002&$-$1.316&B5$\,$III&0.05&0.52&\nodata&&O5C020010\\
HD$\,$52266&219.133&$-$0.680&O9.5$\,$IVn&0.22&1.8&\nodata&&O5C027010\\
HD$\,$69106&254.519&$-$1.331&B0.5$\,$IVnn&0.14&1.5&80&10&O5LH03010$-$50\\
HD$\,$71634&273.326&$-$11.524&B7$\,$IV&0.09&0.32&\nodata&&O5C090010\\
HD$\,$72754 (FY Vel)&266.828&$-$5.815&B2$\,$I:pe&0.31&3.9&\nodata&&O5C03E010\\
HD$\,$75309&265.857&$-$1.900&B1$\,$IIp&0.18&2.9&65&3&O5C05B010\\
HD$\,$79186 (GX Vel)&267.366&2.252&B5$\,$Ia&0.23&1.9&\nodata&&O5C092010\\
HD$\,$88115&285.317&$-$5.530&B1.5$\,$IIn&0.12&3.7&145&3&O54305010$-$60\\
HD$\,$91824&285.698&0.067&O7$\,$V&0.22&3.0&61&6&O5C095010\\
HD$\,$91983&285.877&0.053&B1$\,$III&0.14&3.0&61&7&O5C08N010\\
HD$\,$93205&287.568&$-$0.706&O3$\,$Vf+&0.34&3.3&105&6&O4QX01010$-$40\\
HD$\,$93222&287.738&$-$1.016&O7$\,$IIIf&0.32&3.6&77&6&O4QX02010$-$40\\
HD$\,$93843&288.243&$-$0.902&O5$\,$IIIf&0.24&3.5&107&10&O5LH04010$-$40\\
HD$\,$94454&295.693&$-$14.725&B8$\,$III&0.19&0.30&74&7&O6LJ0H010\\
HD$\,$94493&289.016&$-$1.177&B1$\,$Ib&0.15&3.4&\nodata&&O54306010$-$20\\
HD$\,$99857&294.779&$-$4.940&B0.5$\,$Ib&0.27&3.5&83&10&O54301010$-$60\\
&&&&&&&&O54301020\\
&&&&&&&&O54301030\\
&&&&&&&&O54301040\\
&&&&&&&&O54301050\\
&&&&&&&&O54301060\\
HD$\,$99872&296.692&$-$10.617&B3$\,$V&0.29&0.24&66&7&O6LJ0I020\\
HD$\,$102065&300.027&$-$17.996&B2$\,$V&0.28&0.18&59&6&O4O001010$-$30\\
HD$\,$103779&296.848&$-$1.023&B0.5$\,$Iab&0.17&4.3&86&6&O54302010$-$20\\
HD$\,$104705 (DF Cru)&297.456&$-$0.336&B0$\,$Ib&0.17&5.0&92&6&O57R01010,~30\\
HD$\,$106343 (DL 
Cru)&298.933&$-$1.825&B1.5$\,$Ia&0.23&3.3&\nodata&&O54310010$-$20\\
HD$\,$108002&300.158&$-$2.482&B2$\,$Ia/Iab&0.18&4.2&77&7&O6LJ08020\\
HD$\,$108639&300.218&1.950&B0.2$\,$III&0.26&2.4&88&7&O6LJ0A020\\
HD$\,$109399&301.716&$-$9.883&B0.7$\,$II&0.19&2.9&\nodata&&O54303010$-$20\\
HD$\,$111934 (BU Cru)&303.204&2.514&B1.5$\,$Ib&0.32&2.3&\nodata&&O5C03N010\\
HD$\,$112999&304.176&2.176&B6$\,$Vn&0.17&0.45&96&7&O6LJ0C010$-$20\\
HD$\,$114886&305.522&$-$0.826&O9$\,$IIIn&0.32&1.8&92&7&O6LJ0D020\\
HD$\,$115071&305.766&0.153&B0.5$\,$Vn&0.40&2.7&71&7&O6LJ0E010$-$20\\
HD$\,$115455&306.063&0.216&O7.5$\,$III&0.40&2.6&81&7&O6LJ0F010$-$20\\
HD$\,$116781&307.053&$-$0.065&B0$\,$IIIne&0.31&2.2&\nodata&&O5LH05010$-$40\\
HD$\,$116852&304.884&$-$16.131&O9$\,$III&0.14&4.5&70&6&O5C01C010\\
&&&&&&&&O63571010\\
&&&&&&&&O8NA03010$-$20\\
HD$\,$120086&329.611&57.505&B2$\,$V&0.04&0.99&\nodata&&O5LH06010$-$50\\
HD$\,$121968&333.976&55.840&B1$\,$V&0.11&3.1&38&6&O57R02010$-$20\\
HD$\,$122879&312.264&1.791&B0$\,$Ia&0.29&3.3&90&7&O5C037010\\
&&&&&&&&O5LH07010$-$40\\
&&&&&&&&O6LZ57010\\
HD$\,$124314&312.667&$-$0.425&O6$\,$Vnf&0.43&1.4&74&7&O54307010$-$20\\
HD$\,$140037&340.151&18.042&B5$\,$III&0.08&0.77&\nodata&&O6LJ04010\\
HD$\,$142315&348.981&23.300&B9$\,$V&0.10&0.15&\nodata&&O5C03Y010\\
HD$\,$142763&31.616&46.960&B8$\,$III&0.01&0.28&\nodata&&O5C040010\\
HD$\,$144965&339.044&8.418&B2$\,$Vne&0.27&0.51&70&7&O6LJ05010\\
HD$\,$147683&344.858&10.089&B4$\,$V+B4$\,$V&0.28&0.37&58&7&O6LJ06020\\
HD$\,$147888 ($\rho$~Oph~D)&353.648&17.710&B3$\,$V&0.42&0.12&44&7&O59S05010\\
HD$\,$148594&350.930&13.940&B8$\,$Vnn&0.18&0.19&\nodata&&O5C04A010\\
HD$\,$148937&336.368&$-$0.218&O6.5$\,$I&0.55&2.2&\nodata&&O6F301010$-$20\\
HD$\,$152590&344.842&1.830&O7$\,$V&0.37&3.6&64&7&O5C08P010\\
&&&&&&&&O8NA04010$-$20\\
HD$\,$156110&70.996&35.713&B3$\,$Vn&0.03&0.62&\nodata&&O5C01K010\\
HD$\,$157857&12.972&13.311&O6.5$\,$IIIf&0.37&3.1&86&7&O5C04D010\\
HD$\,$165246&6.400&$-$1.562&O8$\,$Vn&0.33&1.9&\nodata&&O8NA05010$-$20\\
HD$\,$175360&12.531&$-$11.289&B6$\,$III&0.12&0.24&\nodata&&O5C047010\\
HD$\,$177989&17.814&$-$11.881&B0$\,$III&0.11&6.0&52&6&O57R04010$-$20\\
&&&&&&&&O57R03010$-$20\\
HD$\,$185418&53.604&$-$2.171&B0.5$\,$V&0.38&1.2&105&6&O5C01Q010\\
HD$\,$192639&74.903&1.479&O7$\,$Ibf&0.56&2.1&98&2&O5C08T010\\
HD$\,$195965&85.707&4.995&B0$\,$V&0.19&1.1&91&7&O6BG01010$-$20\\
HD$\,$198478 (55 Cyg)&85.755&1.490&B3$\,$Ia&0.43&1.3&\nodata&&O5C06J010\\
HD$\,$198781&99.946&12.614&B0.5$\,$V&0.26&0.69&65&7&O5C049010\\
HD$\,$201345&78.438&$-$9.544&O9$\,$V&0.14&2.2&147&6&O5C050010\\
&&&&&&&&O6359P010\\
HD$\,$202347&88.225&$-$2.077&B1.5$\,$V&0.11&0.95&116&3&O5G301010,~40$-$50\\
HD$\,$203374&100.514&8.622&B2$\,$Vn&0.43&0.34&87&10&O5LH08010$-$60\\
HD$\,$203532&309.461&$-$31.739&B3$\,$IV&0.24&0.22&49&6&O5C01S010\\
HD$\,$206267&99.292&3.738&O6.5$\,$V&0.45&0.86&65&2&O5LH09010$-$40\\
HD$\,$206773&99.802&3.620&B0$\,$V:nnep&0.39&0.82&94&4&O5C04T010\\
HD$\,$207198&103.138&6.995&O9.5$\,$Ib$-$II&0.47&1.3&66&2&O59S06010$-$20\\
HD$\,$208440&104.031&6.439&B1$\,$V&0.27&1.1&75&4&O5C06M010\\
HD$\,$208947&106.550&8.996&B2$\,$V&0.16&0.56&\nodata&&O5LH0A010$-$40\\
HD$\,$209339&104.579&5.869&B0$\,$IV&0.24&1.2&90&4&O5LH0B010$-$40\\
&&&&&&&&O6LZ92010\\
HD$\,$210809&99.849&$-$3.130&O9$\,$Iab&0.28&4.3&87&7&O5C01V010\\
HD$\,$210839 
($\lambda$~Cep)&103.829&2.611&O6$\,$Infp&0.49&1.1&72&2&O54304010$-$20\\
HD$\,$212791&101.644&$-$4.303&B3ne&0.18&0.62&\nodata&&O5C04Q010\\
HD$\,$218915&108.064&$-$6.893&O9.5$\,$Iabe&0.21&5.0&86&6&O57R05010,~30\\
HD$\,$219188&83.031&$-$50.172&B0.5$\,$IIIn&0.09&2.1&103&1&O6E701010\\
&&&&&&&&O8DP01010\\
&&&&&&&&O8SW01010\\
HD$\,$220057&112.131&0.210&B3$\,$IV&0.17&0.77&65&7&O5C01X010\\
HD$\,$224151&115.438&$-$4.644&B0.5$\,$II$-$III&0.34&1.3&252&10&
O54308010$-$20\\
HDE$\,$232522&130.701&$-$6.715&B1$\,$II&0.14&6.1&\nodata&&O5C08J010\\
HDE$\,$303308&287.595&$-$0.613&O3$\,$Vf&0.33&3.8&86&6&O4QX04010$-$40\\
\enddata
\tablenotetext{a}{$B-V$ color excesses and distances to the stars were either 
taken from listings of the same stars in Bowen et al (2008) or Jenkins
(2009), or else they were computed by using the same procedures that they
invoked.} \tablenotetext{b}{The molecular hydrogen rotational temperature
from $J=0$ to 1 that was adopted as an indicator for the kinetic temperature of
the intervening gas.}
\tablenotetext{c}{ Reference for the source of the $T_{01}$ value given in the 
previous column: (1) Savage et al. (1977); (2)  Rachford et al. 
 (2002); (3)  Andr\'e et al. (2003); (4)  Pan et al. (2005); 
(5) Lee et al. (2007); (6) Burgh et al. (2007); (7)  Sheffer et al. 
 (2008); (8) Rachford et al. (2009); (9) Burgh et al. 
 (2010); (10) ``J.~M.~Shull (2009) in preparation'' listed in Burgh et al. 
 (2010); (11) Jensen et al. (2010).}

\end{deluxetable}
\clearpage
\placefigure{phase_diag}
\clearpage

\section{C~I Results}\label{results}

Descriptions of our analysis and the mathematical details of the interpretation of 
the C~I absorption multiplets covered by {\it HST\/} were presented by JT01.  
Except for some enhancements in technique and the use of more up to date 
atomic data discussed in Appendix~\ref{improvements}, we have implemented 
once again the methods of JT01.

Briefly, after we normalized the intensity profiles to an assumed continuum (that 
usually varies smoothly with wavelength), we converted them to apparent optical 
depths (Savage \& Sembach 1991).  For the average spread in 
radial velocity of the C~I in a typical sightline, the individual lines in any given 
multiplet overlap each other.  This introduces confusion in the interpretation of 
the optical depths.  However, we can unravel this confusion by observing 
different multiplets, because the locations of different transitions with respect to 
each other change, thus allowing one to resolve ambiguous mixtures of opacities.  
JT01 devised a way to construct a system of linear equations that could be solved 
to reveal the apparent column densities $N(\rm{ C~I})$, $N({\rm C~I^*})$, and 
$N({\rm C~I^{**}})$ as a function of velocity.\footnote{The notation adopted 
here is consistent with that of JT01: $N({\rm C~I})$ refers to the column density of 
atomic carbon in its $^3P_0$ ground fine-structure state,  while $N({\rm 
C~I^*})$ and $N({\rm C~I^{**}})$ refer to the column densities of the excited 
$^3P_1$ and $^3P_2$ levels, respectively.  The quantity $N({\rm C~I_{ total}})$ 
equals the sum of the column densities in all three levels.  Strictly speaking, we 
measure {\it apparent\/} column densities [$N_a$ in the notation of Savage \& 
Sembach (1991)], which differ from true 
column densities because the recorded intensities are smoothed by the 
instrumental line spread function.  In the interest of simplicity, we will refer to 
such apparent column densities as simply $N$ and treat them is if they were true 
column densities.  Possible errors in this assumption and our avoidance of
cases where they are large
are discussed in \S\protect\ref{distortions}.} Once this has been done, we 
can evaluate the quantities $f1\equiv N({\rm C~I^*})/N({\rm C~I_{ total}})$  and 
$f2\equiv N({\rm C~I^{**}})/N({\rm C~I_{ total}})$ , which are useful 
representations of the excitation conditions when we want to understand not 
only the physical conditions in any given absorbing region, but also possible 
combinations of contributions from differing regions that overlap each other at a 
particular velocity.

\begin{figure*}[h!]
\plotone{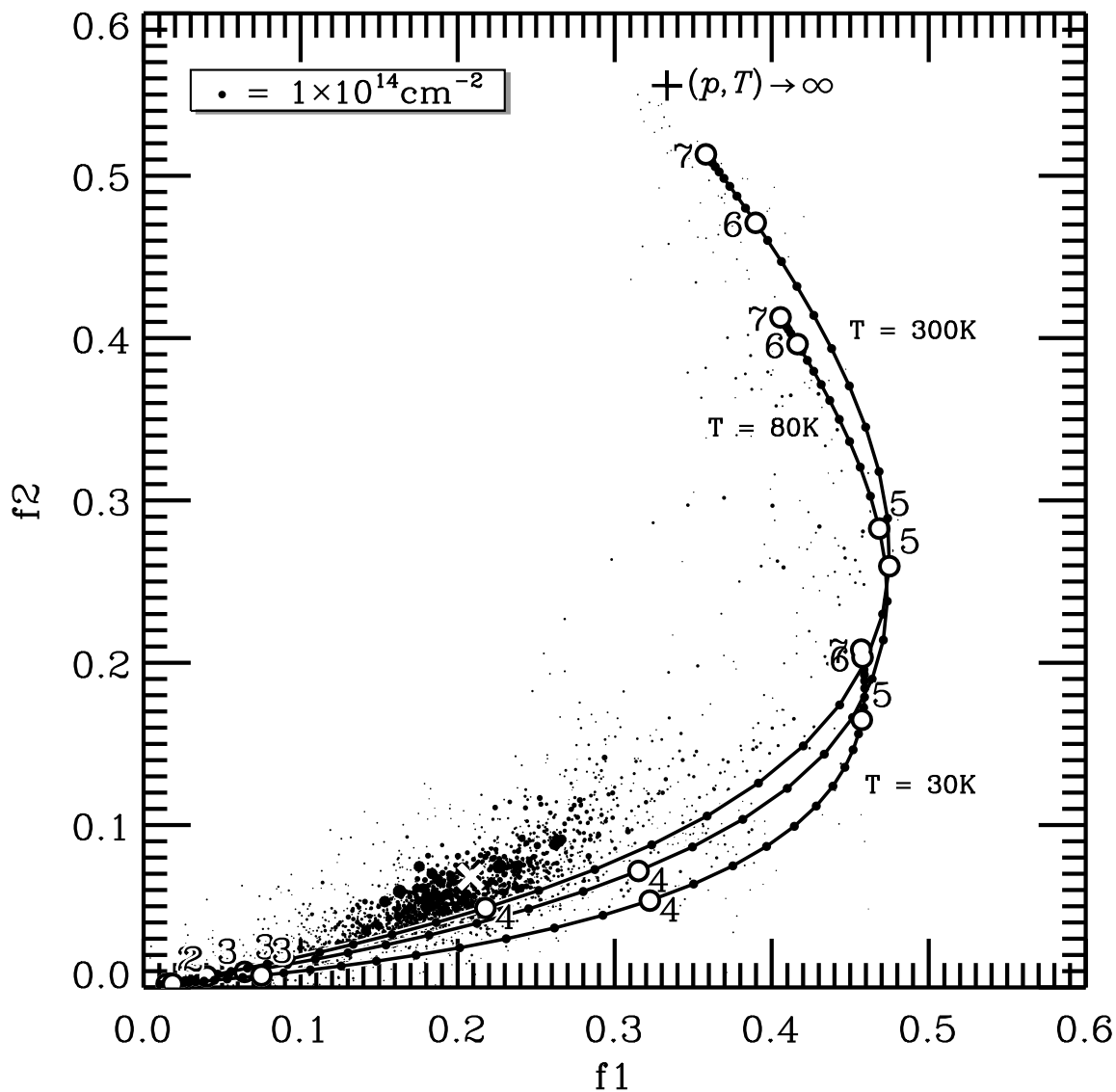}
\caption{Measurements of $f1$ and $f2$ for all velocity bins, each having a width 
of $0.5\,{\rm km~s}^{-1}$, for the 89  stars in the survey that had uncertainties 
$\sigma(f1)$ and $\sigma(f2)$ less than 0.03.  The area of each dot is 
proportional to the respective value of $N({\rm C~I_{total}})$, with a 
normalization in size as shown in the box at the upper left portion of the plot.  The 
white $\times$ located at $f1=0.209$, $f2=0.068$ represents the ``center of 
mass'' of all of the dots.  The curves indicate the expected level populations for 
three different temperatures, assuming the gas mixture as specified in 
\S\protect\ref{particle_mix}, with different values of $\log (p/k)$ indicated with 
dots (adjacent dots represent differences of 0.1~dex).  The large open circles on 
the curves indicate integer values of  $\log (p/k)$ with accompanying numbers to 
indicate their values.  Populations that are proportional to the degeneracies of the 
levels are indicated by the + sign labeled ``$(p,\,T)\rightarrow\infty$.'' 
\label{all_v_f1f2}}
\end{figure*}

As explained originally by Jenkins \& Shaya (1979) 
and once again by JT01, the balance of collisional excitations (and de-excitations) 
against the spontaneous radiative decay of the excited levels establishes an 
equilibrium value for the level populations that depends on the local density and 
temperature (and to a much lesser extent, the composition of the gas).  As the 
densities increase at any given temperature, the locations of points on a diagram 
of $f1$ vs. $f2$ trace an upward arching curve (see Fig.~\ref{all_v_f1f2}) that 
stretches from the origin for low densities to a point at very high densities that 
approaches a Boltzmann distribution for the levels at the temperature in question.

When the absorptions from two or more regions are superposed at a single 
velocity, the outcome for $f1$ and $f2$ is at the ``center of mass'' for the values 
that apply to the individual contributors, with respective weights equal to their 
values of $N({\rm C~I_{total}})$.  This outcome is not without some ambiguity, 
since various combinations of conditions in any ensemble of different clouds can 
produce the same result.  We will address this issue later in \S\ref{admixtures} 
when we make a simplifying assumption about such mixtures, and in 
\S\ref{pileup} we will discuss the consequences of possible averaging effects that 
are difficult to recognize.  Some additional complexity emerges when one 
considers the effects of optical pumping by starlight photons, which we will cover 
in \S\ref{derivations}.

Figure~\ref{all_v_f1f2} shows the outcome for all of our measurements of ($f1$, 
$f2$) at each velocity interval that showed acceptable results, and for every star 
in the survey.  The area of each dot in this diagram is proportional to $N({\rm 
C~I_{total}})$.  It is clear that practically all of the points fall above the equilibrium 
calculations for $f1$ and $f2$, even for a 300$\,$K temperature that is well 
above the nominal values for the CNM.  From this we conclude that either there is 
always a mixture of two or more regions with vastly differing conditions for every 
velocity channel or that for some reason(s) the curves are incorrect or 
inappropriate (we will touch upon this issue later in 
\S\ref{misleading_conclusion}).  A generalized picture of how we interpret some 
plausible admixtures will be presented in \S\ref{admixtures}.
\placefigure{all_v_f1f2}

\clearpage
\clearpage

\section{Properties of the C~I-Bearing Gas}\label{properties}

\subsection{Corrections for the Ionization of 
Carbon}\label{ionization_corrections}

It is important to realize that C~I in the gas that we observe is a minor constituent, 
since the ionization potential of neutral carbon (11.26$\,$eV) is below that of 
neutral hydrogen (13.6$\,$eV).  Most of the carbon atoms in H~I regions are 
singly ionized.  The relative proportion of carbon in the neutral form can vary by 
enormous factors according to the local density of the gas and the strength of the 
local radiation field that is responsible for ionizing the atoms.  As we attempt to 
derive an evenhanded picture of the pressure distribution for the general neutral 
gas, rather than just a value that is weighted in proportion to the amount of C~I 
that is present, we must devise a means for asssessing how much C~II 
accompanies the C~I.  In essence, we use C~II as an indicator for the total amount 
of the neutral material.

Direct measures of $N({\rm C~II})$ are very difficult to carry out.  The only 
available transition in the wavelength bands covered by our survey is the one at 
1334.53$\,$\AA.  This line is strongly saturated, and the only way to measure 
$N({\rm C~II})$ with this feature is by sensing the strength of its damping wings 
 (Sofia et al. 2011).  However, this measure applies to C~II at all 
velocities, rather than at the velocities where we are able to make use of 
information from C~I.  While there exists a very weak intersystem transition at 
2325.40$\,$\AA, this feature is outside the wavelength coverage of most of our 
observations and also requires a very high signal-to-noise ratio for a reliable 
detection (Sofia et al. 2004).

To overcome our inability to measure directly $N({\rm C~II})$ as a function of 
velocity, we instead used O~I as a proxy for C~II.   The very weak O~I intersystem 
line at 1356$\,$\AA\ is ideal for tracing all but the smallest column densities of 
material per unit velocity.  For velocity intervals over which the C~I absorptions 
could be measured reliably, we found that only on very rare occasions was the 
O~I line too weak to measure.  For such instances we had to use the weakest line 
of S~II at 1250$\,$\AA\ as a substitute for O~I.\footnote{ Over very restricted 
velocity intervals there was an intermediate range of column densities per unit 
velocity where the O~I line was too weak to observe (apparent optical depth 
$\tau_{\rm a}<0.05$) and the S~II line was badly saturated ($\tau_{\rm a}>2.5$).  
In this range, we adopted a geometric mean as an approximate compromise 
between the respective upper and lower limits.}  For deriving $N$(C~II), we 
assumed that C, O and S were depleted below their respective protosolar 
abundances (Lodders 2003) by amounts equal to $-0.162$~dex, 
$-0.123$~dex and $-0.275$~dex, respectively, which corresponds to a moderate 
depletion strength ($F_*=0.5$) in the generalized representation of Jenkins 
 (2009).  The assumed abundance of O relative to C could 
be in error by about 0.05$\,$dex if actual the depletion strength is either 
$F_*=0.0$ or 1.0 instead of 0.5.  Also, a few new determinations of $N({\rm 
C~II})$ by Sofia et al. (2011) based on fitting the damping 
wings of the strong line at 1334.53$\,$\AA\ instead of using the weak intersystem 
line suggest that the abundances of C in the ISM may be about 0.3$\,$dex lower 
than stated above, but this change would create a uniform offset that would 
apply to all of our cases.

One can imagine that some of the C~II-bearing gas may be situated in fully ionized 
regions intersected by our sight lines, especially if most of the ionizing radiation in 
the ISM is below the ionization potential of singly ionized carbon (24.38$\,$eV).  
At velocities where we rely on O~I as a proxy, we are confident that our estimate 
for the amount of C~II applies only to neutral gas, since H~II regions are devoid of 
O~I because the ionizations of O and H are strongly coupled by a charge exchange 
reaction with a large rate constant (Field \& Steigman 1971; Chambaud et al. 1980;
Stancil et al. 1999).  The same is not true for S~II; its behavior should be similar to
that of C~II (singly ionized S and C have ionization potentials within 1$\,$eV of each
other).  However, we find that the continuity over velocity between the O~I and S~II
profiles is usually good, which argues against the existence of much contamination
from H~II regions.

\subsection{Kinetic Temperature of the C~I-bearing Gas}\label{T}

The relationship between the thermal pressure and the fine-structure excitations has
a weak dependence on temperature.  Thus, even though this effect is small, we must
still try to reduce the ambiguities in the measurement of $p$ when either
$n$(H) or $T$ is unknown.  Fortunately, we can make use of the fact that 
molecular hydrogen usually accompanies C~I and thus we can utilize 
measurements of the $J=0$ to 1 rotation temperature of H$_2$ to indicate the 
most probable kinetic temperature of the material for which we are making 
pressure measurements.  Columns (7) and (8) of Table~\ref{sightlines_table} 
show the outcomes of such rotation temperature measurements, $T_{01}$, along 
with the sources in the literature where such measurements were listed.  In cases 
where $T_{01}$ was not measured, we had no alternative but to adopt an 
arbitrary number $T=80\,$K, which is  close to the median value of all of the 
measurements.  However, variations in $T_{01}$ from one sightline to the next 
have {\it rms\/} deviations of only 30$\,$K, and deviations of this magnitude alter 
the outcome for $\log (p/k)$ by only about 0.06~dex.

We acknowledge the presence of an unavoidable limitation that $T_{01}$ shows 
simply an average over all velocities, and it is weighted in proportion to the local 
density of hydrogen molecules, $n({\rm H}_2)$, instead of $n({\rm C~I_{total}})$.  
In some circumstances variations in temperature across different regions along 
our sightlines might compromise the accuracy of our results, but this is probably 
not a very important effect.

\subsection{The Particle Mix}\label{particle_mix}

The composition of the gas has a small, but nonnegligible effect on the expected 
outcomes for $f1$ and $f2$.  For instance, for $ 2.8 \lesssim \log (p/k)\lesssim 
3.8$ the inferred pressure for a given ($f1$, $f2$) for pure atomic hydrogen is 
about 0.1~dex lower than for the equivalent number density of pure H$_2$ with 
$T_{01}=80\,$K.  Thus, in order to minimize the error in the interpretation of the 
measurements, it is good to adopt an estimate for the most probable mix of gas 
constituents.  Any deviations in the true conditions from whatever we adopt for 
the molecular fraction, $f({\rm H}_2)=2n({\rm H}_2)/[2n({\rm H}_2)+n({\rm 
H~I})]$, will result in an error for $\log (p/k)$ of less than 0.1~dex.

We estimate that approximately half of the material in our lines of sight arise from 
the WNM, which is free of H$_2$, while the remaining half (CNM) that we can see 
with C~I has an appreciable molecular content.  If we assume that the CNM has 
$f({\rm H}_2)=0.60$, then an overall value  $f({\rm H}_2)=0.42$ would apply to 
the entire sightline.  The latter value is very close to the median outcome for 
$f({\rm H}_2)$ found by Rachford et al. (2009) in their {\it FUSE\/} 
survey of sightlines similar to the ones in the present study.  We therefore adopt 
the assumption that $f({\rm H}_2)=0.60$ in our C~I-bearing gas, and the mix of 
ortho- and para-H$_2$ is governed by the determination of $T_{01}$ (which is 
set to 80$\,$K if unknown).   Another constituent is helium, whose ratio to 
hydrogen in atoms and molecules is assumed to be the protosolar value 0.094 
given by Lodders (2003).   In their normal concentrations 
in the CNM, electrons and protons have a negligible influence on $f1$ and $f2$.

\subsection{Local Starlight Intensity}\label{starlight}

In \S\ref{ionization_corrections} we explained how we derive the amount of C~II 
that accompanies the C~I.  This determination at any particular velocity in a given 
sightline has two applications.  First, as we indicated earlier, it represents our best 
estimate for the total amount of neutral material associated with the C~I.  Second, 
the ratio of C~II to C~I allows us to estimate the local radiation density, which in 
turn will be useful for applying a correction to the equilibrium $f1$ and $f2$ 
values that allows for effects of optical pumping of the fine-structure levels (see 
\S\ref{initial_estimates} and \S\ref{pumping} for details).  As we will show later in 
\S\ref{basic_dist} and \S\ref{origins}, the radiation density outcomes are by 
themselves of special interest in our overall outlook on the distribution of thermal 
pressures.

The ionization balance of carbon atoms in the neutral ISM is given by the relation
\begin{eqnarray}\label{C_ionization}
n({\rm C~I_{total}})(I/I_0)\Gamma_0({\rm C~I})&=
n({\rm C~II})[\alpha_e({\rm C~II},T)n(e)&\nonumber\\
&+\alpha_g({\rm C~II},n(e),I,T)n({\rm H})]~,&
\end{eqnarray}
where the photoionization rate $\Gamma_0({\rm C~I})=2.0\times 10^{-10}{\rm 
s}^{-1}$  (Weingartner \& Draine 2001a) if the radiation field density $I$ 
is equal to  a value $I_0$ specified by Mathis et al. (1983) for the average
intensity of ultraviolet starlight in our part of the Galaxy, $\alpha_e({\rm C~II},
T)$ is the radiative plus dielectronic recombination coefficient of C~II with free
electrons as a function of temperature $T$  (Shull \& Van Steenberg 1982), and
$\alpha_g({\rm C~II},n(e),I,T)$ is the C~II recombination rate due to collisions
with dust grains (and subsequent transfer of an electron) normalized to the local
hydrogen density  (Weingartner \& Draine 2001a).\footnote{In the notation of 
Weingartner \& Draine (2001a), this electron transfer rate from grains is 
expressed as $\alpha_g({\rm C}^+,\psi,T)$, where $\psi=GT^{\onehalf} n(e)^{-1}$ 
and $G=1.13$ for the interstellar radiation field of Mathis et al. (1983).} To
solve for $n(e)$, we assume that free electrons are created by both the
photoionization of some heavy elements, amounting to $2\times 10^{-4}n(H)$,
supplemented by electrons liberated from the cosmic-ray ionization of hydrogen at
a rate $\zeta_{\rm CR}=2\times 10^{-16}{\rm s}^{-1}$  (Indriolo et al. 2007;
Neufeld et al. 2010).\footnote{For the column densities of hydrogen considered
here, the average ionization from x-rays is almost negligible by comparison
(Wolfire et al. 1995).} In order to estimate the density of electrons created by
these cosmic-ray ionizations, one must calculate the balance between the creation
of free protons with a density $n(p)$ against their recombination with free
electrons and also electrons on dust grains using an equation analogous to
Eq.~\ref{C_ionization},
\begin{eqnarray}\label{H_ionization}
\zeta_{\rm CR}n({\rm H})=n(p)[\alpha_e({\rm H~II},T)n(e)&&~~~~~~~~~\nonumber\\
+\alpha_g({\rm 
H~II},n(e),I,T)n({\rm H})]~.~~~~~~~~~~
\end{eqnarray} 
As with the case for C~II, we obtain a formula for $\alpha_g({\rm H~II},n(e),I,T)$ 
from Weingartner \& Draine  (2001a).

There may be some shortcomings in our simple formulation in 
Eq.~\ref{C_ionization}.  Welty et al. (1999) found inconsistencies in the 
determinations of electron densities using the ratios of neutral and ionized forms 
of different elements, and these problems are not resolved when the grain 
recombination processes are included (Weingartner \& Draine 2001a).  Either the
rates incorporated into Eq.~\ref{C_ionization} are inaccurate or other kinds of 
reactions may be important, such as charge exchange with protons,  or the 
formation and destruction of CO (van Dishoeck \& Black 1988) or other C-bearing
diatomic molecules (Prasad \& Huntress 1980; van Dishoeck \& Black 1986, 1989).

If we use our determination of $N$(C~II) discussed in 
\S\ref{ionization_corrections},  compare it with $N({\rm C~I_{total}})$, and apply 
the equilibrium condition expressed by Eq.~\ref{C_ionization} to determine the 
local starlight density, we should obtain a reasonably accurate result for $I/I_0$ 
provided that there is not a large amount of C~II arising from the WNM at exactly 
the same velocity.  If such an additional contribution is present, we will 
overestimate the radiation density.  Likewise, this overestimate of $N$(C~II) 
accompanying the C~I will give a disproportionately large figure for the total 
amount of neutral material for the particular measurement at hand.

\section{Admixtures of Different Kinds of Gas}\label{admixtures}

It is clear that the 2-dimensional distribution of the points shown in 
Fig.~\ref{all_v_f1f2} represents a complex mixture of high and low pressure gas.  
Some insight on the possible nature of this mixture may be gained by stripping 
away the information about the 2-dimensional scatter of the points and focusing 
on just a composite value of ($f1$, $f2$) for all of the measurements, that is, a 
single ``center of mass'' of all of the points in the diagram.  The location of this 
value is shown by a white $\times$ in the figure.   The question we now ask is 
whether or not some simple, generalized distribution function for pressures can 
reproduce the observed composite ($f1$, $f2$) pair.

\begin{figure*}
\plotone{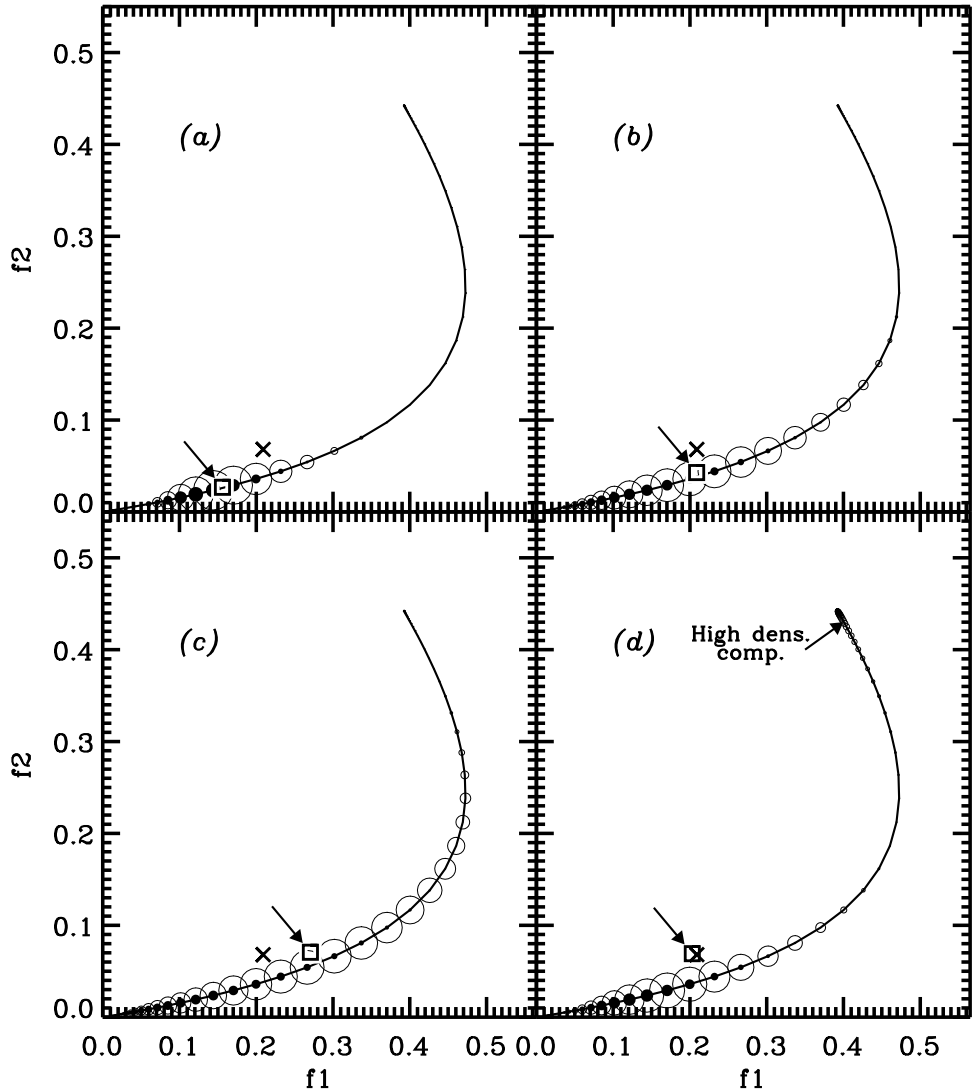}
\caption{A schematic demonstration of the behavior of the composite values of 
$f1$ and $f2$ when the pressure distribution follows a simple lognormal 
behavior.   The distribution is approximated by a discrete collection of H~I packets 
spaced 0.1~dex apart in pressure, illustrated by black dots strung along the 
equilibrium curve, with the area of each dot indicating the amount of H~I.  After 
factoring in the ionization equilibrium equation for carbon atoms, the amounts of 
C~I are strongly biased in favor of higher densities.  The amounts of C~I are 
indicated by the areas enclosed by open circles.  The ``center of mass'' for the C~I 
packets appears at the location of the square with an arrow pointing toward it, 
while the observed composite ($f1$, $f2$) shown in Fig.~\protect\ref{all_v_f1f2} 
is indicated with an $\times$.  Panels ({\it a\/}) through ({\it c\/}) show lognormal 
distributions with 3 successively increasing values of $\sigma$, while panel ({\it 
d\/}) shows the outcome when a small, additional contribution of very high 
density gas is present (very small circles at the top of the equilibrium curve). 
\label{f1f2_lognormal}}
\end{figure*}

A very elementary but plausible pressure relationship to propose is 
a lognormal distribution, which is appropriate for situations where random 
pressure fluctuations arise from turbulence
% *** V\'azquez Semadeni
(V\'azquez Semadeni 1994; Nordlund \& Padoan 1999; Kritsuk et al. 2007) -- see
\S\ref{basic_dist}.  Panels ($a$) through ($c$) in Figure~\ref{f1f2_lognormal} show
reconstructions of the combinations of $f1$ and $f2$ for such a distribution for
three different values for the width of the distribution in $\log (p/k)$ and a
single value for the location of the peak.  One might initially suppose that the
curvature of the high pressure tail in this distribution along an arc that traces
the single-region ($f1$, $f2$) combinations could pull the distribution's
composite ($f1$, $f2$) to a location above the curve at low pressures.  However,
Figure~\ref{f1f2_lognormal} illustrates that this curvature appears to be
insufficient to make our model lognormal composite values rise as high (in $f2$)
as the measured ones without passing beyond the composite measurement of $f1$. 
Instead, it appears that a more complex picture is called for, one that requires the
use of a bimodial distribution of pressures.  The simplest such model is to propose
the existence of a separate, small contribution from material at pressures $\log (p/k)
> 5.5$ and a temperature $T>80\,$K.   Panel ($d$) of the figure shows that such a
contribution can solve our problem with the elevated composite ($f1$, $f2$).   This
high pressure contribution seems to be present in a majority of the sightlines, since
most  of the individual points that are shown in Fig.~\ref{all_v_f1f2} are pulled
above the curves. 
\clearpage
\placefigure{f1f2_lognormal}
\clearpage
\begin{figure*}
\plotone{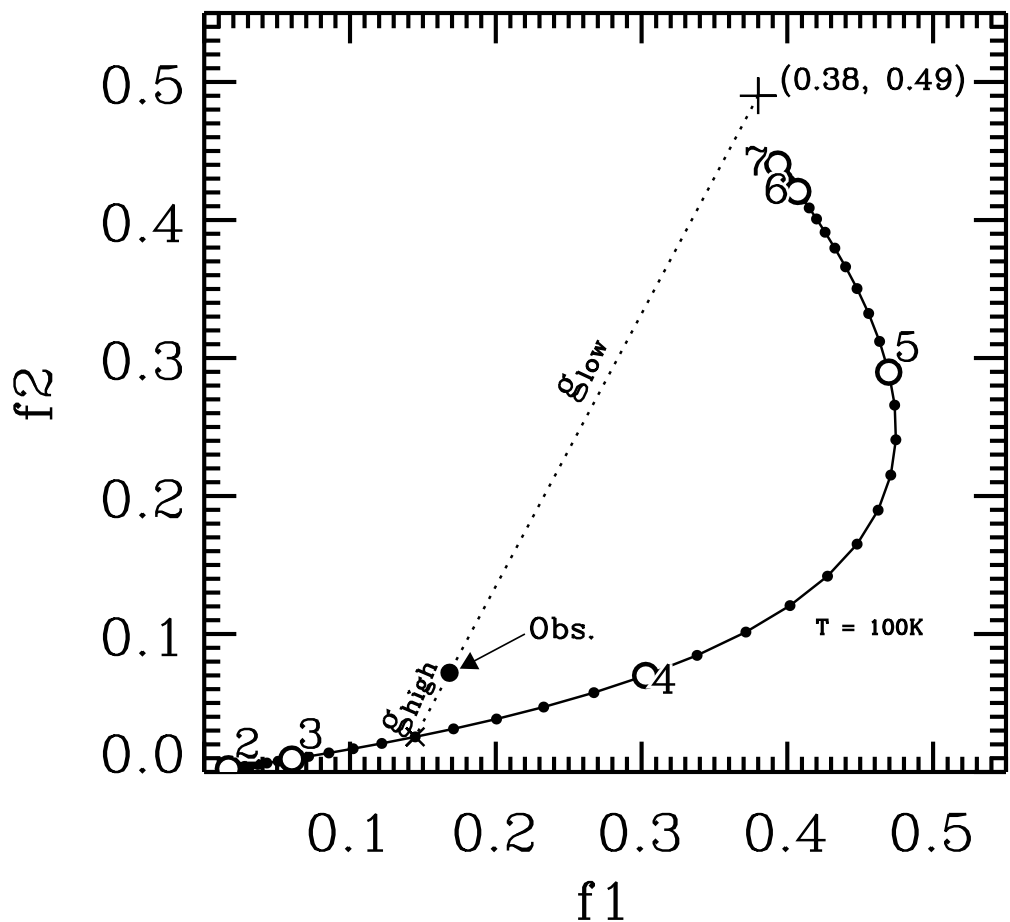}
\caption{A demonstration of how an observed ($f1$, $f2$) at a given velocity is 
decomposed into a superposition of low and high pressure regions.  As in 
Fig.~\protect\ref{all_v_f1f2}, the equilibrium track is marked with a scale in $\log 
(p/k)$, with open circles and accompanying numbers showing integer values of 
this quantity.  An assigned location of (0.38, 0.49) in this diagram applies to the 
high pressure component, and a line that projects from this point through the 
observed ($f1$, $f2$) intersects the equilibrium curve (for a given temperature) 
at a point $\times$ that corresponds to the pressure of the low pressure 
component (in this case $\log (p/k)_{\rm low} = 3.5$).  The relative distances 
along the projection line indicate fractions of C~I in the two components: in this 
depiction, the length of the segment above ``Obs.'' indicates that fraction of gas at 
normal (low) pressures is $g_{\rm low}=0.90$, and the remaining fraction in the 
high pressure component is indicated by the length of the lower line segment, 
yielding $g_{\rm high}=0.10$. \label{fig3}}
\end{figure*}
\clearpage

The exact properties of the distribution of high pressure material is not known, 
but for ($f1$, $f2$) outcomes not far above the lower portions of the equilibrium 
curves, such details do not matter much for the low pressure gas.  We need only 
to know the general vicinity in the upper portion of the diagram where such 
material resides.  Henceforth, we adopt (0.38, 0.49) for a fiducial $f1$ and $f2$ 
for the high pressure component (which corresponds approximately to 
$T=300\,$K, $n({\rm H~I})=4000\,{\rm cm}^{-3}$), and consider that from one 
observation to the next, the relative proportion of this gas is some small fraction 
of the total that can vary from one case to the next.  The consequences of the 
high-pressure reference point differing from reality will be discussed briefly in 
\S\ref{overview}.

\placefigure{fig3}

Figure~\ref{fig3} is a schematic illustration of how we geometrically decompose 
an observed combination of $f1$ and $f2$ into a superposition of the two 
proposed types of gas, one at a very high pressure (but with poorly known 
physical conditions) and the other at a normal, low pressure.   The basic strategy is 
to find where a projection from the assumed high density locus (0.38, 0.49) 
through the observed combination of $f1$ and $f2$ extends to a specific point on 
the equilibrium curve drawn for an appropriate temperature, as defined by 
$T_{01}$, if available.  We regard the pressure that corresponds to this 
intersection point to represent the proper result for the low density gas.  The ratio 
of C~I in the high pressure gas to the overall total is given by the quantity 
$1-g_{\rm low}$ (or simply $g_{\rm high}$) shown in the diagram. 

Two different considerations governed our choice of the fiducial high pressure 
($f1$, $f2$) to be situated near the top of the points shown in 
Fig.~\ref{all_v_f1f2}.  One is based on the plausibility that for most of the 
ordinary lines of sight the amount of this gas (in terms of the total gas, not just 
C~I) is probably a small fraction of all of the gas.  At the highest pressures, C~I is 
more conspicuous (because of the shift in the ionization balance toward the 
neutral form of C), which in turn leads to a smaller quantity for the inferred 
amount of singly-ionized carbon.  The other consideration is a practical one: the 
numerical results for the decompositions into low and high pressure gas are more 
stable when the high pressure point is well removed from the measured values of 
$f1$ and $f2$ -- this becomes important for outcomes at moderately high 
pressures above the main distribution of low pressures.

In the next several sections, we will concentrate on the measurements of the low 
pressure component, using the method just described.  Later, in 
\S\ref{high_pressure_comp}, we will turn our attention to the small amount of 
gas at high pressures and discuss its possible significance in our understanding of 
processes in the diffuse ISM.

\section{Derivations of Thermal Pressures}\label{derivations}

\subsection{Initial Estimates of Conditions}\label{initial_estimates}

For each measurement of ($f1$, $f2$), we apply the construction demonstrated 
in Fig.~\ref{fig3} to determine the quantities $\log (p/k)_{\rm low} $ and $g_{\rm 
low}$.  This determination is based on the initial assumption that the radiation 
field intensity $I(\lambda)$ in the gas in the low pressure regime is equal to the 
average Galactic value $I_0(\lambda)$ (see \S\ref{starlight}).  Any deviation of 
the true intensity from this relationship will result in an error in $\log (p/k)$ 
because an incorrect calculation of the optical pumping rate was applied.  
(Different pumping rates have virtually no effect on the value of $g_{\rm low}$ 
since the equilibrium values of $f1$ and $f2$ simply shift along the curve that 
represents different pressures.)

In order to obtain a more accurate result, we must evaluate how much the true 
intensity $I(\lambda)$ differs from $I_0(\lambda)$, so that our pumping 
correction will be more accurate.  We make the simplifying assumption that the 
relative distribution of intensity over $\lambda$ does not change appreciably 
from one location to the next, but that the overall level of radiation is a free 
parameter that can vary.  We then estimate an initial approximation for the value 
of this parameter, which we call $I/I_0$.  To derive this estimate we make use of 
the ionization equilibrium equation, Eq.~\ref{C_ionization}, to solve for $I/I_0$ 
for gas at any particular velocity increment, after replacing $n$(C~II) by $N$(C~II) 
(as determined according to the method described in 
\S\ref{ionization_corrections}), $n({\rm C~I_{total}})$ by $g_{\rm low}N({\rm 
C~I_{total}})$ and $n$(H) by our initial approximate value of $p/(kT_{01})$.   In 
making the substitution for $n({\rm C~I_{total}})$, we assume that it is safe to 
declare that virtually all of the C~II is associated with the low pressure component 
of C~I, but this condition could be violated if the radiation density experienced by 
the high pressure gas is many orders of magnitude higher than that of the gas at 
ordinary pressures.

\subsection{Convergence to Final Values}\label{convergence}

After evaluating the new radiation intensity level $I/I_0$, we are in a position to 
repeat the calculation of $\log (p/k)_{\rm low}$ using a better representation for 
the shifts in the expected values of $f1$ and $f2$ caused by optical 
pumping.\footnote{The shift in the outcome for $\log(p/k)$ depends not only on 
the strength of the pumping field intensity, but also on $p/k$ itself.  Figure~6 of 
Jenkins \& Shaya (1979) shows the ($f1,f2$) 
equilibrium tracks for $I/I_0=1$ and 10, but in terms of our revision of the 
pumping rates derived in \S\ref{pumping}, these tracks are equivalent to 
present-day values for $I/I_0$ equal to about 1.5 and 15.  Representative values 
for $\log(p/k)$ and $\log(I/I_0)$  are listed for each sight line in Columns (6) and 
(7) of Table~\ref{obs_quantities_table}.}   However, the new value for the 
pressure will have an impact on the density $n$(H) used in the equation for the 
ionization equilibrium, hence the calculation of this balance must be repeated in 
order to derive a modified number for radiation enhancement factor $I/I_0$, one 
that is better suited for a more accurate derivation of the pressure.  We cycle 
through the alternation between pressure and ionization calculations many times 
until the densities and intensities converge to stable solutions.

\section{Overview of Sightlines}\label{overview}

Table~\ref{obs_quantities_table} presents a number of properties of the C~I and 
(inferred) C~II data that we obtained for the sightlines that were suitable for 
study.  The numbers in this table give general indications integrated over velocity; 
they were not used in the analysis of the pressure distribution, which relied on the 
more detailed results that we obtained for the explicit velocity channels.

We show our estimates of the total column densities of C~II in Column~(3) of the 
table.  They compare favorably with the few direct determinations reported in the 
literature (see note $f$); rms deviations between our values and others amount 
to 0.22$\,$dex.  When we consider that the direct measurements of $N$(C~II) 
have quoted errors of order 0.1$\,$dex, the magnitudes of the disagreements 
indicate that our values are probably uncertain by about 0.20$\,$dex.   The 
largest deviations seem to occur for cases where the other determinations are 
higher than ours, which may indicate that we are not registering some C~II in 
ionized gas because we are mostly using O~I as an indicator (see 
\S\ref{ionization_corrections}).

The relative coverages of velocities where $f1$ and $f2$ could be measured, 
weighted by their respective values of $N$(C~II), are given for each sightline in 
Column~(4) of the table.  These quantities vary by large factors from one case to 
the next.  For the entire survey that spanned a total length of 180$\,$kpc, the 
total $N({\rm C~II})=3.8\times 10^{19}{\rm cm}^{-2}$, while that within our 
sampled velocity intervals represents $N({\rm C~II})=2.3\times 10^{19}{\rm 
cm}^{-2}$ (61\%).  It is difficult to gauge the real fraction of the gas for which our 
measurements apply (i.e., CNM vs. CNM + WNM) because some of the WNM 
material can overlap in velocity the CNM that we sampled.  Based on approximate 
interpolations of the velocity profiles of gas that is relatively free of C~I, we 
estimate the fraction to be in the general vicinity of 15\%, which means that on 
average our determinations could be systematically low by about $-0.07\,$dex.

Uncertainties in $g_{\rm low}$ and $\log(p/k)_{\rm low}$ are probably 
dominated by deviations of the real high-pressure conditions from those that 
apply to our adopted location for $f1,f2=(0.38,0.49)$, as we outlined in 
\S\ref{admixtures}.  One can estimate the magnitudes of such deviations by 
examining plausible alternative geometrical constructions of the type depicted in 
Fig.~\ref{fig3}.  For example, if conditions in the high pressure gas are closer to 
$T=100\,$K,  $\log(p/k)=5.3$ and $g_{\rm high}\approx 0.1$ (i.e., twice the 
general average), an apparent value of $\log (p/k)_{\rm low}= 3.5$ may be 
$0.1\,$dex higher than the true value.  The magnitude of this effect scales in 
proportion to $g_{\rm high}$, and it is diminished for higher values of $\log 
(p/k)_{\rm low}$.

In the light of our remarks about various forms of sampling bias in 
\S\ref{observations} (item 1), it should come as no surprise that values of $\log 
(I/I_0)$ shown in Column (7) of the table are all greater than zero.  This is a 
consequence of the CNM being preferentially located in the vicinity of hot stars, 
rather than in random locations in the Galactic disk.  This preference seems to 
overcome the effects of attenuation of starlight by dust.  However, to some 
limited extent our intensity outcomes could be elevated in a systematic fashion by 
the presence of unrelated WNM gas that is at the same velocity as the C~I.  This 
extra gas would mislead us into thinking the carbon atoms in the regions of 
interest are more ionized than in reality.
\placetable{obs_quantities_table}

\clearpage
\begin{deluxetable}{
l   % name
c  % N(C I_total)
c   % N(C II) total
c   % percentage N(C II) covered
c   % < frac low obs.>
c   % < log (p/k)_low>
c   % median log Intens.
}
\tabletypesize{\footnotesize}
\tablecolumns{7}
\tablewidth{0pt}
\tablecaption{Observed and Calculated Quantities over all Velocities in the 
Sightlines\label{obs_quantities_table}}
\tablehead{
\colhead{Target} &\colhead{$\log N({\rm C~I_{total}})$\tablenotemark{a}} &
 \colhead{Calc. $\log$} & \colhead{Percent  C~II} &
\multicolumn{2}{c}{Weighted Averages\tablenotemark{d}} &
\colhead{Median}\\
\cline{5-6}
\colhead{Star} & \colhead{$({\rm cm}^{-2})$} & \colhead{$N({\rm 
C~II})$\tablenotemark{b}$({\rm cm}^{-2})$} &
\colhead{Observed\tablenotemark{c}} & \colhead{$g_{\rm low}$} &
\colhead {$\log (p/k)_{\rm low}$} 
& \colhead{$\log (I/I_0)$\tablenotemark{e}}\\
\colhead{(1)}& \colhead{(2)}& \colhead{(3)}& \colhead{(4)}&
 \colhead{(5)}& \colhead{(6)}& \colhead{(7)}
}
\startdata
    CPD-59D2603\dotfill& $\gtrapprox$14.64&17.87&25.4&0.92&3.63&0.31\\
          HD$\,$108\dotfill & $\gtrapprox$14.98&17.81&64.4&0.95&3.82&0.30\\
         HD$\,$1383\dotfill & $\gtrapprox$14.80&17.84&44.5&0.96&3.49&0.27\\
         HD$\,$3827\dotfill & 13.57$\pm$0.07&17.22&  7.3&0.92&3.51&0.16\\
        HD$\,$15137\dotfill &14.62$\pm$0.01&17.69&74.3&0.98&3.61&0.32\\
        HD$\,$23478\dotfill & $>$14.75&17.35&46.4&0.93&3.62&0.46\\
        HD$\,$24190\dotfill & 14.45$\pm$0.01&17.49&84.5&0.97&3.64&0.62\\
HD$\,$24534 (X Per) \dotfill & $>$14.81&17.49\tablenotemark{f}&
5.0&0.95&4.17&1.16\\
        HD$\,$27778 (62 Tau)\dotfill & $>$14.98&17.48\tablenotemark{f}&
53.4&0.90&3.54&0.49\\
        HD$\,$32040\dotfill & 13.26$\pm$0.03&16.58&18.1&0.97&3.07&0.11\\
        HD$\,$36408\dotfill & $\gtrapprox$14.17&17.33&54.5&0.93&3.82&0.81\\
        HD$\,$37021 ($\theta^1$ Ori B)\dotfill & 
13.64$\pm$0.05&17.78\tablenotemark{f}&
67.5&0.39&3.83&1.83\\
        HD$\,$37061 ($\nu$~Ori)\dotfill & 
13.97$\pm$0.03&17.90\tablenotemark{f}&
86.3&0.59&4.28&1.80\\
        HD$\,$37903\dotfill & 14.24$\pm$0.05&17.75&48.8&0.79&4.61&1.37\\
        HD$\,$40893\dotfill & 14.67$\pm$0.01&17.79&91.7&0.98&3.40&0.37\\
        HD$\,$43818 (11 Gem)\dotfill & 
14.76$\pm$0.01&17.89&91.9&0.98&3.49&0.35\\
        HD$\,$44173\dotfill & 13.66$\pm$0.05&16.98& 4.1&0.83&3.45&0.45\\
        HD$\,$52266\dotfill & 14.34$\pm$0.01&17.62&63.0&0.98&3.39&0.48\\
        HD$\,$69106\dotfill & 14.30$\pm$0.01&17.37&81.1&0.96&3.55&0.45\\
        HD$\,$71634\dotfill & 14.13$\pm$0.02&17.18&68.3&0.89&4.08&0.88\\
        HD$\,$72754 (FY Vel)\dotfill & 
$\gtrapprox$14.40&17.54&37.0&0.89&3.86&0.83\\
        HD$\,$75309\dotfill & 14.39$\pm$0.02&17.51&74.3&0.97&3.41&0.46\\
        HD$\,$79186 (GX Vel)\dotfill & 
$\gtrapprox$14.57&17.76&69.8&0.96&3.33&0.40\\
        HD$\,$88115\dotfill & 14.03$\pm$0.05&17.56&29.4&0.98&3.55&0.51\\
        HD$\,$91824\dotfill & $\gtrapprox$14.45&17.49&63.6&0.93&3.62&0.78\\
        HD$\,$91983\dotfill & 14.54$\pm$0.01&17.55&76.3&0.96&3.53&0.51\\
        HD$\,$93205\dotfill & 14.56$\pm$0.02&17.83&51.5&
\nodata\tablenotemark{g}&\nodata\tablenotemark{g}&0.58\\
        HD$\,$93222\dotfill & 14.36$\pm$0.01&17.82&72.6&0.95&4.41&0.82\\
        HD$\,$93843\dotfill & 14.15$\pm$0.05&17.67&12.8&0.88&4.13&0.66\\
        HD$\,$94454\dotfill & $>$14.29&17.50&30.5&0.93&3.60&0.68\\
        HD$\,$94493\dotfill & 14.26$\pm$0.02&17.51&59.1&0.97&3.53&0.49\\
        HD$\,$99857\dotfill & 14.59$\pm$0.01&17.71&69.9&0.94&3.65&0.55\\
        HD$\,$99872\dotfill & $\gtrapprox$14.41&17.45&79.4&0.94&3.62&0.86\\
       HD$\,$102065\dotfill & 
$\gtrapprox$14.22&17.40&53.9&0.96&3.62&0.61\\
       HD$\,$103779\dotfill & 14.23$\pm$0.03&17.60&44.8&1.00&3.29&0.33\\
       HD$\,$104705 (DF Cru)\dotfill & 
14.25$\pm$0.01&17.58&66.7&0.98&3.48&0.53\\
       HD$\,$106343(DL Cru)\dotfill & 
14.34$\pm$0.01&17.62&58.1&0.96&3.51&0.49\\
       HD$\,$108002\dotfill & 
$\gtrapprox$14.45&17.58&51.3&0.96&3.41&0.34\\
       HD$\,$108639\dotfill & 14.31$\pm$0.01&17.75&76.3&0.96&3.49&0.55\\
       HD$\,$109399\dotfill & 14.27$\pm$0.02&17.60&37.9&0.97&3.69&0.59\\
       HD$\,$111934 (BU Cru)\dotfill & 
$\gtrapprox$14.48&17.70&43.5&0.99&3.71&0.63\\
       HD$\,$112999\dotfill & 14.23$\pm$0.02&17.42&82.7&0.97&3.61&0.53\\
       HD$\,$114886\dotfill & 14.73$\pm$0.01&17.75&66.5&0.95&3.62&0.37\\
       HD$\,$115071\dotfill & 14.69$\pm$0.01&17.87&74.3&0.94&3.64&0.74\\
       HD$\,$115455\dotfill & 14.63$\pm$0.02&17.85&61.3&0.97&3.52&0.58\\
       HD$\,$116781\dotfill & 14.28$\pm$0.02&17.75&32.2&0.98&3.42&0.47\\
       HD$\,$116852\dotfill & 14.15$\pm$0.02&17.45&66.0&0.95&3.72&0.80\\
       HD$\,$120086\dotfill & 13.20$\pm$0.08&16.98& 3.6&0.99&3.72&0.50\\
       HD$\,$121968\dotfill & 13.38$\pm$0.06&16.87&47.0&0.95&3.81&1.27\\
       HD$\,$122879\dotfill & 14.42$\pm$0.01&17.75&76.3&0.99&3.59&0.49\\
       HD$\,$124314\dotfill & 
$\gtrapprox$14.66&17.85&66.0&0.97&3.54&0.58\\
       HD$\,$140037\dotfill & $\gtrapprox$14.03&17.24& 4.8&$\approx 
1.0$&5.19&0.95\\
       HD$\,$142315\dotfill & 13.78$\pm$0.06&17.33& 7.3&0.93&3.38&0.50\\
       HD$\,$142763\dotfill & 13.38$\pm$0.06&17.03&24.9&0.93&3.47&0.70\\
       HD$\,$144965\dotfill & $>$14.28&17.52& 9.4&0.82&3.80&0.76\\
       HD$\,$147683\dotfill & 
$\gtrapprox$14.80&17.64&39.7&0.89&3.89&0.57\\
HD$\,$147888 ($\rho$~Oph)& $>$14.51&17.80\tablenotemark{f}&
26.9&0.68&3.98&1.20\\
       HD$\,$148594\dotfill & 
$\gtrapprox$14.12&17.54&39.3&0.79&4.51&1.53\\
       HD$\,$148937\dotfill & 
$\gtrapprox$14.87&17.96&90.6&0.95&3.84&0.67\\
       HD$\,$152590\dotfill & 14.61$\pm$0.01&17.76\tablenotemark{f}&
77.0&0.87&3.71&0.77\\
       HD$\,$156110\dotfill & 13.88$\pm$0.04&17.08&45.6&0.99&3.83&0.57\\
       HD$\,$157857\dotfill & 14.61$\pm$0.01&17.77&87.2&0.95&3.50&0.46\\
       HD$\,$165246\dotfill & 
$\gtrapprox$14.33&17.73&59.4&0.94&3.58&0.79\\
       HD$\,$175360\dotfill & 13.98$\pm$0.01&17.29&76.6&1.00&3.18&0.39\\
       HD$\,$177989\dotfill & 
$\gtrapprox$14.66&17.44&72.5&0.96&3.59&0.35\\
       HD$\,$185418\dotfill & $>$14.72&17.71&66.8&0.97&3.41&0.23\\
       HD$\,$192639\dotfill & 
$\gtrapprox$14.74&17.85&80.0&0.97&3.68&0.52\\
       HD$\,$195965\dotfill & 
$\gtrapprox$14.48&17.42&60.2&0.96&3.56&0.32\\
       HD$\,$198478 (55 Cyg)\dotfill & 
$\gtrapprox$14.84&17.80&84.5&0.92&3.68&0.43\\
       HD$\,$198781\dotfill & 
$\gtrapprox$14.56&17.57&32.0&0.92&3.49&0.37\\
       HD$\,$201345\dotfill & 13.97$\pm$0.02&17.42&63.4&0.97&3.50&0.32\\
       HD$\,$202347\dotfill & 14.61$\pm$0.01&17.33&79.5&0.95&3.76&0.20\\
       HD$\,$203374\dotfill & 
$\gtrapprox$14.98&17.68&81.1&0.94&3.63&0.31\\
       HD$\,$203532\dotfill & $>$14.65&17.40& 4.2&0.85&4.40&1.26\\
       HD$\,$206267\dotfill & 
$\gtrapprox$15.29&17.85&72.4&0.93&3.64&0.30\\
       HD$\,$206773\dotfill & 
$\gtrapprox$14.70&17.55&71.8&0.96&3.55&0.19\\
       HD$\,$207198\dotfill & $\gtrapprox$15.24&17.81\tablenotemark{f}&
87.6&0.94&3.63&0.20\\
       HD$\,$208440\dotfill & 
$\gtrapprox$14.84&17.68&78.8&0.94&3.66&0.35\\
       HD$\,$208947\dotfill & 
$\gtrapprox$14.63&17.36&53.2&0.97&3.60&0.33\\
       HD$\,$209339\dotfill & 14.76$\pm$0.01&17.61&87.1&0.96&3.69&0.34\\
       HD$\,$210809\dotfill & 
$\gtrapprox$14.70&17.69&36.8&0.94&3.66&0.29\\
       HD$\,$210839 ($\lambda$ Cep)\dotfill & 
$\gtrapprox$14.95&17.79&63.8&0.86&4.16&0.47\\
       HD$\,$212791\dotfill & 14.28$\pm$0.03&17.44&29.9&0.97&3.54&0.38\\
       HD$\,$218915\dotfill & 14.56$\pm$0.01&17.59&68.9&0.97&3.58&0.33\\
       HD$\,$219188\dotfill & 13.92$\pm$0.04&17.08&77.6&0.99&2.97&0.01\\
       HD$\,$220057\dotfill & $>$14.71&17.46&25.6&0.94&3.51&0.35\\
       HD$\,$224151\dotfill & 
$\gtrapprox$14.62&17.80&56.1&0.97&3.80&0.16\\
      HDE$\,$232522\dotfill & 
$\gtrapprox$14.60&17.66&58.5&0.97&3.53&0.35\\
      HDE$\,$303308\dotfill & 14.69$\pm$0.00&17.83&72.9&
\nodata\tablenotemark{g}&\nodata\tablenotemark{g}&0.59\\
\enddata
% The following note is referenced in the text, so if the letter changes, make 
% a corresponding change in the text.
\tablenotetext{a}{Integrated over all velocities where C~I absorption seems to be 
visible (not just over the restricted regions where the lines are strong enough to 
yield good measurements of $f1$ and $f2$).  Sometimes there was evidence that 
unresolved saturations were evident at certain velocities, as indicated by a test 
that is discussed in \S\ref{distortions}.  When this occurred over very limited 
portions of the profile, we indicate a mild inequality with ``$\gtrapprox$.''  When 
a substantial portion of the profile exhibited such behavior, we indicate a more 
severe inequality by ``$>$.''  When errors in the column densities are given, they 
indicate only the quantifiable errors arising from noise or uncertainties in the 
continuum levels.  These errors indicate the relative quality of the measurements, 
but they are not fully realistic because they do not take into account uncertainties 
in our adopted $f$-values or occasional flaws in the MAMA detector used by 
STIS.}
\tablenotetext{b}{The computed amount of C~II at all velocities based on the 
absorption profiles of O~I or S~II; see \S\protect\ref{ionization_corrections} for 
details.  These amounts compare favorably with the observed amounts for a few 
stars in note $f$.}
\tablenotetext{c}{The relative amount of C~II, as represented by its proxy O~I 
(and sometimes S~II), within the velocity interval where determinations of $f1$ 
and $f2$ were good enough to be considered for pressure measurements, 
compared to the amount seen at all velocities, as shown in the previous column.}
\tablenotetext{d}{Calculated according to the following: $\sum [g_{\rm 
low}N({\rm C~I_{total}})] /\sum N({\rm C~I_{total}})$ and $\log \sum [(p/k)_{\rm 
low}N({\rm C~II})] /\sum N({\rm C~II})$.}
\tablenotetext{e}{Our estimate for the local density of UV radiation from starlight 
that is more energetic than the ionization potential of neutral carbon, compared 
to an adopted standard $I_0$ based on a level specified by  Mathis et al. 
 (1983) for the average intensity of ultraviolet starlight 
in the Galactic plane at a Galactocentric distance of 10~kpc.  This estimate is based 
on our evaluation of the ionization equilibrium of C, as expressed in 
Eq.~\protect\ref{C_ionization}.}
% The following note is referenced in the text, so if the letter changes, make 
% a corresponding change in the text.
\tablenotetext{f}{Compare with actual measurements of $\log N$(C~II) using the 
intersystem C~II] line at 2325$\,$\AA: From Sofia et al. (1998) HD$\,$24534:
$17.51~(+0.11,\,-0.16)$.  From Sofia et al.  (2004) HD$\,$2778: $<17.34$; HD$\,
$37021: $17.82~(+0.12,\,-0.18)$; HD$\,$37061: $18.13~(+0.04,\,-0.06)$; 
HD$\,$147888: $18.00~(+0.07,\,-0.09)$; HD$\,$152590: 
$18.21~(+0.08,\,-0.10)$; HD$\,$207198: $17.98~(+0.11,\,-0.14)$.  However, 
recent measurements of the damping wings for the allowed transition at 
1334.53$\,$\AA\ reported by Sofia et al. (2011) indicate 
that these column densities may be too large by a factor of about 2.}
\tablenotetext{g}{Gas within a component at large negative velocities has 
conditions very near the high pressure reference mark.  Hence the projection 
onto the low pressure arc is meaningless.}
\end{deluxetable}

Of particular interest are the characteristic sizes of the regions containing the C~I 
that we are able to study.  Within any velocity bin, we can determine a value for 
the local density of gas particles, composed of atomic hydrogen, helium atoms, 
and hydrogen molecules.  Once again, if we assume that $f({\rm H}_2)=0.6$ (see 
\S\ref{particle_mix}) and ${\rm He/H}=0.09$, it follows that the local density of 
hydrogen nuclei is given by $n({\rm H})=p/(0.79kT)$.  The longitudinal thickness 
occupied by the gas is then equal to the column density of these nuclei, $N({\rm 
H})$, divided by $n$(H).  We obtain $N$(H) by multiplying the amount of carbon, 
measured by the methods outlined in \S\ref{ionization_corrections},  by the 
general expectation for $({\rm H/C})=5040$ in the ISM.  A sum over velocity of all 
of the length segments gives the overall thickness of the C~I-bearing region(s) in 
any particular line of sight.

\begin{figure}[b!]
\epsscale{1.0}
\plotone{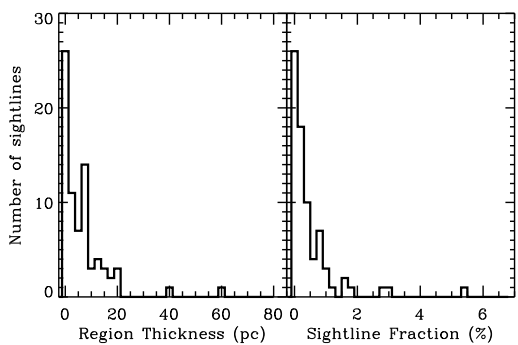}
\caption{Histograms that show ({\it left\/}) the total thicknesses of the regions and 
({\it right\/}) their relative occupation fractions in the sightlines that we are able to 
measure in the survey.\label{filling_factor}}
\end{figure}

\placefigure{filling_factor}

Figure~\ref{filling_factor} shows the distribution of region thicknesses for all of 
the lines of sight in our survey, which generally have dimensions of less than 
20$\,$pc.  The occupation fractions are quite small, generally less than about 2\%.

For the benefit of future investigations that may require more detailed 
information about individual sight lines, Table~\ref{MRT} provides a 
machine-readable summary that lists for each velocity channel the measured 
values of $f1$, $f2$, and $N({\rm C~I}_{\rm total})$, along with the calculated 
values of $N({\rm C~II})$, $g_{\rm low}$, $\log (p/k)_{\rm low}$, and $\log 
(I/I_0)$.

\section{Behavior with Velocity}\label{behavior_velocity}

The radial velocities that we measure in the C~I profiles arise from various 
kinematical phenomena, such as differential velocities caused by rotation or 
density waves in the Galaxy, coherent motions caused by discrete dynamical 
events such as supernova explosions, mass loss from stars, the collision of infalling 
halo gas with material in the Galactic plane, and random motions arising from 
turbulence.    With the exception of differential Galactic rotation, all of these 
effects can transform some of their energy into an increase of the thermal 
pressures.  In their limited sample of only 21 stars, JT01 found elevated pressures 
in gases whose velocities deviated away from the range that was expected for 
differential Galactic rotation.  We now revisit this issue for our present, much 
larger sample of sightlines to further substantiate the evidence for a coupling 
between the thermal pressures and unusual dynamical properties of the gas.

\begin{figure*}
\epsscale{2.2}
\plotone{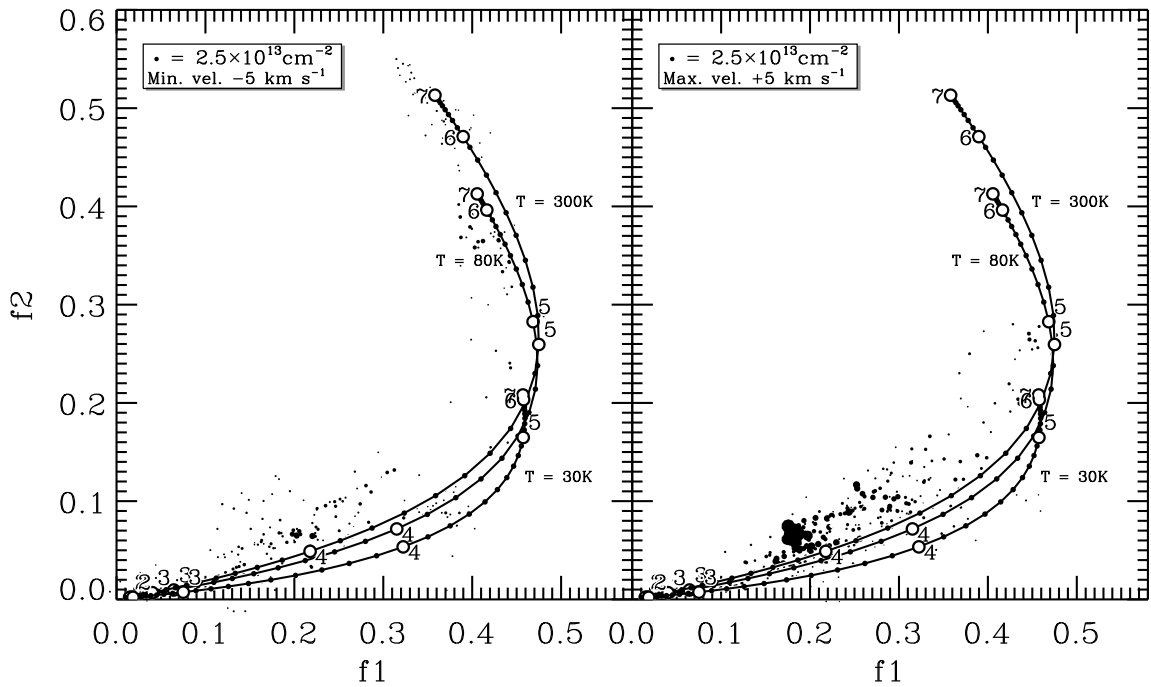}
\caption{Presentations similar to Fig.~\protect\ref{all_v_f1f2}, except that the 
measurements include velocities only below the minimum value permitted by 
differential Galactic rotation, but with an extra margin of $5\,{\rm km~s}^{-1}$, 
i.e., $v<\min(v_{\rm gr})-5\,{\rm km~s}^{-1}$ ({\it left-hand panel\/}) or more 
than $5\,{\rm km~s}^{-1}$ above the maximum permitted velocities, i.e., 
$v>\max(v_{\rm gr})+5\,{\rm km~s}^{-1}$ ({\it right-hand panel\/}).  The dot 
diameters in these diagrams are twice as large as those in 
Fig.~\protect\ref{all_v_f1f2} for a given column density of ${\rm 
C~I_{total}}$.\label{outlier_v_f1f2}}
\end{figure*}

\placefigure{outlier_v_f1f2}

Figure~\ref{outlier_v_f1f2} shows the measurements of $f1$ and $f2$ at 
velocities that are either above or below the respective line-of-sight velocity 
ranges permitted by differential Galactic rotation, assuming that the rotation 
curve is flat at $254\,{\rm km~s}^{-1}$ and the distance to the Galactic center is 
8.4$\,$kpc (Reid et al. 2009).  An extra margin of $5\,{\rm km~s}^{-1}$ is added
to the exclusion zone for permitted velocities, so that we are more certain of 
showing material that genuinely disturbed in some manner.  When we compare 
the results of Fig.~\ref{outlier_v_f1f2} to those shown in Fig.~\ref{all_v_f1f2}, it 
is clear that gases at high velocity do not have nearly as strong a central 
concentration near $f1\approx 0.2$ and $f2\approx 0.07$.  The ``center of 
mass'' ($f1,\,f2$) locations for all of the points shown in Fig.~\ref{outlier_v_f1f2} 
are (0.265, 0.163) for negative velocities and (0.228, 0.078) for positive 
velocities.  By comparison, for all measurements shown in Fig.~\ref{all_v_f1f2}, 
we found the balance point to be at (0.209, 0.068).  These differences are 
principally caused by a greater prominence of a more highly dispersed population 
of points in $f1$ and $f2$, and they should come as no surprise since they 
demonstrate the expected coupling of the dynamics of the gas to the observed 
enhancements in the thermal pressures.
\placetable{MRT}

\clearpage
\begin{deluxetable}{
l   % Target Star
r   % velocity
c   % f1
c   % f1 error
c   % f2
c   % f2 error
c   % N(C I)_total
c   % N(C II)
c   % g_low
c   % log(p/k)_low
c   % log(I/I_0)
}
\tabletypesize{\footnotesize}
\tablecolumns{11}
\tablewidth{0pt}
\tablecaption{Observed and Calculated Quantities in Specific Velocity 
Channels\label{MRT}}
\tablehead{
\colhead{Target} & \colhead{Velocity} & \colhead{$f1$} &
\colhead{$f1$} & \colhead{$f2$} & \colhead{$f2$} &
\colhead{$N({\rm C~I}_{\rm total})$} & \colhead{$N({\rm C~II})$} &
\colhead{$g_{\rm low}$} & \colhead{$\log (p/k)_{\rm low}$} &
\colhead{$\log (I/I_0)$}\\
\colhead{Star} & \colhead{(${\rm km~s}^{-1}$)} & \colhead{} & \colhead{Error} & 
\colhead{} & \colhead{Error} &
\colhead{(cm$^{-2}$)} & \colhead{(cm$^{-2}$)} & 
\colhead{} & \colhead{} & \colhead{}
}
\startdata
CPD-59D2603&   4.0& 0.213& 0.016& 0.068& 0.026& 2.13e+013& 1.34e+016& 0.930& 3.63& 0.37\\
CPD-59D2603 &   4.5& 0.201& 0.013& 0.065& 0.020& 2.81e+013& 1.44e+016& 0.933& 3.60& 0.28\\
CPD-59D2603&   5.0& 0.200& 0.011& 0.065& 0.017& 3.33e+013& 1.56e+016& 0.934& 3.60& 0.25\\
CPD-59D2603&   5.5& 0.207& 0.010& 0.065& 0.015& 3.83e+013& 1.76e+016& 0.936& 3.62& 0.26\\
CPD-59D2603&   6.0& 0.218& 0.009& 0.064& 0.013& 4.46e+013& 2.02e+016& 0.945& 3.67& 0.29\\
CPD-59D2603&   6.5& 0.224& 0.008& 0.065& 0.011& 5.30e+013& 2.46e+016& 0.944& 3.68& 0.31\\
CPD-59D2603&   7.0& 0.227& 0.008& 0.066& 0.009& 6.26e+013& 3.03e+016& 0.943& 3.69& 0.33\\
CPD-59D2603&   7.5& 0.230& 0.008& 0.070& 0.008& 7.17e+013& 3.80e+016& 0.938& 3.70& 0.37\\
CPD-59D2603&   8.0& 0.233& 0.008& 0.074& 0.007& 7.88e+013& 4.63e+016& 0.931& 3.70& 0.41\\
CPD-59D2603&   8.5& 0.222& 0.008& 0.079& 0.007& 8.10e+013& 5.13e+016& 0.910& 3.65& 0.40\\
CPD-59D2603&   9.0& 0.200& 0.007& 0.081& 0.007& 7.65e+013& 5.18e+016& 0.894& 3.56& 0.35\\
CPD-59D2603&   9.5& 0.181& 0.007& 0.081& 0.009& 6.53e+013& 4.85e+016& 0.885& 3.47& 0.32\\
 HD102065\dotfill &   8.5& 0.188& 0.019& 0.071& 0.016& 2.42e+012& 5.98e+014& 0.908& 3.51& 0.09\\
 HD102065\dotfill &   9.0& 0.241& 0.012& 0.072& 0.010& 3.87e+012& 6.54e+014& 0.934& 3.71& 0.14\\
 HD102065\dotfill &   9.5& 0.275& 0.008& 0.067& 0.006& 5.93e+012& 7.52e+014& 0.966& 3.83& 0.14\\
 HD102065\dotfill &  11.0& 0.223& 0.004& 0.050& 0.003& 1.64e+013& 1.56e+016& 0.973& 3.62& 0.61\\
 HD102065\dotfill &  11.5& 0.203& 0.005& 0.048& 0.002& 1.92e+013& 1.99e+016& 0.967& 3.55& 0.59\\
 HD102065\dotfill &  12.0& 0.187& 0.005& 0.046& 0.002& 2.14e+013& 2.40e+016& 0.964& 3.49& 0.57\\
 HD102065\dotfill &  12.5& 0.176& 0.005& 0.043& 0.002& 2.38e+013& 2.83e+016& 0.965& 3.44& 0.54\\
 HD102065\dotfill &  13.0& 0.177& 0.004& 0.043& 0.002& 2.70e+013& 3.13e+016& 0.965& 3.45& 0.55\\
 HD102065\dotfill &  13.5& 0.187& 0.004& 0.046& 0.002& 3.06e+013& 3.67e+016& 0.967& 3.49& 0.59\\
 HD102065\dotfill &  14.0& 0.204& 0.004& 0.051& 0.002& 3.32e+013& 4.35e+016& 0.964& 3.54& 0.66\\
 HD102065\dotfill &  16.0& 0.318& 0.005& 0.099& 0.003& 1.47e+013& 2.83e+016& 0.925& 3.85& 1.06\\
 HD102065\dotfill &  16.5& 0.315& 0.005& 0.101& 0.004& 1.00e+013& 1.87e+016& 0.918& 3.84& 1.03\\
 HD102065\dotfill &  17.0& 0.298& 0.007& 0.091& 0.006& 6.78e+012& 1.05e+016& 0.931& 3.81& 0.95\\
 HD102065\dotfill &  17.5& 0.281& 0.010& 0.074& 0.008& 4.54e+012& 3.97e+015& 0.954& 3.80& 0.74\\
 HD102065\dotfill &  18.0& 0.270& 0.014& 0.059& 0.012& 3.07e+012& 2.74e+015& 0.983& 3.78& 0.72\\
 HD102065\dotfill &  18.5& 0.273& 0.021& 0.053& 0.018& 2.11e+012& 2.55e+015& 0.998& 3.77& 0.80\\
 HD102065\dotfill &  19.0& 0.283& 0.030& 0.055& 0.026& 1.45e+012& 2.15e+015& 1.000& 3.79& 0.89\\
\enddata
\tablenotetext{~}{(This table is available in its entirety in a machine-readable form 
in the online journal.  A portion is shown here for guidance regarding its form and 
content.)}
\end{deluxetable}
\clearpage

Models of pressurized clouds behind weak shocks in the ISM computed by Bergin 
et al. (2004)  reveal that the outcomes for ($f1,\,f2$) are centered on 
values of approximately (0.36,$\,$0.18), (0.40,$\,$0.25) and (0.37,$\,$0.35) for 
the post-shock condensations behind shocks with velocities of 10, 20 and 
$50\,{\rm km~s}^{-1}$, respectively (see their Fig.~8; in these cases the resultant 
ram pressures were $1.4\times 10^4$, $5.8\times 10^4$ and $3.6\times 
10^5{\rm cm}^{-3}$K for a preshock density of $1\,{\rm cm}^{-3}$).    These 
results for $f1$ and $f2$ are well removed from the densest clustering of 
measurements shown in Fig.~\ref{all_v_f1f2}, but they do seem consistent with 
the more sparse population of points having $f2>0.15$, which is more strongly 
emphasized in the unusual velocity ranges represented by the two panels of 
Fig.\ref{outlier_v_f1f2}.

Figure~\ref{outlier_v_f1f2} shows clearly that a moderate number of the 
measurements in the positive-velocity regime exhibit higher pressures than usual, 
but not to the great extremes revealed by the negative velocity gas.  We offer a 
simple interpretation for why this happens.  We propose that a significant fraction 
of the high pressure material arises from stellar mass-loss outflows that eventually 
collide with the ambient medium, creating dense, expanding shells that are at high 
pressures (Castor et al. 1975; Weaver et al. 1977).   Another 
possibility is that small clouds surrounding the stars are pressurized and 
accelerated by either the momentum transfer arising from photoevaporation 
 (Oort \& Spitzer 1955; Kahn 1969; Bertoldi 1989) or 
the momentary surge in pressure of a newly developed H~II region.   These 
phenomena associated with our target stars should be visible to us if they contain 
C~I.   The shells should be intercepted by our sight lines regardless of whether 
they (or possibly small clouds inside them) are large and at moderately high 
pressures or very small and at much higher pressures.  The foreground portions 
of such shells are responsible for the negative velocity gas that we can view in the 
star's spectrum.  For positive velocity gas the situation is different.  Here, we rely 
entirely on the random chance of seeing either one of the large-scale events (of 
non stellar origin) mentioned at the beginning of this section, or else the rear 
portion of some region or shell that is created by some foreground star or stellar 
association that is unrelated to the star that we are viewing.  This being the case, 
there may be a vanishingly small chance that we will intercept a highly pressurized 
shell with a small diameter, but the chances increase for larger shells that have 
lower pressures at their boundaries.    This observational bias against small, high 
pressure events at positive velocities could explain why we see no points at $\log 
(p/k) > 5$ in the right-hand panel of the figure.

In the next section (and in \S\ref{origins}), we will reinforce the picture that high 
pressures arise from the increased dynamical activity near bright stars.  We will 
show evidence that there is a strong correlation between pressures and the local 
intensities of ionizing radiation.

\section{Interpretations of the Results}\label{interpretation}

\subsection{Basic Distribution Functions}\label{basic_dist}

After evaluating the conditions within each velocity interval for all of the lines of 
sight, we are in a position to look at the composite outcome of all of the results of 
the dominant low-pressure component.  All of the presentations in this section 
will show distributions expressed in terms of the amount of hydrogen in a given 
condition.  In order to do so, we must convert our original measurement weights 
based on $N({\rm C~I_{total}})$ into ones that account for the equivalent column 
densities of C~II that we derived from our determinations of O~I (and on rare 
occasions S~II) at identical velocities, as discussed in 
\S\ref{ionization_corrections}.  Once again, we convert from $N$(C~II) to $N$(H) 
by multiplying the amount of carbon by $({\rm H/C})=5040$ in the ISM.  As we 
indicated earlier (\S\ref{starlight}), WNM material at the same velocity as the 
CNM will tend to inflate somewhat the derived value of $N$(H) associated with 
the C~I that is used for determining the pressure.

\begin{figure*}
\plotone{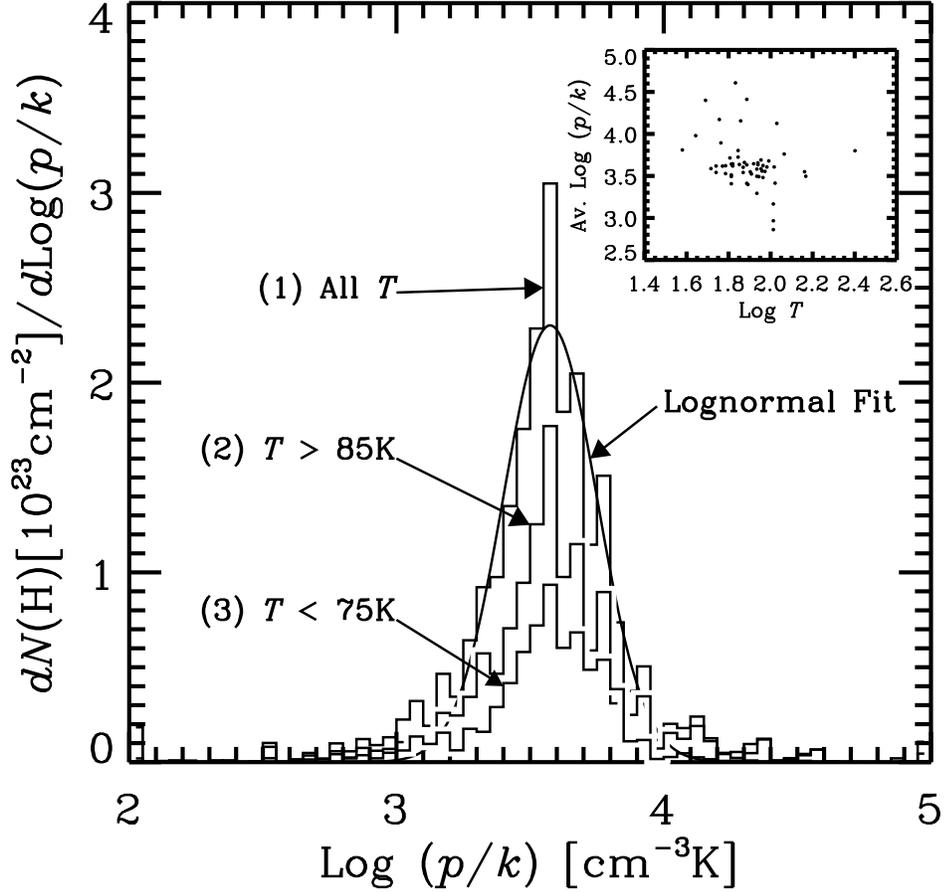}
\caption{The distribution of thermal pressures, normalized to the estimated 
amount of hydrogen present, for three different temperature conditions, as 
indicated by the $J=0$ to 1 rotation temperatures of H$_2$: (1) all of the gas 
(tallest profile), (2) gas for which $T_{01} > 85\,$K (middle profile) and (3) gas for 
which $T < 75\,$K (shortest profile).  The median temperature for all cases is 
$77\,$K, i.e., a value that is between the two limits.  Sight lines where $T_{01}$ 
measurements do not exist are included in condition (1) but excluded from 
conditions (2) and (3).  A best-fit lognormal distribution for condition (1) is shown 
by the solid curve, and it is represented by Eq.~\protect\ref{log_normal_fit}.  The 
inset shows a scatter plot of the C~I weighted average $\log (p/k)$ given in 
Column (6) of Table~\protect\ref{obs_quantities_table} vs. $T_{01}$ (if known), 
as listed in Column (7) of Table~\protect\ref{sightlines_table}.\label{fig4}}
\end{figure*}

\placefigure{fig4}

The histogram distributions shown in Fig.~\ref{fig4} reveal that the pressure 
distribution function is not strongly influenced by the temperature of the gas, as 
deduced from the measurements of $T_{01}$ of H$_2$.   Nevertheless, the 
evidence that we have suggests an inverse correlation of pressures with 
temperature, although the scatter in this relationship is large.  For the points 
shown in the inset of the figure, the Spearman rank order correlation coefficient 
is $-0.29$.  This determination is significantly different from a zero correlation for 
the population at the 97.5\% confidence level for 58 pairs of measurements.  The 
dispersion of the results is so large that it is difficult to assign a value for the 
apparent polytropic index of the gas, but the sign of the trend is consistent with a 
slope of less than one for the CNM thermal equilibrium track near the minimum 
pressure shown in Fig.~\ref{phase_diag}.

The central portion of the distribution of thermal pressures (for all $T_{01}$) 
follows closely a lognormal distribution given by
\begin{eqnarray}\label{log_normal_fit}
dN({\rm H})/d\log(p/k)=&\nonumber\\
2.30\times 10^{23}&\exp\left[-{(\log(p/k)-3.58)^2\over 
2(0.175)^2}\right]\,{\rm cm}^{-2}~.
\end{eqnarray}
This lognormal relationship is shown by the smooth curve in Fig.~\ref{fig4} (and 
will be shown again later in a log-log representation by the smooth gray curve in 
Fig.~\ref{hist_pok}).  Outside the range $3.2<\log(p/k)<4.0$ it understates the 
observed amount of material in the wings of the profile (this is not evident in 
Fig.~\ref{fig4}, but is clearly shown later in Fig.~\ref{hist_pok}). 

\begin{figure}[t!]
\epsscale{1.0}
\plotone{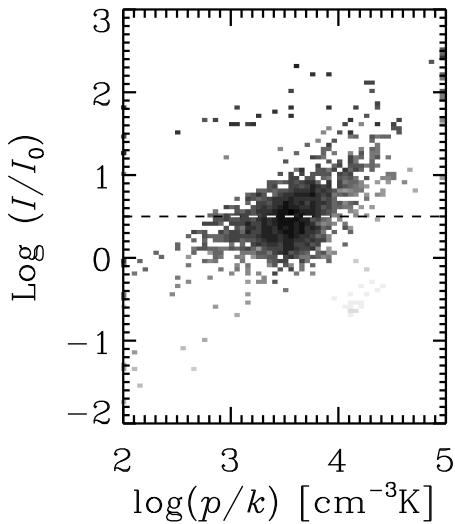}
\caption{A gray-scale representation of the logarithm of the amount of H~I gas 
that we found as a function of $\log(I/I_0)$, i.e., the logarithm of the 
enhancement of the starlight density above average, against the thermal 
pressure, expressed in terms of $\log (p/k)$.  The dashed line shows the cutoff 
equal to $\sqrt{10}$ times the average field that we established for defining the 
low intensity distribution shown in 
Fig.~\protect\ref{hist_pok}.\label{log_pok_bin}}
\end{figure}

\begin{figure*}[t!]
\epsscale{2.2}
\plotone{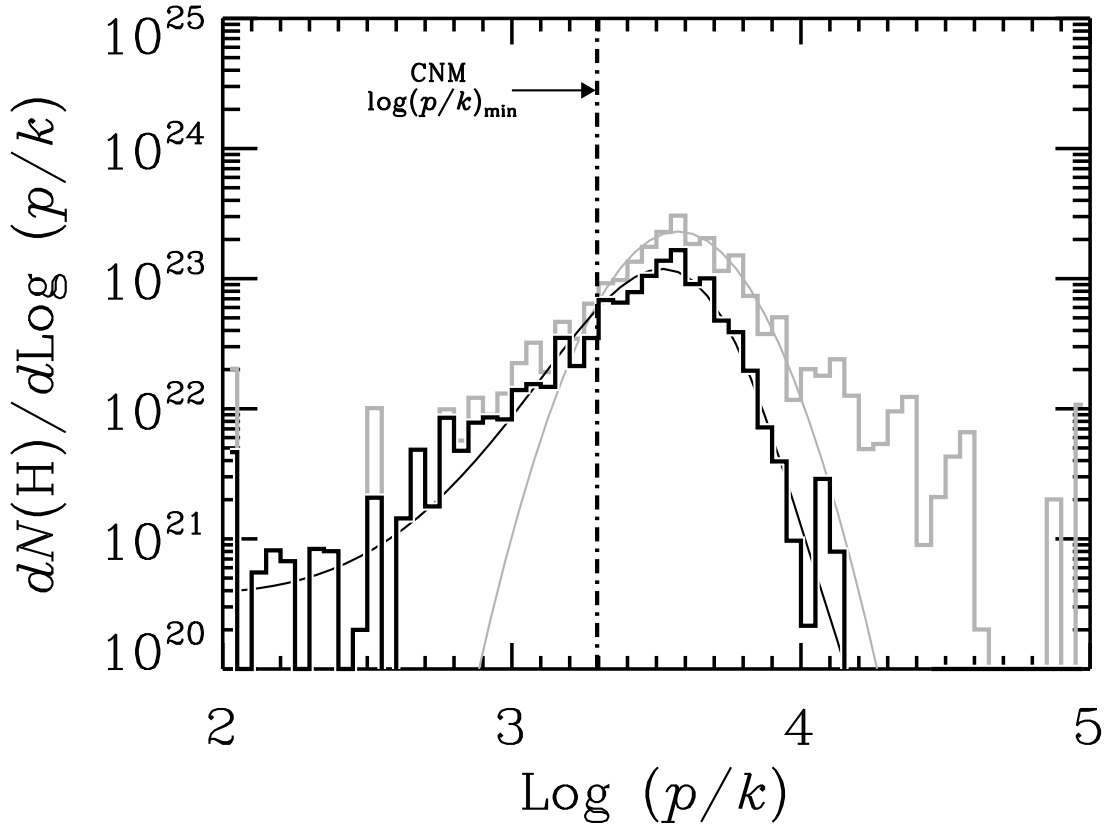}
\caption{Log-log presentations of the distribution of thermal pressures for two 
cases: all of the gas sampled by C~I is shown by the gray histogram, while a subset 
of the material for which $I/I_0<10^{0.5}$ is shown by the black histogram.  This 
intensity cutoff limits the sample to all of the outcomes that appear below the 
dashed line in Fig.~\protect\ref{log_pok_bin}.  The thin curves show how well the 
analytical expressions given in  Eqs.~\protect\ref{log_normal_fit} (gray) and 
\protect\ref{polynomial_fit} (black) fit the results.  The vertical dot-dash line 
labeled CNM $\log(p/k)_{\rm min}$ corresponds to the minimum pressure that is 
allowed for a static CNM, as shown by a similar line with the same designation in 
Fig.~\protect\ref{phase_diag}.\label{hist_pok}}
\end{figure*}

\placefigure{log_pok_bin}

Figure~\ref{log_pok_bin} shows that the outcomes for the starlight densities and 
the thermal pressures are not independent of each other.  In regions that are 
close to stars that emit UV radiation ($I/I_0\approx 10$), we find that with a few 
exceptions the average pressures increase to values in the general vicinity of $\log 
(p/k) \sim 4$.  This enhancement supports the viewpoint that turbulent energies 
are greater in the general vicinity of young stars, a phenomenon that may be 
related to changes in the morphology of H~I near stellar associations that were 
found by Robitaille et al. (2010).

In order to obtain a representation for the pressures in the general ISM somewhat 
removed from the bright stars, we will limit further study of the distribution to 
only those cases where $I/I_0<10^{0.5}$, a limit that is depicted by the dashed 
line in Fig.~\ref{log_pok_bin}.  We consider that any gas elements that are above 
that line represent localized regions that are exceptionally close to sources of 
mechanical energy and are thus not representative of the general, diffuse ISM.

Figure~\ref{hist_pok} shows in a log-log format the distribution of thermal 
pressures for $I/I_0<10^{0.5}$ (black histogram) compared with the distribution 
for all intensities (gray histogram).   In terms of $\log(p/k)$ [and using a linear 
representation of $dN({\rm H})/d\log(p/k)$], the distribution for the low-intensity 
results has a mean of 3.47, a standard deviation of 0.253, a skewness of $-1.8$, 
and a kurtosis\footnote{Our definition of kurtosis includes a subtraction of 3 from 
the fourth moment divided by $\sigma^4$, thus making the kurtosis of a 
Gaussian distribution equal to zero.  Sometimes in the literature, e.g. Federrath et 
al. (2010), this ``$-3$'' term is omitted.}
of 6.2.   The influence of the excess of low pressure outcomes, as evidenced by 
the negative skewness, causes the standard deviation listed here to be larger than 
the value 0.175 given in Eq.~\ref{log_normal_fit}, which would apply to just the 
central portion of the profile.

\placefigure{hist_pok}

For the convenience of those who wish to 
reproduce a reasonably good representation of the low-intensity data in 
analytical form, we supply an empirical polynomial fit,
\begin{eqnarray}\label{polynomial_fit}
dN({\rm H})/d\log(p/k)=1.16\times 10^{23}\exp(-0.0192z\nonumber\\
-0.00387z^2+2.39\times 10^{-5}z^3\nonumber\\
+6.24\times 10^{-7}z^4-6.77\times 10^{-9}z^5)\,{\rm cm}^{-2}~,
\end{eqnarray}
where the dimensionless quantity $z=(p/k)^{\onehalf}-60$ (for $p/k$ expressed 
in terms of ${\rm cm}^{-3}$K).  This empirical fit is shown by the thin, black curve 
in Fig.~\ref{hist_pok}.   If this distribution function is converted into a linear 
representation, i.e., $N({\rm H})$ as a function of $p/k$, we find that for $p/k < 
5500\,{\rm cm}^{-3}\,$K it does not differ appreciably from a Gaussian function 
with mean value of $p/k=3700\,{\rm cm}^{ -3}\,$K and a standard deviation of 
$1200\,{\rm cm}^{-3}\,$K.  The distribution is somewhat higher than this 
Gaussian function above $5500\,{\rm cm}^{-3}\,$K.

Our mean value quoted above is 0.22$\,$dex higher than the value $2240\,{\rm 
cm}^{-3}$K that we listed earlier (JT01).  There are three independent reasons 
that can account for nearly all of this difference.  First, we used revised rates for 
the collisional excitation and radiative decay of the upper fine-structure levels of 
C~I, as discussed in \S\ref{atomic_phys_param}.  This accounts for an elevation of 
typical determinations of pressures of about 0.05$\,$dex.  Second, our new 
estimate for the strength of the optical pumping of the levels has been reduced 
(\S\ref{pumping}), with the result that a typical pressure measurement should be 
raised by another 0.05$\,$dex.  Third, our earlier specification for the mean value 
of $p/k$ was for a temperature (40$\,$K) that gave the lowest inferred pressure 
for a given level of C~I excitation, while the present result uses either actual 
temperatures measured from the H$_2$ rotational excitations or a median value 
of 80$\,$K if an explicit measurement for a sight line is not available.  Under most 
circumstances, the inferred pressure at 80$\,$K is about 0.1$\,$dex higher than 
that for 40$\,$K.    Taken together, these three effects can account for an 
elevation of our new pressures over the old ones by 0.2$\,$dex.

As a final point, we add a cautionary note that the errors in our determinations 
for $\log (p/k)$ become much larger than usual when their values fall below 3.0.  
The reason for this is that the changes in $f1$, the major discriminant for 
pressures, become very small at low pressures, as shown by the shrinkage in the 
spacing between the 0.1~dex markers in Figs.~\ref{all_v_f1f2} and \ref{fig3}.  

\subsection{Volume-Weighted Distributions}\label{vol_weighted}

Up to now, the distribution functions that we have shown 
(Figs.~\ref{fig4}$-$\ref{hist_pok}) have been weighted in proportion to our 
calculated hydrogen column densities, which is equivalent to a sampling by mass.  
In many cases, investigators showing results of computer simulations of ISM 
turbulence express their outcomes according to the counts of volume cells having 
different pressures.  In order to make a conversion from a mass-weighted 
distribution to a volume-weighted one, we must make a simplifying assumption 
that we are viewing an ensemble of gases that has internal random pressure 
fluctuations that change with time, but that is otherwise approximately uniform in 
nature and that can be characterized as having an equation of state with a single 
value for the polytropic index $\gamma$ (equal to the slope of $\log p$ vs. $\log 
n$ or the ratio of specific heats $c_p/c_v$).  In this situation, the changes in 
pressure cause the volumes of mass parcels to change in proportion to 
$p^{-1/\gamma}$, and this factor must be applied to the mass-weighted 
distribution function to obtain the volume-weighted one.

The smooth curves in Fig.~\ref{volume_dist} show how the mass-weighted 
distribution in $\log (p/k)$ for the low starlight intensities would appear after 
being converted to volume weighted ones for three different assumed values of 
$\gamma$.  These three examples illustrate the behavior of the gas under the 
conditions (1) $\gamma=0.7$, which is a good approximation of the slope of 
thermal equilibrium curve for the CNM shown in Fig.~\ref{phase_diag}, (2) the 
relationship for $\gamma=1.0$ that corresponds to an isothermal gas, and (3)  a 
condition $\gamma=5/3$ that indicates that the gas is undergoing adiabatic 
compressions and expansions (and assuming that the gas has a purely atomic 
composition).  The divergent behavior of the curves at the far left portion of this 
diagram probably arises from either deviations caused by small number statistics 
for the samples at the extremely low pressures or the fact that the errors in $\log 
(p/k)$ become larger than usual at the low pressure extreme.  It is important to 
emphasize that in a turbulent cascade the notion that the gas has a single 
polytropic index on all length scales is an oversimplification; we will explore this 
issue in more detail in \S\ref{crossing_times}.

\subsection{Pileup in Velocity Bins}\label{pileup}

As we discussed in \S\ref{results}, any outcomes for $f1$ and $f2$ at a particular 
velocity may represent a composite result for two or more regions that are seen 
in projection along the line of sight.  In \S\ref{admixtures} we explained how we 
separated contributions from small amounts of gas at extraordinarily large 
pressures, well away from the dominant regime of low pressures.  However, we 
have yet to address the possibility that two or more regions at somewhat 
different pressures along the lower, nearly straight portion of the $f1-f2$ 
equilibrium curve can create an apparent outcome that represents a proper 
C~I-weighted mean, but without revealing the true dispersion of pressures from 
the contributors.  If such superpositions are taking place frequently, they will tend 
to decrease the width our observed overall distribution shown in 
Figs.~\ref{fig4}$-$\ref{hist_pok}. 

One way to gain an insight on this possibility is to examine how deviations from 
the mean $\log (p/k)$ scale with the corresponding amounts of C~I.   If we 
imagine a simple picture where all of the C~I exists within independent parcels, 
each with some single, representative value $N_0({\rm C~I_{total}})$, we would 
expect to find that the dispersion of any collection of measurements having some 
multiple $n$ times $N_0({\rm C~I_{total}})$ would show us a standard deviation 
equal to $\sigma_{\rm true}/\sqrt{n}$, where $\sigma_{\rm true}$ is the real 
dispersion in $\log(p/k)$ for the individual packets that are seen in projection.

\begin{figure*}[t]
\epsscale{2.0}
\plotone{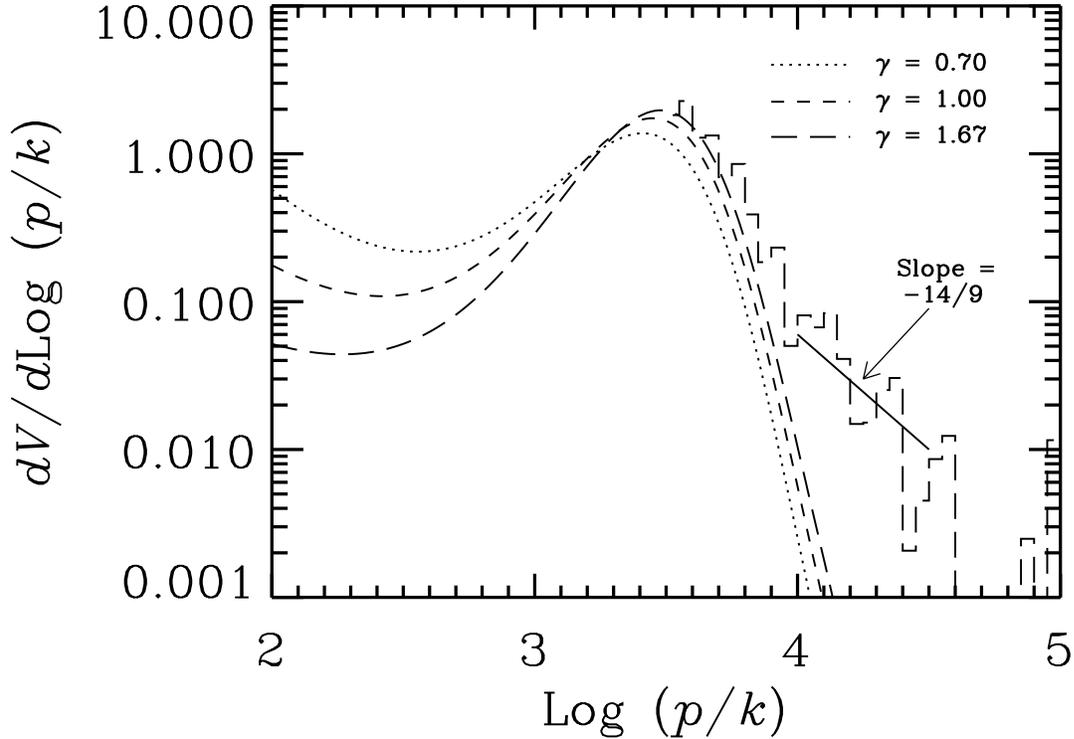}
\caption{({\it three smooth curves:\/}) The results of a conversion of the 
mass-weighted distribution curve for starlight intensity levels $I/I_0<10^{0.5}$ 
(the black, smooth curve shown in Fig.~\protect\ref{hist_pok}) to volume 
weighted ones for three assumed values for $\gamma$, corresponding to cases 
where the gas behaves as if it were in thermal equilibrium ($\gamma=0.70$), 
isothermal ($\gamma=1.0$) and adiabatic ($\gamma=1.67$).  ({\it 
Histogram-style trace:\/}) The volume-weighted distribution for all intensity levels 
for $\log (p/k)>3.5$, assuming $\gamma=1.67$.  This distribution is relevant to a 
discussion that is presented in \S\protect\ref{coherent_disturbances} about the 
possible creation of higher than normal pressures by expanding supernova 
remnants.  In all four cases, the curves are normalized such that their integrals 
over all $\log (p/k)$ equal 1.0.\label{volume_dist}}
\end{figure*}

\placefigure{volume_dist}

\begin{figure*}[t]
\plotone{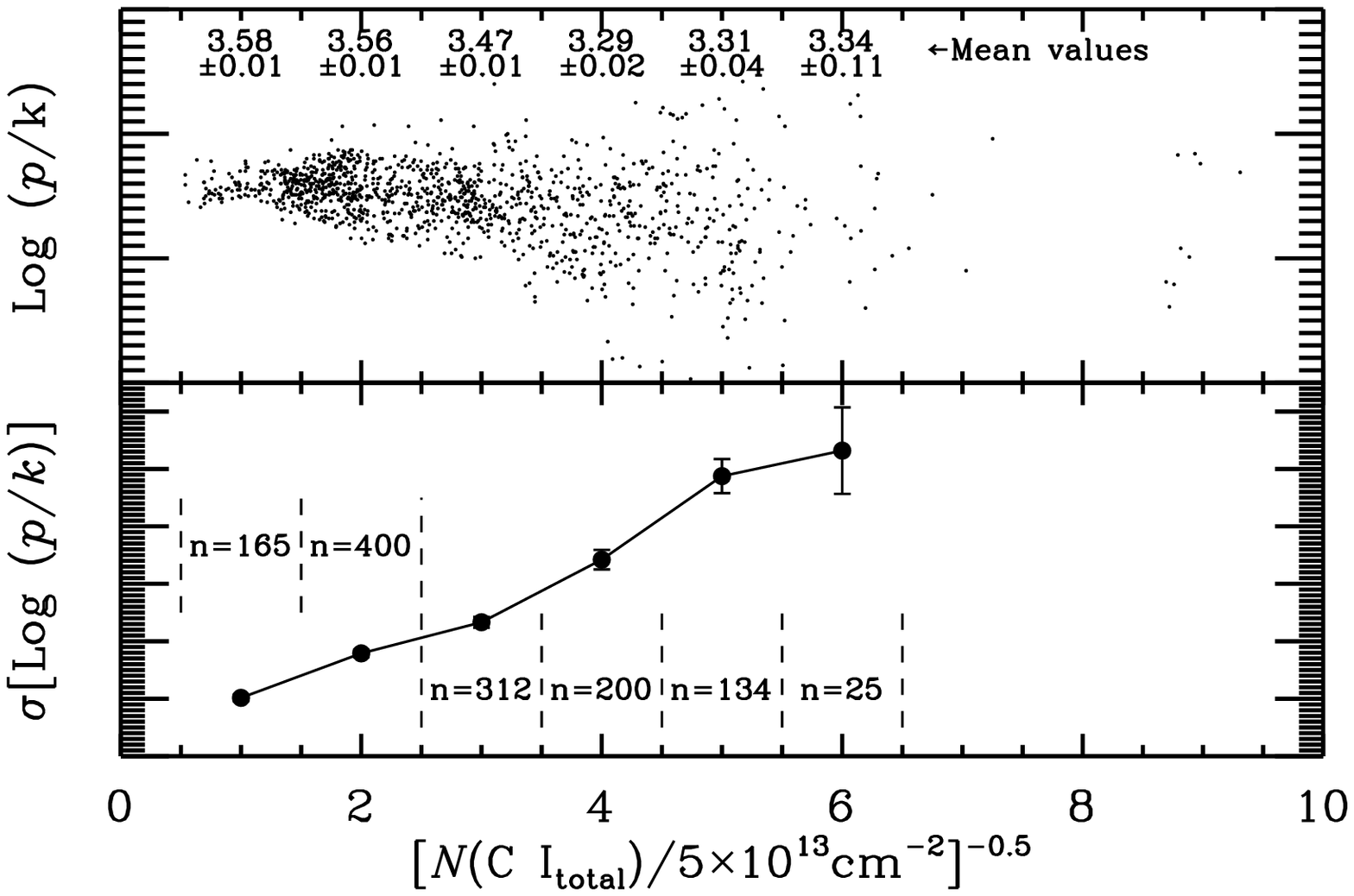}
\caption{{\it Top panel:\/} Individual measurements of $\log(p/k)$ in the low 
pressure regime as a function of the inverse square-root of the column density of 
C~I$_{\rm total}$ for all velocity channels of width $0.5\,{\rm km~s}^{-1}$ that 
had $\log I/I_0<10^{0.5}$.  The mean values of $\log(p/k)$ are listed near the top 
of the panel for successive intervals centered on integral values of $[N({\rm 
C~I_{total}})/5\times 10^{13}\,{\rm cm}^{-2}]^{-0.5}$.  {\it Lower panel:\/} The 
standard deviations in $\log(p/k)$ for measurements within the intervals, 
showing an almost linear progression up to a value $\sigma[\log(p/k)]\approx 
0.5$, at which point the column density of independent packets of gas have 
characteristic values $N_0({\rm C~I_{total}})=2\times 10^{12}{\rm cm}^{-2}$.  
The ``$n=$'' designations show the number of points that were used to evaluate 
$\sigma[\log (p/k)]$ in each bin.\label{sigma_p}}
\end{figure*}

\placefigure{sigma_p}

The upper panel of Figure~\ref{sigma_p} shows the apparent outcomes for values 
of $\log(p/k)$ as a function of $N({\rm C~I_{total}})^{-0.5}$; it is clear that as the 
column densities decrease (i.e., moving toward the right of the plot), the vertical 
dispersions increase.  For data segregated within successive bins having a width of 
$\sqrt{2}\times 10^{-7}{\rm cm}$, the lower panel indicates that the {\it rms\/} 
dispersion indeed seems to scale in direct proportion to $N({\rm 
C~I_{total}})^{-0.5}$, but only up to about $N({\rm 
C~I_{total}})^{-0.5}=5\sqrt{2}\times 10^{-7}{\rm cm}$ (indicating that $N_0({\rm 
C~I_{total}})\approx 2\times 10^{12}{\rm cm}^{-2}$).  Thus, on the one hand, 
one could imagine that $\sigma_{\rm true}$ could be as large as around 0.5, 
instead of our overall measured value of 0.253.  On the other hand, the proposed 
model for superpositions may be only a product of our imagination: perhaps 
coherent regions with larger values of $N({\rm C~I_{total}})$ have a real 
tendency be less easily perturbed by various external forces that cause pressure 
deviations away from some mean value.  In essence, the trend shown in 
Fig.~\ref{sigma_p} may reflect a real physical effect rather than a trend caused by 
random superpositions of unrelated, small gas clouds.

In principle, a trivial explanation for the effect shown in Fig.~\ref{sigma_p} might 
be that as $N({\rm C~I_{total}})$ decreases the measurement errors in 
$\log(p/k)$ increase.  However, as we explain later in \S\ref{err_f1f2}, the 
$1\sigma$ errors in $f1$ and $f2$ should be equal to 0.03 or less for the 
measurements to be accepted.  At normal pressures this amount of error is 
equivalent to a change in $\log (p/k)$ equal to only 0.1~dex, far smaller than the 
observed dispersion that is shown for low column density cases in 
Fig.~\ref{sigma_p}.  One reason that we are able to maintain small errors for 
lower column densities is that our system of weighting the measurements causes 
a shift of emphasis from weak atomic transitions to stronger ones as $N({\rm 
C~I}_{\rm total})$ decreases.

\section{Discussion}\label{discussion}

\subsection{Distribution Width and Shape}\label{shape}

\subsubsection{Overall Shape}\label{overall_shape}

The highly symmetrical appearance of our distribution for all of the material that 
we sampled in the regime of ordinary pressures (which we identified as the ``low 
pressure component'' in \S\ref{admixtures}) is an illusion that arises from the 
projection of the irregularly shaped distribution depicted in 
Fig.~\ref{log_pok_bin} onto the $x$-axis that represents $\log(p/k)$.  The 
distribution function reverts to one with a strong negative skewness when we 
limit our consideration to conditions where $I/I_0<10^{0.5}$ (below the dashed 
line in Fig.~\ref{log_pok_bin}).  This behavior is inconsistent with turbulence in an 
isothermal gas, which should show a pure lognormal density (and pressure) 
volume-weighted distribution function
% *** V\'azquez-Semadeni
 (V\'azquez Semadeni 1994; Nordlund \& Padoan 1999; Kritsuk et al. 2007). 

\subsubsection{Deviations to Low Pressures}\label{deviations_low}
 
A substantial fraction of the material (29\%)  -- that which is depicted to the left of 
the vertical dash-dot line in Fig.~\ref{hist_pok} -- is detected at pressures below 
those permissible for a static CNM, $(p/k)_{\rm min}=1960\,{\rm cm}^{-3}\,$K, as 
defined by the ``standard model'' for the thermal equilibrium curve presented by 
Wolfire et al. (2003) that we show in 
Fig.~\ref{phase_diag}.\footnote{With a parametric formulation discussed in the 
next paragraph using our value for $I/I_0=1.0$ and $\zeta_{\rm CR}=2\times 
10^{-16}{\rm s}^{-1}$, $(p/k)_{\rm min}$ decreases very slightly to $1860\,{\rm 
cm}^{-3}$K.}  From this we conclude that either their curve does not apply to the 
media we are viewing or else that rarefactions caused by turbulence create 
momentary excursions below the curve.  The recovery toward normal pressures 
for regions that reach anomalously low pressures in some cases might be inhibited 
by temporary, locally high values of magnetic pressure (Mac Low et al. 2005).

A valid question to pose is whether or not we could understand the existence of 
low pressures by large changes in some of the parameters that influence the value 
of $(p/k)_{\rm min}$ within some localized regions.  Wolfire et al. (2003) 
expressed a simple equation [their Eq.~(33)] that gives some guidance on this 
possibility.  We restate their equation terms of our variables by substituting 
$0.674(I/I_0)$ for their normalized ISM intensity $G_0^\prime$ at a 
Galactocentric distance of 8.5$\,$kpc.  The reason for this substitution is that we 
have adopted a more recent measure of a standard intensity $I_0$ 
 (Mathis et al. 1983) that is lower than the one they chose to use 
[taken from Draine (1978)].  Also, we set their parameter 
for the total ionization rate (multiplied by $10^{16}{\rm s}^{-1}$) 
$\zeta_t^\prime=2.0$, since we have adopted a cosmic ray ionization rate 
$\zeta_{\rm CR}=2\times 10^{-16}{\rm s}^{-1}$ (see \S\ref{starlight}) and 
assumed that the x-ray and EUV ionization rates are very small in comparison.  
Our restatement of their equation takes the form,
\begin{equation}\label{pmin_eq}
(p/k)_{\rm min}=5730(Z_d^\prime I/I_0){Z_g^\prime\over 1+2.08(Z_d^\prime 
I/I_0)^{0.365}}~,
\end{equation}
where $Z_d^\prime$ is equal to the normalized ratio of interstellar dust grains to 
polycyclic aromatic hydrocarbons (PAH), and $Z_g^\prime$ is the normalized gas 
phase abundance of heavy elements that are responsible for radiative cooling of 
the gas (chiefly C and O).  The two quantities $Z_d^\prime$ and $Z_g^\prime$ 
are generally assumed to be equal to 1.0 for conditions in our part of the Galaxy.  
If $\log(I/I_0)=-0.35$, we find from Eq.~\ref{pmin_eq} that $(p/k)_{\rm 
min}=1000\,{\rm cm}^{-3}$K.  However, the distribution of outcomes shown in 
Fig.~\ref{log_pok_bin} indicate most of our pressure measurements apply to 
regions with $\log(I/I_0)>-0.35$.  Another way to reduce $(p/k)_{\rm min}$ to 
$1000\,{\rm cm}^{-3}$K would be to have $I/I_0=1.0$ but with either a ratio of 
the dust grain to the PAH concentration $Z_d^\prime$ as low as 0.45 times the 
normally assumed value or a reduction of $Z_g^\prime$ to 0.54 times the normal 
amount.  Even with the possible deviations discussed here that would make  
$(p/k)_{\rm min}$ reach as low as $1000\,{\rm cm}^{-3}$K, we still have 
measurable amounts of gas below the pressure threshold for a stable CNM.

\subsubsection{Comparisons with Expectations of the Effects of 
Turbulence}\label{comparisons_turbulence}

The magnitude and skewness of the fluctuations in thermal pressure give an 
indication of the strength and character of the turbulence in the ISM 
 (Padoan et al. 1997b).  For instance, the one-dimensional simulations of 
Passot \& V\'azquez-Semadeni (1998) illustrated how $\gamma$ 
changes the sign of the skewness of the distribution: $\gamma<1$ makes the 
distribution shallower on the high pressure side of the peak and steeper on the 
low pressure side, while the opposite is true for $\gamma>1$.  The influence of 
the polytropic index on the shape of the distribution can also be seen in the results 
of three-dimensional simulations performed by Li, et al. (2003) and Audit 
\& Hennebelle (2010).   Studies of turbulence in an isothermal medium 
by Federrath et al.  (2008) indicated that the shape of 
the distribution may also be governed by the character of the driving force: 
solenoidal (divergence-free) driving forces result in a symmetrical distribution 
(close to lognormal), while compressive (curl-free) driving can create a negative 
skewness.  In short, the appearance of our pressure distribution seems to favor 
either $\gamma > 1$ (i.e, somewhere between isothermal and adiabatic 
behavior), turbulence that is compressive in nature, or some combination of the 
two.

One important application of our determination of the dispersion of thermal 
pressures is to estimate the strength of the turbulence using a quantitative 
comparison based on computer MHD simulations.  Padoan et al. (1997a, b)
introduced a scaling relation between the rms dispersion $\sigma_s$ for a
log-normal distribution of the quantity $s=\ln p$ in terms of a simple function of
the Mach number $M$,
\begin{equation}\label{sigma_s}
\sigma_s=[\ln(1+b^2M^2)]^{0.5}~.
\end{equation}
Investigators that have adopted this formalism generally find that their 
simulations seem to support the validity of a scaling with M shown in  
Eq.~\ref{sigma_s}, but values for the constant $b$ appear to vary from one study 
to the next.  Federrath et al. (2008, 2010) and Brunt (2010) have summarized the
outcomes for $b$ for many different cases reported in the literature: extremes in
$b$ have ranged from 0.3 to 1.0, depending on the conditions in the computations.
  Simulations carried out by Federrath et al. (2008, 2010) indicate that whether
or not the forcing of the turbulence is solenoidal or compressive can have a
strong influence on $b$.  Lemaster \& Stone (2008) have shown that 
magnetic fields have only a small effect on the relationship between $\sigma_s$ 
and $M$.

We can derive a characteristic turbulent Mach number for the C~I-bearing gas by 
taking our dispersion for $\ln p$, adopting a value for $b$, and solving for $M$ in 
Eq.~\ref{sigma_s}.  Here, it is appropriate to use a volume weighted distribution 
of pressures, since that is the conventional way of describing the outcomes of the 
simulations.  A best-fit of a log-normal distribution to the portion $\log (p/k)>3$ 
of the isothermal curve for low $I/I_0$ shown in Fig.~\ref{volume_dist} yields 
$\sigma_s=0.46$.   From the analysis of the possible effects of averaging in 
velocity bins that we presented in \S\ref{pileup}, we acknowledge that the true 
dispersion of $\log (p/k)$ could be as large as 0.5, leading to $\sigma_s=0.5\ln 
10=1.2$.   For our choice of $b$, we adopt the finding by Brunt  (2010) that $b=
0.48^{+0.15}_{-0.11}$, which was based on the observed density and velocity
variances in cold gas with large turbulent Mach numbers in the Taurus molecular
cloud.   This value is near the middle of the range of those derived from computer
simulations of MHD turbulence mentioned in the above paragraph.  With this value
for $b$, we solve for $M$ in Eq.~\ref{sigma_s} and derive $M=1.0^{+0.3}_{-0.2}$
for our lower value of $\sigma_s$ and $M=3.7^{+1.1}_{-0.9}$ for the larger one. 
For our representative values  $f({\rm H_2})=0.60$ (\S\ref{particle_mix}) and $T=
80\,$K (\S\ref{T}), the isothermal sound speed $c_s=0.50\,{\rm km~s}^{-1}$.  For
the smallest value of $M$ minus its error, we expect the velocity dispersion
$\sigma_v=0.8c_s=0.40\,{\rm km~s}^{-1}$, and for the largest $M$ plus its error
we expect that $\sigma_v=4.8c_s=2.4\,{\rm km~s}^{-1}$.

Over a wide dynamic range in linear separations, the velocity differences for 
packets of material in the ISM have been observed to scale in proportion to these 
separations to a fixed power (Larson 1979, 1981; Heithausen 1996; Brunt \& Heyer
2002a; Brunt \& Kerton 2002). We can factor in our values of $\sigma_v$ into this
relationship to estimate the largest characteristic scales for the turbulent
motions, which in turn indicate the largest cloud sizes (or energy injection
scales).  For the power-law relationship, we adopt the recent finding of Heyer \&
Brunt (2004), \begin{equation}\label{sigma_v}
\sigma_v=(0.96\pm0.17)r_{\rm pc}^{0.59\pm 0.07},
\end{equation}
where $r_{\rm pc}$ is the linear separation in pc.  Solving for $r_{\rm pc}$ using 
our velocity dispersions in this equation yields $r_{\rm 
pc}=0.23^{+0.04}_{-0.02}$ for $\sigma_v=0.40\,{\rm km~s}^{-1}$ and $r_{\rm 
pc}=4.7^{+3.7}_{-1.6}$ for $\sigma_v=2.4\,{\rm km~s}^{-1}$.

In Fig.~\ref{filling_factor} we showed the distribution of thicknesses of our 
C~I-bearing clouds for the different lines of sight in our survey.  The median of all 
the values for the entire collection is 5.5$\,$pc.  On the one hand, this median 
value is close to the upper end of our range of $r_{\rm pc}$, which may indicate 
that our larger value of $\sigma_s$, i.e., the largest possible dispersion found in 
\S\ref{pileup}, represents the correct value for the deviations of thermal 
pressures.  On the other hand, the smaller dimensions that apply to the direct 
measurement $\sigma_s=0.46$ may simply indicate that we are usually viewing a 
superposition of many independent, smaller clouds along each line of sight.  We 
caution that the trend expressed in Eq.~\ref{sigma_v} is evaluated from 
$^{12}{\rm CO}$ $J=1-0$ emission-line data for molecular clouds, which may 
differ from the relationship for the more diffuse regions that we have sampled. 

\subsection{Time Constants}\label{time_constants}

Since fluid elements in a turbulent medium have physical properties that change 
with time, it is important to establish the time intervals that are required for 
various measurable quantities to converge nearly to their equilibrium values.   
There are three different contexts where we compare two (or more) states of any 
particular constituent: (1) the ratio of C~I fine-structure populations, (2) the 
balance between neutral and ionized forms of the carbon atoms and (3) the 
$J=0$ to 1 rotational temperature $T_{01}$ of H$_2$.  A fourth time-variable 
quantity of interest is the kinetic temperature of the gas, which not only 
influences the other three quantities that we measure but also the manner in 
which the gas responds to disturbances.  We will compute the characteristic 
$e$-folding times for these processes in the following subsections, and later we 
will compare them to the eddy turnover times for different size scales.

In a general context, we can imagine atoms or molecules in two possible states 
with equilibrium number densities $n_{\rm 0,eq}$ in some lower level and 
$n_{\rm 1,eq}$ in an upper one.  In equilibrium,
\begin{equation}\label{equilib}
R_{01}n_{\rm 0,eq}=R_{10}n_{\rm 1,eq}~,
\end{equation}
where $R_{01}$ and $R_{10}$ are the upward and downward conversion rates, 
respectively.  We can propose a solution for the time behavior of the lower level, 
$n_0$, to take the form
\begin{equation}\label{proposed_behavior}
n_0(t)=n_{\rm 0,eq}+(n_{\rm 0,i}-n_{\rm 0,eq})e^{-\gamma t}
\end{equation}
as the concentration of $n_0$ adjusts itself from some initial density $n_{\rm 
0,i}$ to its equilibrium value $n_{\rm 0,eq}$.  This form must agree with the 
condition
\begin{eqnarray}\label{dndt}
{dn_0(t)\over dt}&=&n_1(t)R_{10}-n_0(t)R_{01}\nonumber\\
&=&n_{\rm tot}R_{10}-n_0(t)(R_{10}+R_{01})~,
\end{eqnarray}
where $n_{\rm tot}=n_0(t)+n_1(t)$ is the (constant) sum of the number densities 
of the two levels.  If we insert the proposed time behavior 
(Eq.~\ref{proposed_behavior}) into the $n_0(t)$ term of Eq.~\ref{dndt} and 
compare it to an explicit differentiation of Eq.~\ref{proposed_behavior} with 
time, we can equate the $e^{-\gamma t}$ terms to reveal that
\begin{equation}\label{gamma}
\gamma=R_{10}+R_{01}~.
\end{equation}
(The sum of the remaining terms without $e^{-\gamma t}$ equals zero.)  In 
essence, any departure from the equilibrium level distribution, either positive or 
negative in sign, will decay to the equilibrium condition in an exponential fashion 
with an $e$-folding time constant given by Eq.~\ref{gamma}.  In the following 
three subsections, we apply this rule to population ratios discussed in items (1) to 
(3) at the beginning of this section.

\subsubsection{C~I Fine Structure Levels}\label{fsl_timedep}

As a simplification, we consider only the first two levels and ignore the existence 
of the third (highest) one.  Here, $R_{01}$ equals the sum of the upward rate 
constants for various collision partners times their respective densities.  $R_{10}$ 
equals the sum of the downward rate constants times the densities plus also the 
spontaneous decay probability $A_{10}$.  If $f1$ is small, $A_{10}=7.93\times 
10^{-8}{\rm s}^{-1}$
% *** Galav\'is
(Galav\'is et al. 1997) dominates over the collisional excitation (and 
de-excitation) terms.  The inverse of $A_{10}$ equals 146~days.  If $f1$ is not 
small, the collisional terms make $R_{10}+R_{01}$ even larger and thus reduce 
the time constant to less than 146~days.  Clearly, even for the more complex 
situation for the interactions with the highest of the three fine-structure levels, 
the time constants are extraordinarily short ($A_{21}^{-1}=44\,{\rm days}$).

\subsubsection{The Photoionization and Recombination of Carbon 
Atoms}\label{ioniz_timedep}

Since the equilibrium concentrations of neutral carbon are much smaller than the 
ionized forms in all cases that we consider, as is evident from 
Fig.~\ref{phase_diag},  it is clear that the ionization rate 
$R_{01}=(I/I_0)\Gamma_0$ dominates over the various recombination terms 
shown in Eq.~\ref{C_ionization} that make up $R_{10}$.  
Figure~\ref{log_pok_bin} shows us that $I/I_0=1$ is about the lowest value of 
the radiation density that we encounter.  Hence, the longest time constant that 
we expect to apply is simply $\Gamma_0^{-1}=5\times 10^9{\rm s}=160\,{\rm 
yr}$, and this time shortens in proportion to the increase in $I$ above the 
reference value $I_0$.

\begin{figure*}
\plotone{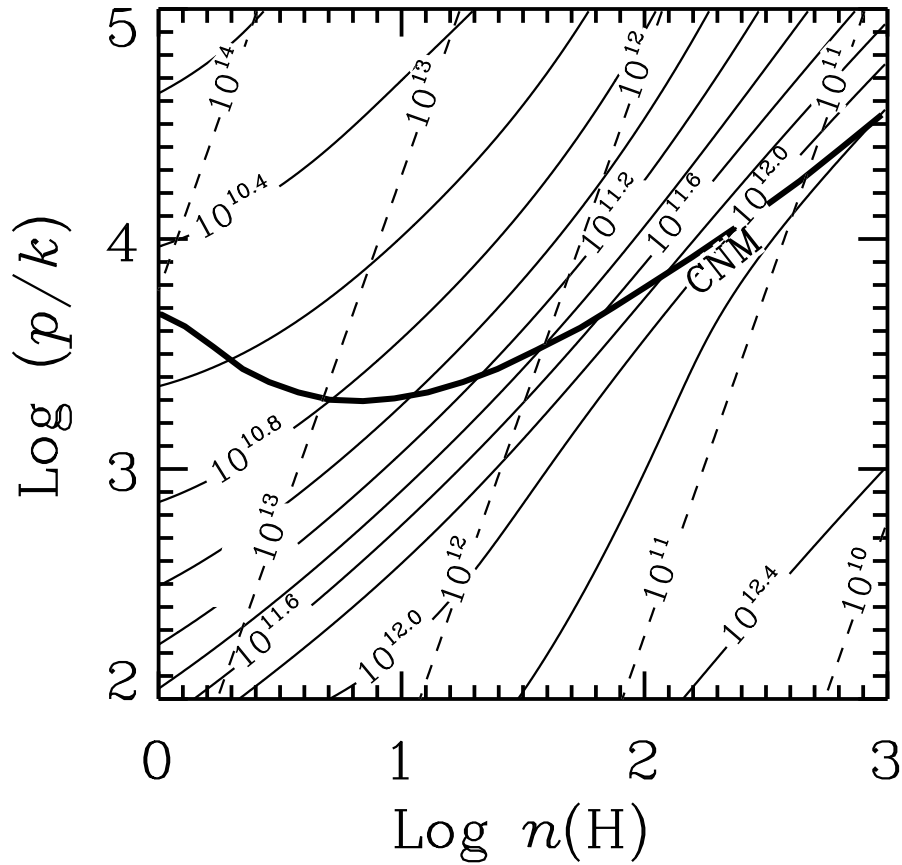}
\caption{Time constants (in seconds) for (1) the relaxation of the $T_{01}$ 
rotation temperature of H$_2$ to the local kinetic temperature (solid contours), 
as a result of ortho-para conversions of the lowest two levels due to collisions 
with protons and (2) cooling times $t_{\rm cool}$ as given in 
Eq.~\protect\ref{t_cool} (dashed lines).  The thermal equilibrium curve for the 
CNM that appears in Fig.~\protect\ref{phase_diag} is shown by the thick curve.  
This diagram was constructed assuming that the gas has $f({\rm H}_2)=0.6$ and 
He/H=0.09 (see \S\protect\ref{particle_mix}).\label{r_t_01}}
\end{figure*}

\placefigure{r_t_01}

\subsubsection{The $J=0$ to 1 Rotation Temperature $T_{01}$ of 
H$_2$}\label{T_01_timedep}
 
Cecchi-Pestellini et al. (2005) have performed 
detailed calculations of the time-dependent H$_2$ level populations in turbulent 
media that have short-lived pockets of hot gas that can leave an imprint on the 
rotation temperatures.  Here, we focus on a much simpler discussion for $T_{01}$ 
of the two lowest rotational levels, since they are relevant to our determinations 
of kinetic temperatures.  We evaluate the characteristic time for changes in the 
population ratio of $J=0$ to that of $J=1$ when there is a sudden change in the 
kinetic temperature, but we neglect any of the effects of repopulating the lower 
levels from cascades from higher levels of excitation. 

The rate coefficient for ortho-para transitions caused by neutral hydrogen 
impacts onto H$_2$ is extremely low at the temperatures that we consider [for 
$T<300\,$K, $k_{01}<10^{-16}{\rm cm}^3{\rm s}^{-1}$ (Sun \& Dalgarno 1994)].  For
protons, however, the rate constants are much larger: $k_{10}=2.0\times 10^{-
10}{\rm cm}^3{\rm s}^{-1}$ (Gerlich 1990), and $k_{01}=9\exp(-171/T)k_{10}$. 
The solid contours in Fig.~\ref{r_t_01} show how the time constants vary over
the $\log (p/k)$ vs. $\log n({\rm H})$ diagram when we combine the rate
constants with determinations of $n(p)$ using Eq.~\ref{H_ionization}.

\subsubsection{The Kinetic Temperature}\label{kinetic_T_timedep}

Unlike the cases that we considered in \S\S\ref{fsl_timedep}-\ref{T_01_timedep}, 
for the kinetic temperature we must work with a continuous variable instead of a 
population ratio of two states of some constituent.  Thus, a somewhat different 
tactic is needed to assess the characteristic relaxation time.  Wolfire et al. 
 (2003) have evaluated the isobaric cooling time for the 
ISM and find that 
\begin{equation}\label{t_cool}
t_{\rm cool}=7.40\times 10^{11}\left({T\over 80\,{\rm 
K}}\right)^{1.2}\left({p/k\over 3000\,{\rm cm}^{-3}{\rm K}}\right)^{-0.8}{\rm s}~.
\end{equation}
They state that this formula is valid to within a factor 1.35 over the temperature 
range  $55<T<8500\,$K.  By itself, the coefficient in Eq.~\ref{t_cool} applies to 
conditions very close to our median temperature and pressure in the survey 
($T_{01}=80\,$K  and $p/k=3000\,{\rm cm}^{-3}\,$K), and it is not much 
different from the relaxation time for $T_{01}$ ($3.95\times 10^{11}{\rm 
s}=12,500\,{\rm yr}$) at the same temperature and pressure.  We depict values 
of $t_{\rm cool}$ by the nearly straight, dashed contours in Fig.~\ref{r_t_01}.

\subsection{Turbulent Eddy Crossing Times}\label{crossing_times}

As the length scales become smaller, the dwell times for conditions become 
shorter.  It then follows that these shortened durations could curtail physical stasis 
in certain respects.  To estimate the time scales $\Delta t=r/\Delta v$ for changes 
to occur in a turbulent eddy with a characteristic radius $r$, we can once again 
make use of the power-law relationship between velocity shears $\Delta v$ and 
length scales $r_{\rm pc}$  (as we did in \S\ref{comparisons_turbulence}), but 
this time we adopt the findings taken from observations at smaller scales.   A slight 
reduction of the slope seems to occur for these shorter length scales: Falgarone et 
al. (1992) conclude that $\Delta v\approx r_{\rm pc}^{0.4}$ for 
$10^{-2}<r_{\rm pc}<1$, and their result agrees with that of Brunt \& Heyer 
 (2002b) and Heyer \& Brunt  (2004) at a common scale $r_{\rm 
pc}=1$.  This velocity trend for the shorter lengths is consistent with the 
theoretical study of turbulence by Boldyrev et al. (2002), and we will 
adopt it for our investigation of time scales.  We recognize, however, that there 
are isolated observations, such as those carried out by Sakamoto (2002), 
Sakamoto \& Sunaka (2003) and Heithausen (2004, 2006), that 
show some specific regions where the velocity differences measured over 
$r_{\rm pc}\sim 10^{-3}-10^{-1}$ are almost one order of magnitude above this 
velocity-size relationship; see Fig.~10 of Falgarone et al. (2009).  Also, 
observations of CO emission by Hily-Blant et al. (2008) demonstrate that 
isolated concentrations of turbulent energy over small scales create occasional, 
large velocity deviations that go well beyond the tails of a Gaussian distribution.  
Finally, Shetty et al. (2010) indicate that projection effects in the 
position-position-velocity (PPV) data overestimate the power-law slope (by one to 
a few tenths) and underestimate the velocity amplitudes (by about a factor of 
two) in a 3D physical position-position-position (PPP) representation of a 
turbulent medium.  With these points in mind, we make an extrapolation of the 
trend $\Delta v=r_{\rm pc}^{0.4}\,{\rm km~s}^{-1}$ toward very small scales to 
yield $\Delta t=r_{\rm pc}^{0.6}\,$Myr, but acknowledge that in some 
circumstances this form for $\Delta t$ may significantly overestimate the time 
span for rapid changes in conditions.

For a length scale $r_{\rm pc}=0.00046$ (or 95$\,$AU), we expect that $\Delta 
t= 0.01\,$Myr (i.e., $10^{11.5}{\rm s}$).  Along the CNM equilibrium curve 
shown in Fig.~\ref{r_t_01}, this time equals $t_{\rm cool}$ at $\log(p/k)=3.85$. 
The crossing time is about equal to the $e$-folding time for $T_{01}$ at a slightly 
lower pressure, $\log(p/k)=3.6$.   Thus, in short, for scale sizes smaller than 
about 100$\,$AU (but perhaps a few thousand AU for some of the more active 
regions) we can expect that turbulent fluctuations at the pressures that we are 
considering will depart from the CNM thermal equilibrium curve 
($\gamma\approx 0.7$)  and exhibit an effective $\gamma$ that is somewhere 
between 0.7 and the adiabatic value of 5/3 (for a pure atomic gas).   Over smaller 
scales (or lower pressures) $T_{01}$ may depart from the local kinetic 
temperature.  Over all of the practical size scales, the equilibrium results for the 
C~I fine-structure excitations and C ionization should apply, but with the provision 
that their outcomes depend on the instantaneous temperature. 

\subsection{Possible Effects from Coherent, Large Scale 
Disturbances}\label{coherent_disturbances}

Up to now, we have considered the effects of compressions and rarefactions 
caused by random turbulent motions.  However, the injection of mechanical 
energy over macroscopic scales in the Galactic disk can also create deviations in 
pressure.  Supernova blast waves constitute a principal source of this energy in 
the ISM.   We know that there are strongly elevated pressures inside clouds that 
have recently been overtaken by a supernova blast wave, as shown by the 
enhanced C~I fine-structure excitations that appear in the spectra of stars within 
and behind the Vela supernova remnant (Jenkins et al. 1981, 1984, 1998; Jenkins \&
Wallerstein 1995; Wallerstein et al. 1995; Nichols \& Slavin 2004). We have good
reason to expect that pressure increases with somewhat smaller amplitudes should
persist even within remnants that are no longer identifiable because they are so
old or disrupted.

In a more general context, at random locations in the disk of the Galaxy the 
outcomes for the thermal pressure enhancements arising from the effects of 
supernovae are expected to be appreciably different for the various broad 
temperature regimes in the ISM, as shown by several different computer 
simulations (de Avillez \& Breitschwerdt 2005a; Mac Low et al. 2005).  The
simulations have many free parameters that influence the properties of the
average pressures and the shapes of the distribution functions.  For this reason, we
will restrict our attention to a very simple test of the plausibility that, beyond the
limited range of fluctuations caused by turbulence, there is a broader, underlying
spread of pressures caused coherent, large scale mechanical disturbances arising from
supernova explosions.

We can adopt a tactic similar to one developed by Jenkins et al. (1983) in their
comparison of C~I pressures to a prediction based on the theory of the three-phase ISM
advanced by McKee \& Ostriker (1977).  Small, neutral clouds that are overtaken by an
expanding supernova blast wave should rapidly (and adiabatically) adjust their
internal thermal pressures to equal that of the hot medium well inside the remnant's 
boundary.  We can now make a simple prediction of what would happen if these 
clouds actually defined the trend of the pressure distribution well above the mean 
pressure and then compare this outcome with our observations.

If the radius $r$ of any remnant in the adiabatic phase grows in proportion to 
$t^\eta$ and its volume-weighted average internal pressure $p$ is proportional 
to $r^\alpha$, we find that
\begin{equation}
dp/dt=(dp/dr)(dr/dt)\propto r^{\alpha-1+(\eta - 1)/\eta} .
\end{equation}
A time-averaged occupation volume $V(p)$ is then given by
\begin{equation}
V(p)\propto r^3/(dp/dt)=r^{3-\alpha+1/\eta} ,
\end{equation}
which gives an overall volume filling factor per unit $\log p$ that is proportional 
to $V(p)p=p^{-14/9}$ for  $\alpha=-3$ and $\eta=3/5$ (McKee \& Ostriker
1977), as long as the remnants do not overlap each other, which should be true at
pressures well above the median pressure.

The histogram-style trace in Figure~\ref{volume_dist} shows our thermal 
pressure distribution on the assumption that the overtaken clouds contract 
adiabatically, i.e., with $\gamma=5/3$.  Unlike the smooth curves shown in this 
figure, this distribution represents our entire dataset, i.e., not just the instances 
where $I/I_0<10^{0.5}$.  Our reason for this choice is that we wish to avoid a 
bias against regions of higher than normal starlight intensity, because the 
locations of supernova remnants are correlated with those of associations of 
early-type stars.  As Fig.~\ref{volume_dist} shows, the pressure distribution has a 
slope that is roughly consistent with $dV/d\ln p\propto p^{-14/9}$.

\subsection{High Pressure Component}\label{high_pressure_comp}

In \S\ref{admixtures} we proposed that some small fraction of all of the gas that 
we observed has an extraordinarily high pressure ($p/k\gtrsim 3\times 
10^5\,{\rm cm}^{-3}\,$K, $T > 80\,$K), in order to nudge the $f2$ outcomes to 
locations above the normal equilibrium tracks shown in Fig.~\ref{all_v_f1f2}.  We 
now explore various explanations for the excesses in $f2$, starting with ones that 
do not imply the presence of small amounts of gas at high pressures.   Later, on 
the premise that the existence of the high pressure material is indeed real, we 
review some suggestions made by other investigators on its possible origin.

\subsubsection{A Misleading Conclusion?}\label{misleading_conclusion}

Before we fully accept our interpretation that the anomalously high $f2$ 
measurements imply the existence of small amounts of high pressure material, 
well separated from the main pressure distribution function presented in 
\S\ref{basic_dist},  we should briefly investigate possible errors in the 
interpretation of the outcomes in $f1$ and $f2$.  One possibility is that the 
excitation cross sections or the decay rates for the excited levels have systematic 
errors that underestimate the populations in the $^3P_2$ state (C~I$^{**}$)  
relative to those in the $^3P_1$ level (C~I$^*$).  We feel that this is unlikely, 
since earlier calculations of these quantities that appeared in the literature (i.e., 
the ones adopted by JT01\footnote{Appendix~\ref{atomic_phys_param} 
discusses our current updates for the atomic parameters.}) did not yield outcomes 
that predicted greater values of $f2$ for their respective $f1$ counterparts.  
However, on more fundamental grounds we do not feel qualified to comment on 
the accuracy of the atomic physics calculations, so we will not pursue this issue 
further.

Another possibility for misleading results could be errors in our determinations of 
$f1$ and $f2$.  Two possibilities for the origin of such errors could either be 
errors in the adopted $f$-values for the C~I transitions or our under-appreciation 
of the effects of misleading apparent optical depths caused by unresolved, 
saturated substructures in the absorption line profiles.  For the former of the two, 
we feel that our investigation discussed in Appendix~\ref{fval_validation} 
provides some assurance that we are not experiencing systematic errors in the 
relative strengths of weak multiplets versus the strong ones.  However, our 
derived $f$-values rely on the correctness of the published relative line strengths 
within multiplets.  These relative strengths have a direct influence on our derived 
values of $N({\rm C~I}^*)$ and $N({\rm C~I}^{**})$, relative to each other and to 
$N({\rm C~I})$.

As for the possibility that we are being misled by incorrect optical depth 
measurements, we feel that the precautions that we discuss in \S\ref{distortions} 
for screening out such cases provide adequate safeguards.  Moreover, it is 
reassuring to see that for individual determinations of the apparent fraction of C~I 
in the high pressure component, $g_{\rm high}$, in each velocity bin (i.e., not the 
overall averages shown in Table~\ref{obs_quantities_table}), there is no trend 
with $N({\rm C~I_{total}})$, an effect that we would have expected to see if the 
phenomenon were driven by the strengths of the absorption lines.

Still another effect to examine is the possibility that there is an anomalous means 
for exciting the fine-structure levels.  Positively charged collision partners will give 
proportionally stronger excitations of the second excited level of C~I, as indicated 
by the differences in cross sections for protons compared to neutrals -- see Fig.~1 
of Silva \& Viegas (2002).  If ambipolar diffusion (ion-neutral slip) created 
by MHD shocks and Alfv\'en waves create enough suprathermal protons (and 
heavy element ions) to further excite the C~I, they might help to explain the larger 
outcomes for $f2$.  While this is a qualitatively attractive explanation, in a 
quantitative sense it seems to fail: the required fractional concentration of the 
positively charged collision partners, greater than about 30\%, seems to be 
unreasonably large (e.g., the conditions $n({\rm H~I})=2\,{\rm cm}^{-3}$, 
$T({\rm H~I})=600\,$K, $n(p)=0.6\,{\rm cm}^{-3}$, $T(p)=20,000\,$K should 
give $f1=0.23$ and $f2=0.066$, which is not far from our measured average 
shown by the white $\times$ in Fig.~\ref{all_v_f1f2}.  Smaller ion fractions fail to 
do so however).

\subsubsection{The Amount of the High Pressure 
Component}\label{amount_hipress}

We now move on to the premise that we advocated earlier in \S\ref{admixtures} 
that the anomalously large values of $f2$ arise from a small admixture of high 
pressure gas in virtually all of the cases that we examined.  In terms of $N({\rm 
C~I})$, the overall fraction $g_{\rm high}$ is usually about 5\%.  However it is 
important to note that, except in the presence of exceptionally strong ionization 
field strengths, this outcome must arise from much smaller proportions of H~I 
because the neutral fraction of carbon increases with pressure, making small 
amounts of high pressure gas far more conspicuous.   For instance, we can expect 
a factor 100 enhancement in fractional amount of C~I, $n({\rm 
C~I_{total}})/[n({\rm C~II})+ n({\rm C~I_{total}})]$, when the gas is at $\log 
(p/k)=6$, $T=300\,$K over that which would apply to material with more 
conventional physical conditions $\log(p/k)=3.6$, $T=80\,$K.  On average, this 
makes the fractional mass contribution of H~I in the high pressure component 
only about $g_{\rm high}/100$, a 0.05\% mass fraction.  We add a caution, 
however, that this fraction could be larger if the actual pressure of the 
high-pressure component is lower than the value stated above.

\subsubsection{Radiation from the Excited Levels of Carbon 
Atoms}\label{fsl_radiation}

Radiative decay of the upper fine-structure level of C~II is an important cooling 
mechanism for the ISM.  The rate of this cooling per unit mass can be monitored 
by either directly observing the emission at 1900$\,$GHz ($157.7\,\mu$m) 
 (Stutzki et al. 1988; Langer et al. 2010; Pineda et al. 2010; Velusamy et al.
2010) or by viewing the C~II$^*$ absorption features at 1037.018 and 1335.708$\,
$\AA, as has been done for both the ISM of our Galaxy (Lehner et al. 2004) and
the distant, damped L$\alpha$ systems in quasar spectra (Wolfe et al. 2003a, b). 
It is worthwhile to ask the question: could the regions that we view with
enhanced pressures make an important contribution to either the absorption or
emission measurements?  It is difficult to formulate a precise answer, since we
do not fully understand the nature of these regions.  On the one hand, the factor
100 diminution in the H~I content mentioned in the previous section is
approximately offset by a factor 100 enhancement in the collision rate for exciting the
upper level of C~II.   In this circumstance, as long as we are still below the critical
density for establishing the C~II$^*$ population, our typical value of $g_{\rm high}$
of 5\% would be approximately the correct answer for the enhancement of the emission
intensity (in the optically thin limit) or for the increase in $N$(C~II$^*$) over that
from the gas at normal pressures.  On the other hand, if the high pressure regions are
located at sites where the photoionizing radiation level is much higher than elsewhere,
then the H~I concentration is not strongly diminished but the population of the upper
C~II levels is still very high.  Here, the column densities of C~II$^*$ could be 
considerably higher than 5\% of the total and the regions that hold this material 
could emit a substantial amount of radiation.

It is much easier for us to make a quantitative assessment of the enhancement of 
radiation from the excited levels of C~I because the populations of the two upper 
levels are exactly what we observe.  The value of $f1$ within the high pressure 
gas should be about twice that of the normal gas; hence we can expect that the 
radiation at 492$\,$GHz ($609\,\mu$m) seen toward most of the translucent 
clouds (Heithausen et al. 2001; Bensch et al. 2003) should only be increased by
about 10\%.  We estimate that the value of $f2$ in the high pressure gas is about
15 times as large as that in the low pressure gas, so about 44\% of the radiation
at 809$\,$GHz ($371\,\mu$m) could arise from the high pressure regions.

\subsubsection{Possible Origins of the High Pressure Gas}\label{origins}

As we pointed out in \S\ref{admixtures}, in order to obtain the composite $f1$ 
and $f2$ values that we found for the entire survey, the high pressure 
component had to be a distinct population whose distribution in pressure was 
well removed from the low pressure material.  We demonstrated in 
Fig.~\ref{f1f2_lognormal} that it could not be a diminishing tail resembling a 
power law that extends away from the main, low pressure distribution.  In the 
context of turbulence theory, this poses a challenge in the interpretation of the 
high pressure gas, unless one could propose an explanation for the absence of 
intermediate mass fractions at pressures between the two extremes.

In order to justify the presence of certain molecules in the ISM that require 
endothermic reactions for their production, such as CH$^+$, Joulain et al. 
 (1998) proposed the existence of hot gas concentrations within highly 
confined dissipation regions created by turbulence.  From a computer simulation, 
Pety \& Falgarone (2000)  found that extraordinary physical conditions 
could arise in regions that were selected to have special dynamical conditions, 
such as larger than normal amounts of vorticity or negative divergence.  Further 
studies from theoretical or observational perspectives have been presented by 
Godard et al. (2009) and Hily-Blant et al. (2008). These 
investigators concluded that the volume filling factors for these regions are small 
(a few percent), but not as small as the mass fractions that we reported in 
\S\ref{amount_hipress}.  While the intermittent emergence of extreme conditions 
within highly confined dissipation regions in a turbulent regime is an attractive 
prospect for explaining our high pressure gas, it must nevertheless satisfy our 
requirement for a distinct separation from the pressure fluctuations arising from 
regular turbulent disturbances instead of a continuous, low level extension 
thereof.

In \S\ref{behavior_velocity} we showed evidence that the greatest extremes in 
pressure occurred for gases at unusually large negative velocities, and this 
interpretation fits in well with the concept that the target stars (and their 
neighboring stars) play a role. Indeed, an inspection of 
Table~\ref{obs_quantities_table} shows that in some directions, several adjacent 
sight lines all show elevated pressures compared to the typical pressures derived 
from the full sample.  Two prominent examples are stars near or within the Carina 
and Orion Nebulae.  These are dynamically disturbed regions, and they are also 
regions of significantly enhanced starlight density.  This supports the notion that 
the stars somehow raise the pressures in their surroundings and that high values 
of $I/I_{0}$ indicate both recent, enhanced star formation and a more highly 
pressurized ISM. 

\begin{figure*}
\epsscale{1.7}
\plotone{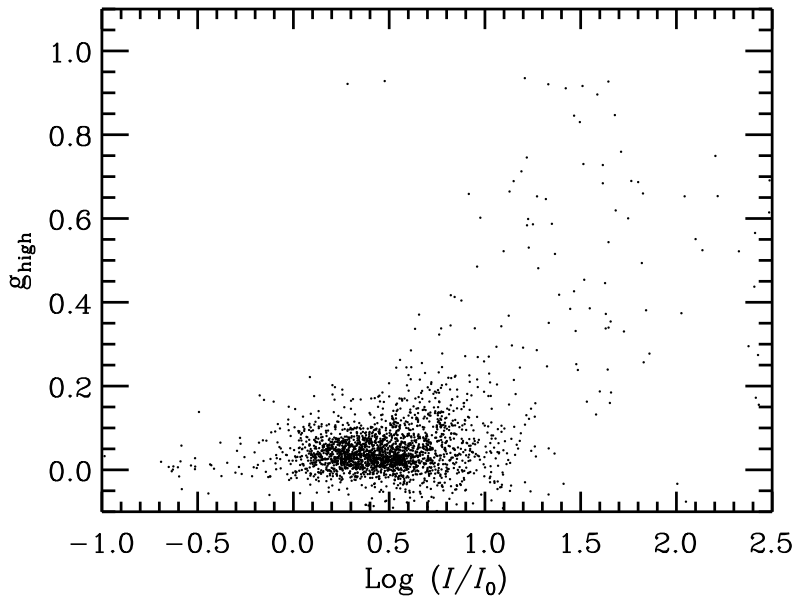}
\caption{The relationship between the fractional amount of high pressure gas, 
$g_{\rm high}$, and the starlight intensity relative to the Galactic average, 
$I/I_0$.\label{fhigh_vs_logi}}
\end{figure*}

Figure~\ref{fhigh_vs_logi} indicates that the quantities $g_{\rm high}$ and 
$I/I_0$ also seem to be connected to each other.  Generally, we can see that 
cases where $g_{\rm high}>0.2$ appear to require that $\log (I/I_0) > 0.5$ and 
that there were very few outcomes that had $g_{\rm high} < 0.2$ that had $\log 
(I/I_0) > 1.5$.  It is unclear whether the dominant cause for pressurization is from 
interactions with mass-loss ejecta, small clouds experiencing a photoevaporation 
``rocket effect'' (Bertoldi 1989; Bertoldi \& McKee 1990; Bertoldi \& Jenkins 1992),
or the sudden creation of an H~II region, all of which can compress the ambient
material and accelerate it toward us.  An additional possibility is that H~I gas
near the stars is heated more strongly by the photoelectric effect from grains
(Weingartner \& Draine 2001b), which could cause a short-term spike in pressure. 
While these effects (or combinations thereof) may be the dominant cause for the
isolated cases that show large values for $g_{\rm high}$, we still find significant
amounts of high pressure material at large positive velocities and even small
admixtures of high pressure material at all velocities.  These outcomes indicate
that other mechanisms unrelated to the target stars may play role as well.  

\placefigure{fhigh_vs_logi}

Field et al. (2009) have proposed that the recoil of H atoms following the 
photodissociation of H$_2$ at the edge of a molecular cloud can create an 
external pressure that helps to confine the cloud.  They estimated that at locations 
where ambient UV field intensity reaches $I/I_0 \approx 60$ the recoil pressure 
can be approximately $1.3\times 10^5{\rm cm}^{-3}\,$K.  (Our $I/I_0$ is 
approximately the same as their $\chi$ intensity parameter.)  If this mechanism 
could increase the thermal pressure in the outer portions of translucent clouds 
that still have reasonable concentrations of H$_2$ and explain the existence of 
the high pressure component that we observe, we would indeed expect to see a 
positive relationship between $g_{\rm high}$ and $I/I_0$.  

Even though the points in Fig.~\ref{fhigh_vs_logi} at first glance seem to favor the 
recoil hypothesis as a possible explanation for the high pressure component, our 
enthusiasm for accepting this proposition must be tempered by two 
considerations: (1) our fiducial pressure $\log (p/k) \gtrsim 5.5$ (a lower limit 
which is to some degree arbitrary and might be relaxed to a slightly lower level) 
requires a value for $I/I_0$ considerably greater than 60, and (2) the correlation 
seen in Fig.~\ref{fhigh_vs_logi} may be a byproduct of some other physical effect 
that responds to higher than normal intensities and generates the high pressures.  
For instance, we know from Fig.~\ref{log_pok_bin} that values of $\log (p/k)$ in 
the low pressure regime are likewise correlated with the intensity of starlight, and 
the greater influence of intermittent dissipation effects in the more strongly 
driven turbulence may account for the more conspicuous presence of high 
pressure gas.

\section{Summary}\label{summary}

We have presented a comprehensive analysis of {\it HST\/} UV spectra stored in 
the MAST archive for 89 stars that were observed with the E140H mode of STIS, 
with the goal of measuring the populations of the three fine-structure levels of the 
ground electronic state of neutral carbon atoms in the ISM.  The ultimate purpose 
of these measurements was to synthesize a distribution function for the thermal 
pressures in gases that mostly represent the cold neutral medium (CNM) in our 
part of the Galactic disk.  This work builds upon a similar study of 21 stars in 
restricted portions of the sky carried out by JT01 in a special observing program 
dedicated to this purpose.  We have repeated the basic analysis protocol 
developed by JT01 for unraveling the velocity profiles for carbon atoms in the 
separate fine-structure levels from the blended features in many different 
multiplets, but with a few improvements in methodology that are outlined in 
various subsections within Appendix~\ref{improvements}.

We interpret most of the variations in the outcomes for thermal pressures to arise 
from fluctuations caused by interstellar turbulence.  Some additional pressure 
excursions that are large in magnitude but for small mass fractions probably arise 
from the the random passages of infrequent but strong shocks created by either 
stellar mass loss or supernova explosions.

The following conclusions have emerged from our study of C~I fine-structure 
excitations:
\begin{enumerate}
\item The relative populations of the two excited fine-structure levels are 
influenced in different ways by the local physical conditions, since the levels have 
significantly different collisional rate constants and energies.  This feature allows 
us to sense in any one velocity channel the presence of admixtures of gas that 
have markedly different conditions.  While there is a multitude of possibilities for 
explaining any particular combination of level populations, we find that when the 
data are viewed as a whole, the most straightforward interpretation is that 
practically all of the gas in the normal range of pressures ($10^3 \lesssim p/k 
\lesssim 10^4{\rm cm}^{-3}\,$K) is accompanied by very small amounts (of order 
0.05\%) of gas at anomalously large pressures and temperatures ($p/k > 
10^{5.5}{\rm cm}^{-3}\,$, $T>80\,$K).  In a small fraction of cases, the 
proportion of the gas at high pressures is markedly higher than this level, both 
because the amount of C~I is greater and the local radiation density is high (which 
makes more of the carbon atoms singly ionized).
\item For a substantial number of our lines of sight, we can make use of molecular 
hydrogen rotation temperatures $T_{01}$ between $J=0$ and 1 to define the 
local kinetic temperature.  Such temperatures are useful in defining one of the 
free parameters in solutions for the level populations.  As an added benefit, we 
can explore whether or not, in a general statistical sense, the pressure outcomes 
are related in some way with such temperatures.  We find only a weak 
anticorrelation, which indicates that pressure fluctuations do not appear to be the 
dominant cause for temperature changes from one place to the next.
\item Excluding the small amounts of high-pressure gas mentioned earlier, the 
pressures of most of the CNM material show some correlation with the local 
radiation densities, as sensed by the observed ratio of O~I (or sometimes S~II) to 
C~I followed by an application of the equation of ionization equilibrium with 
plausible values for O/C and S/C to derive $N$(C~II).  We interpret this trend as 
arising from the fact that the stars that create this radiation are sources of enough 
mechanical energy to make the pressures higher than normal.
\item The main part of the mass-weighted distribution of pressures in our 
complete sample approximately follows a lognormal distribution with a mean 
value for $\log (p/k)$ equal to 3.58 and a standard deviation  of 0.175~dex.  
However, for $\log (p/k) < 3.2$ or $>4.0$ the amount of material is greater than 
a continuation of the lognormal distribution.
\item In order to sense the distribution of pressures in regions well removed from 
the sources of mechanical disturbance (i.e., the stars that emit large amounts of 
radiation), we have isolated for study only those cases where the radiation 
density is less than $10^{0.5}$ times the overall average level.  Under these 
circumstances, the tail on the high pressure side of the distribution becomes 
suppressed, and the remaining distribution develops a negative skewness.  We 
supply a polynomial expression that fits this distribution (Eq.~\ref{polynomial_fit}) 
that is shown in Fig.~\ref{hist_pok}.  About 23\% of the material in this 
distribution is below the minimum pressure for the thermal equilibrium curve of a 
static CNM in our part of the Galaxy, suggesting that short-term fluctuations in 
pressure can occur without the gas being transformed to a stable warm neutral 
medium (WNM).
\item The thicknesses of the regions that we were able to probe, as measured by 
the hydrogen column densities divided by their space densities, are generally less 
than 20$\,$pc.  The filling fractions for the sightlines are generally less than 1\%.  
The remaining 99\% of a typical sightline is filled with much hotter gas having 
densities that are far too low to create enough C~I for us to measure.
\item We recognize that even with the over-determination of conditions provided 
by the two fine-structure levels, we can still underestimate the dispersion of 
pressures because we are viewing at each velocity an average pressure for the 
superposition of regions that could have vastly different pressures.  We have 
studied how the dispersions scale in proportion to $N({\rm 
C~I_{total}})^{-0.5}$ and find that the ISM could conceivably be composed of 
independent packets of gas with a true {\it rms\/} dispersion in $\log (p/k)$ that 
could be as large as 0.5$\,$dex, which is considerably wider than the distribution 
that we constructed directly from the data.   The characteristic column density of 
each packet would be about $N({\rm C~I}_{\rm total})=2\times 10^{12}{\rm 
cm}^{-2}$.  However, an alternate interpretation, and one that is quite plausible, 
is that small volumes of gas have intrinsic pressure variances that are larger 
than for coherent, larger volumes that might be more resistant to perturbations 
from turbulent forces.    This phenomenon could conceivably produce the same 
linear scaling of pressure dispersions against $N({\rm C~I_{total}})^{-0.5}$ that 
we observed.
\item On the basis of our findings reported in items 4 and 7 above, we derive 
characteristic turbulent Mach numbers for the C~I-bearing gas to range between 
0.8 and 4.8.  Since the speed of sound is about $0.5\,{\rm km~s}^{-1}$ if 
$T=80\,$K, we expect that the 3-dimensional velocity dispersion $\sigma_v$ to 
range between 0.40 and $2.4\,{\rm km~s}^{-1}$.  If we equate these numbers to 
observations of velocity structure functions in the ISM, we find that the 
characteristic size $r$ of the clouds or the outer driving scale of the turbulence is 
probably in the range of approximately $0.2 < r < 4.7\,$pc. 
\item Gas with radial velocities well outside the range of motions expected for 
differential galactic rotation is more likely than usual to exhibit exceptionally large 
pressures.  This link of pressures with kinematics helps to support the 
interpretation that shocks and turbulence play an important role in creating the 
positive excursions in pressure.  Packets of C~I moving at negative velocities show 
larger pressure excursions than for those at positive velocities.  We explain this 
difference in terms of an observational bias that favors our viewing the near sides 
of pressurized shells that are expanding away from our target stars.
\item There is a broad range of time scales that are needed to reach equilibrium 
values for various quantities and physical processes that are relevant to our study.  
From the shortest to the longest they are as follows: (1) C~I fine-structure level 
populations (of order 100~days), (2) the balance between C~I and C~II 
established by the competition between photoionizations and various means of 
recombination (160~yr, or shorter if the radiation density is larger than average), 
(3) the coupling of the $J=0$ to 1 rotation temperature of H$_2$ to the local 
kinetic temperature ($10^4\,$yr for typical conditions: $\log (p/k)=3.5$ and 
$T=80\,$K), and (4) the cooling time for the ISM ($3\times 10^4\,$yr for the 
same conditions).  We compute the eddy turnover times for turbulent eddies 
having a radius $r$ using a relation $\Delta t=r/\Delta v$ with an extrapolation to 
small scales $\Delta v=r_{\rm pc}^{0.4}{\rm km~s}^{-1}$, and we find that the 
only items of consequence for $r$ smaller than about  $100-1000\,$AU are (3) 
and (4).  Over these extremely small scales, delays in the adjustments of H$_2$ 
rotation temperatures will give misleading readings for the local kinetic 
temperatures, but the differences in the two temperatures should be minor, 
especially since we can use $T_{01}$ only to indicate an average temperature 
over many small volumes.  Likewise, any lag in the thermal response of the gas 
will make its polytropic index $\gamma$ closer to the adiabatic value, rather than 
matching the slope of the thermal equilibrium curve for the CNM 
($\gamma\approx 0.7$).  The fact that this may be happening is supported by the 
negative skewness of our distribution in $\log (p/k)$, which indicates that the 
turbulent fluctuations are consistent with $\gamma > 1$.
\item For $\log p/k$ above 4.0, we find a slope in the relationship between the 
logarithms of the volume fractions of the gas and $\log (p/k)$ to be consistent 
with a power-law slope of $-14/9$ that is expected for random penetrations of 
expanding supernova remnants in various stages of development.  
\end{enumerate}

\acknowledgments
This research was supported by program number HST-AR-09534.01-A which was 
provided by NASA through a grant from the Space Telescope Science Institute 
(STScI), which is operated by the Association of Universities for Research in 
Astronomy, Incorporated, under NASA contract NAS5-26555.   All of the C~I 
absorption line data that were analyzed for this paper were taken with the 
NASA/ESA {\it Hubble Space Telescope\/} and were downloaded from the 
Multimission Archive at STScI (MAST). 
{\it Facilities:\/} \facility{HST(STIS)}

\appendix
\section{Improvements in the Analysis over that of JT01}\label{improvements}

Over the time since the publication of our initial survey of thermal pressures 
(JT01), we have had the benefit of some extra opportunities to recognize various 
ways to improve our data reduction and analysis methods.  In the following 
subsections, we discuss these new features in our more refined treatments, and 
we also cover a number of improvements in some basic atomic data that have 
emerged since the earlier study.  We refrain from discussing here all of the 
fundamentals of how the analysis was carried out; JT01 explained this in some 
detail. 

\subsection{Available Transitions and Minor Modifications of Some  
$f$-Values}\label{transitions}

The observing program of 21 stars carried out by JT01 had a wavelength 
coverage that included all of the multiplets out of the ground electronic state of 
C~I at wavelengths longward of 1188$\,$\AA, except for 
Multiplet\footnote{Multiplet numbering system from Moore (1970), also adopted by
Morton (2003).} 3 centered at 1561$\,$\AA.  Their analysis was based on intensity
profiles recorded for all of these multiplets, except for a few cases where either
the lines were too weak to be useful or suffered interference from other atomic
species (see Table~3 of JT01).  They adopted a revised set of $f$-values for the
transitions that gave internally consistent outcomes for their analysis, based on
comparisons of optical depths described in \S5.3 of their paper.

Some of the observations used in the current study have exceptionally high 
signal-to-noise ratios.  We have made use of these results to further revise (or 
determine for the first time) the $f$-values for some transitions that were too 
weak to evaluate accurately in the earlier study, again by implementing the 
analysis of JT01.   We list in Table~\ref{new_f-values} our newly adopted 
$f$-values for the current study.  For lines not listed in this table, we used the 
values given by JT01.  New $f$-values were also determined for Multiplet~3, 
which was covered by observations in the archived data but not in our original 
survey.  New $f$-values determined here are very similar to those derived by 
JT01; both sets differ appreciably from ones published in the literature.  In 
Appendix~\ref{fval_validation} we describe a special study that supports the 
validity of both our new and older $f$-values.

\placetable{new_f-values}

\begin{deluxetable}{
l  %Multiplet
c  %lambda
c  %logflambda
}
\tablewidth{0pt}
\tablecaption{New $f$-values\tablenotemark{a}\label{new_f-values}}
\tablehead{
\colhead{} & \colhead{$\lambda$} & \colhead{}\\
\colhead{Multiplet\tablenotemark{b}} & \colhead{(\AA)} & \colhead{log 
($f\lambda$)}
}
\startdata
2\dotfill&1656.267&2.012\tablenotemark{c}\\
&1656.928&2.392\tablenotemark{c}\\
&1657.008&2.266\tablenotemark{c}\\
&1657.379&1.788\tablenotemark{c}\\
&1657.907&1.914\tablenotemark{c}\\
&1658.121&1.789\tablenotemark{c}\\
3\dotfill&1560.309&2.312\\
&1560.682&2.187\\
&1560.709&1.710\\
&1561.340&1.488\\
&1561.367&0.311\\
&1561.438&2.236\\
7.01\dotfill&1277.190&$-0.021$\\
12\dotfill&1192.451&0.711\\
&1192.835&0.129\\
14\dotfill&1189.447&1.006\\
&1189.631&1.246\\
\enddata
\tablenotetext{a}{Changes from values adopted by JT01}
\tablenotetext{b}{ Multiplet numbering system from Moore (1970), also
adopted by Morton (2003).} \tablenotetext{c}{Adopted from Morton
(2003).}
\end{deluxetable}

\subsection{Improved Quantitative Estimates of the 
Uncertainties}\label{uncertainties}

A critical aspect of our study of the thermal pressures is the proper understanding 
and control of errors, since, if they are large enough, they can mislead us into 
thinking the distribution function is broader than reality.  For this reason, we have 
instituted a number of improvements in developing quantitative estimates for 
various sources of error so that we can more reliably screen out measurements of 
inferior quality and have a better confidence that the remaining errors are 
inconsequential.  The discussions in the following subsections build upon the 
concepts presented by JT01 (see their Sections 4 and 5).

\subsubsection{Relative Errors in the Observed Optical Depths}\label{err_tau}

For absorption features that are moderately or very strong, the principal source 
of uncertainty in any intensity measurement is that produced by random noise 
fluctuations in the counts of photoevents.  In the study of JT01, the effects of 
these errors were propagated through the analysis, serving as a guide on the most 
appropriate weight factors for intensities recorded in different multiplets.  JT01 
also recognized the existence of uncertainties in the adopted background level by 
using a formula (their Eq.~7) that reduced or eliminated the relative weights of 
stronger features that came close to zero intensity.  For very weak lines, however, 
systematic uncertainties in the establishment of the continuum level are also 
important.

In the current study, we now add the continuum uncertainties to the noise errors 
in quadrature.  This has practically no effect for moderately strong features, but it 
now decreases the relative importance of very weak multiplets in the final 
solutions.  As did JT01, we adopted best fitting Legendre polynomials for the 
continua at locations somewhat removed from the features.  To construct the 
probable errors in these continua, we evaluate the expected errors in the 
polynomial coefficients, as described by Sembach \& Savage (1992), and then we
multiply them by two in order to make an approximate allowance for additional
uncertainties caused by the arbitrariness in selecting the most appropriate
polynomial order.  From the sizes of residual errors that seem to be extended over
broad ranges of velocities where no absorption is evident, this global increase
in the estimates for the continuum errors seems to be appropriate.

\subsubsection{Error Estimates for $N$(C~I),  $N$(C~I$\,^*$), and  
$N$(C~I$\,^{**}$)}\label{err_N}

In their Section~5, JT01 discussed the various sources of both random and 
systematic errors in column densities. They estimated the combined magnitudes 
from most of these errors by measuring the amplitudes of fluctuations in $N$ at 
velocities well removed from obvious C~I features.  These empirical 
determinations should be satisfactory for velocities where the amplitudes of the 
absorptions are small, but they underestimate the errors at locations where the 
intensity levels are well below the continuum.  In these circumstances, noise 
deviations can create much larger uncertainties in the combined optical depths 
for different lines within a multiplet, and these enhanced uncertainties propagate 
their way through to the solutions for the column densities.

We have now implemented evaluations of errors that should apply equally well to 
both the weak and strong portions of the absorptions.  Eq.~6\footnote{The 
constant A in this equation should actually be C, where C was defined on the 
preceding page (and used in Eq.~3).} of JT01 was used to construct a design 
matrix from which least-squares solutions would emerge for the three column 
densities at each of the different velocities.  The inverse of this matrix gives the 
expected variances and covariances of these variables, as long as there is a proper 
scaling of the rms errors in all of the optical depths that contribute to the matrix 
terms.  By incorporating the error derivations discussed in the above section 
(\S\ref{err_tau}), we believe that our evaluations of the uncertainties in the 
optical depths $\sigma_{\tau(i)}$ properly include all effects except for some that 
are unquantifiable, such as possible detector artifacts, distortions in optical depths 
caused by unresolved saturated structures (discussed in \S\ref{distortions} 
below), and errors in the adopted $f$-values of the transitions.  The magnitudes 
of the covariances are much smaller than the variances, so the squares of the 
errors can be extracted simply from the terms in the main diagonal of the inverse 
matrix.
 
In order to obtain an approximate validation that the error calculations give 
reasonable results, we can draw upon three observations of the star 
HD$\,$219188 taken at different epochs.  If we exclude the velocity component 
that has been identified by Welty (2007) to vary with time, 
we find that for the expected errors that were calculated according to the 
prescription given above, the dispersion about the weighted mean of individual 
observations of column densities at different velocities yielded a $\chi^2=750$ 
for 536 degrees of freedom.  This evaluation was performed after the outcomes 
with a spacing of $0.5\,{\rm km~s}^{-1}$  had been binned by a factor of 3, so 
that the samples were approximately commensurate with the velocity widths of 
the detector's pixels.  One can surmise from this study that our error calculations 
probably underestimate the true errors by a factor of approximately 
$(536/750)^{\onehalf}=0.85$.  There were no obvious differences between the 
magnitudes of individual $\chi^2$ outcomes within strong C~I absorptions as 
opposed to those outside the features.

Over velocity intervals where we sense that there is no C~I absorption, we detect 
some very low level, smooth deviations that are still present and that add to the 
random short-scale noise caused by photon counting statistical variations.  These 
deviations are probably caused by slight inadequacies in the fitting of Legendre 
polynomials to the true continuum levels, which can exhibit troublesome 
variations for stars with low projected rotational velocities.  These deviations have 
a relatively minor influence on column densities at specific velocities, but they can 
contribute nonnegligible errors in $N({\rm C~I_{total}})$ integrated over all 
velocities, since these errors are coherent from one velocity to the next and thus 
can build up in a linear fashion.  In computing the errors associated with the 
values of $N({\rm C~I_{total}})$ listed in Table~\ref{obs_quantities_table} we 
took such errors into account (by direct addition across velocity channels rather 
than quadrature sums).  We also allowed for the fact that the smaller scale 
statistical errors were coherent over contiguous stretches of 2.6 velocity 
channels, which represent the width of a pixel on the STIS ultraviolet 
detector.\footnote{This channel width should not be confused with the line 
spread function of the spectrograph, which is about twice as large.}

\subsubsection{Errors in $f1$ and $f2$}\label{err_f1f2}

JT01 used their own judgment in making the choices for velocity ranges over 
which the C~I absorptions were deemed to be strong enough to include in the 
presentations of $f1$ and $f2$.   The existence of a few errant $f1$ and $f2$ 
points in Fig.~7 of JT01 shows that this selection was not ideal.  We have thus 
introduced a more rigorous and quantifiable selection criterion for data which are 
to be regarded as acceptable.

If we re-express the column densities for C~I in the three fine-structure levels as 
$N_0\equiv N({\rm C~I})$, $N_1\equiv N({\rm C~I}^*)$, and $N_2\equiv N({\rm 
C~I}^{**})$, along with a designation for the total column density $N_{\rm 
tot}\equiv N_0+N_1+N_2$, then the uncertainties $\sigma(N_0)$, 
$\sigma(N_1)$, $\sigma(N_2)$ contribute to an overall error in $f1$ in the 
following manner:
\begin{eqnarray}
\sigma(f1)&=&\left\{ \left[\sigma(N_0){\partial f1\over\partial N_0}\right]^2 +
\left[\sigma(N_1){\partial f1\over\partial N_1}\right]^2 +
\left[\sigma(N_2){\partial f1\over\partial N_2}\right]^2\right\}^{\onehalf}\cr
&=&N_{\rm tot}^{-2}\left\{\sigma(N_1)^2\left[N_{\rm tot}-N_1\right]^2+ 
N_1^2\left[\sigma(N_0)^2+\sigma(N_2)^2\right]\right\}^{\onehalf}\,.
\end{eqnarray}
Likewise,
\begin{equation}
\sigma(f2)= N_{\rm tot}^{-2}\left\{\sigma(N_2)^2\left[N_{\rm tot}-N_2\right]^2+ 
N_2^2\left[\sigma(N_0)^2+\sigma(N_1)^2\right]\right\}^{\onehalf}\,.
\end{equation}

For a determination of $f1$ and $f2$ at any particular velocity, we insisted that in 
order to be considered, both of their $1\sigma$ errors had to be less than 0.03.

\subsubsection{Sensing Possible Distortions Caused by Unresolved Saturated 
Portions of the Absorption Profiles}\label{distortions}

Even though the spectra considered here had a minimum wavelength resolving 
power $R=\lambda/\Delta\lambda=114,000$ for the E140H grating with the 
standard entrance slit\footnote{$R=200,000$ for the narrow entrance slit and 
half-pixel intensity sampling used for the stars observed by JT01.} (Proffitt et al.
2010), there remains a possibility that for some cases we will encounter collections
of unresolved, saturated absorption features.  In such situations, the apparent
optical depths will underestimate the true optical depths after instrumental
smoothing.  One can normally detect this condition by noting that strong lines show
smaller column densities than weaker ones and then applying a correction scheme
proposed by Jenkins (1996).  Here, however, the correction is no longer
straightforward because the individual apparent optical depths lose their identity
after they have undergone the transformations that are needed to unravel the
overlaps of C~I, C~I$^*$, and C~I$^{**}$ absorptions in each multiplet.

While we were unable to perform corrections to restore the apparent optical 
depths to their true (but smoothed) values, we could nevertheless detect 
circumstances where the representations are likely to be inaccurate.   The 
implementation of Eq.~7 of JT01 serves to limit the influence of stronger lines that 
have low residual intensities.  By varying the threshold intensity parameter $I_t$ 
in this equation, we can shift the response of the solutions for the column 
densities either toward or away from the stronger lines.  In so doing, if we find 
that the column density solutions change, we can surmise the outcomes are not 
stable and that distortions are indeed happening.   For adopting or rejecting a set 
of column densities at any given velocity, we adopted the following test:  If any 
column density derived using twice the standard value for $I_t$ (see JT01 
\S5.2.1) came out to be more than 1.2 times that derived from the standard 
$I_t$, we considered the distortion to be unacceptably large and the result for the 
particular velocity was rejected.  Otherwise, deviations less than the factor of 1.2 
were deemed to be acceptable.  Stars for which the optical depth distortions 
seemed to be evident are identified in Table~\ref{obs_quantities_table} (see note 
$a$ of the table for details).

\subsection{New Atomic Physics Parameters}\label{atomic_phys_param}

With the passage of time, re-evaluations of atomic parameters are carried out 
(and we presume that they are better than the older ones).  Since the time of 
publication of our earlier results (JT01), new reaction rates for the excitation of  
the excited fine-structure levels of C~I and O~I by atomic hydrogen have been 
published by Abrahamsson et al. (2007).  We have incorporated these 
new rates into our interpretations of densities and temperatures from the C~I 
level populations.  (The O~I excititations are discussed in 
Appendix~\ref{fval_validation}.)  Rates for other collision partners are the same 
as those adopted by JT01.  Also, we have now replaced the old spontaneous 
radiative decay rates for the excited levels of the two neutral species by those 
given by Galav\'is et al. (1997).

The effect of these changes is that for the representative conditions 
$\log(p/k)=3.6$ and $T=80\,$K the derived values of the thermal pressure 
increase by 0.05~dex above what we would have obtained from the older 
numbers used by JT01.

\subsection{Revised Optical Pumping Rates}\label{pumping}

As pointed out by de Boer \& Morton  (1974),
the ultraviolet transitions from the C~I electronic ground state that can be used to 
measure the column densities of this atom also act to populate the upper 
fine-structure levels through optical pumping by the ambient starlight radiation 
field.  While this effect is usually small compared to the excitations by collisions, it 
nevertheless is a process that should not be ignored since there are occasions 
when the starlight field can be found to be considerably above average.  Jenkins 
\& Shaya (1979) calculated the rates of optical 
pumping for the average level of starlight radiation computed by Jura 
 (1975a, b)
and Witt \& Johnson (1973).  These rates were used 
by JT01.

Our current analysis of the effects of starlight presented in 
\S\ref{ionization_corrections} adopts as a standard the more recent average 
radiation intensity $I_0(\lambda)$ specified by Mathis et al. (1983).  Here, we
used our ionization equilibrium calculations for carbon atoms to determine the
local starlight intensities in terms of a multiplier $I/I_0$ times this field
strength.  In order to make our analysis consistent with the new standard
intensity, and also to be consistent with our new $f$-values for the C~I
transitions, we have recomputed the matrix for the pumping rates between the
different fine-structure levels.  This new matrix, presented in
Table~\ref{pump_matrix} [cf. Table 3 of Jenkins \& Shaya (1979)],  expresses the
transition rates expected for the standard field density.  For every measurement of
C~I and its fine-structure populations, we adjusted the pumping rates in proportion
to the corresponding value of $I/I_0$ -- see \S\ref{derivations}.

Our application of the new pumping matrix overlooks two effects that can modify 
the the pumping rates in opposite directions.  First, we have ignored weak C~I 
transitions at wavelengths shorter than 1188$\,$\AA.  An approximate 
compensation for this is our neglect of the self shielding of the stronger transitions 
deep inside some of the most strongly absorbing clouds.

The overall effect in implementing the revised pumping rates is to raise the 
median pressure by about 0.05~dex for most of the C~I data.  For measurements 
that apply to cases where $I/I_0<10^{0.5}$, as shown in Fig.~\ref{hist_pok}, the 
change is almost negligible.

\section{A Validation of our Previous Determinations of the C~I
{\boldmath $f$}-Values}\label{fval_validation}

One aspect of our earlier study (JT01) of interstellar C~I that is
highly relevant to the work done here was our re-evaluation of the
relative $f$-values of the C~I lines (for all three fine-structure
levels) from one multiplet to the next.  Initially, we found that the
$f$-values given in the literature did not yield self consistent
outcomes for the predicted line strengths, after we had derived
preliminary values of the C~I, C~I$^*$ and C~I$^{**}$ column densities. 
As a result, we derived a new set of $f$-values that gave the best
internal consistencies for the strengths of the lines for all of the
targets collectively.  These relative $f$-values were based on the
1656$\,$\AA\ multiplet as a calibration, in our belief that theoretical
calculations of the strength of lines in this strong multiplet were
probably the most reliable.

\placetable{pump_matrix}

\begin{deluxetable}{
l  %Initial J
c  %final J = 0
c  %final J = 1 
c  %final J = 2
}
\tablecolumns{4}
\tablewidth{0pt}
\tablecaption{New Optical Pumping Rates
Between the C~I Fine-Structure Levels 
($10^{-10}{\rm s}^{-1}$)\tablenotemark{a}\label{pump_matrix}}
\tablehead{
\colhead{} & \multicolumn{3}{c}{Final $J$}\\
\cline{2-4}
\colhead{Initial $J$} & \colhead{0} & \colhead{1} & \colhead{2}
}
\startdata
0\dotfill&\nodata&5.06&3.99\\
1\dotfill&1.69&\nodata&5.23\\
2\dotfill&0.08&3.14&\nodata\\
\enddata
\tablenotetext{a}{Computed for an optically thin medium in the presence of an 
average ultraviolet radiation field density in the ISM at a Galactocentric radius of 
10~kpc, as specified by  Mathis et al. (1983).}
\end{deluxetable}
\clearpage

Our revised $f$-values showed a steady divergence from the earlier,
published ones as the lines became successively weaker.  For each
$-1~{\rm dex}$ change in a published line strength $\log f\lambda_{\rm
pub.}$, the discrepancy $\log f\lambda_{\rm JT01}-\log f\lambda_{\rm
pub.}$ increased by about $+0.3~{\rm dex}$ (see Fig.~3 of
JT01).  More recent determinations of C~I $f$-values have been
compiled by Froese Fischer \& Tachiev (2004) (henceforth FT04).  A plot
of these newer values against those derived by JT01 still shows the
divergence seen in Fig. 3 of JT01, but with less scatter in the
individual points representing the weakest lines.  Henceforth, we will
adopt the values of FT04 as a proxy for $f$-values determined by
investigators other than JT01.

In our earlier study, we recognized that the sense of the divergence was 
consistent with the
possibility that we could have been misled by the underrepresentations
of smoothed optical depths for the most saturated portions of the strong
lines (while at the same time, optical depths of the weak absorptions
would be properly measured).  Such an effect can occur when the lines
have very narrow component structures that are saturated and unresolved
by the spectrograph (Savage \& Sembach 1991; Jenkins 1996), and indeed
this seems to happen in some cases -- see note $a$ in
Table~\ref{obs_quantities_table}.

To overcome our worry about these possible distortions in the apparent
optical depths, we carried out a number of investigations that indicated
that they were unlikely to be the explanation for the divergence that we
found (see the discussion in Section~5.3.2 of JT01).  Nevertheless, it
is still troubling that theoretical and experimental derivations of the
C~I $f$-values by different investigators, such as those cited by Morton 
 (Morton 2003), consistently gave results that disagreed with the 
values derived by JT01.  Recent evidence that interstellar absorption features 
arising from neutral atoms can occasionally exhibit extraordinarily low turbulent 
velocity dispersions (Dunkin \& Crawford 1999; Price et al. 2001; Knauth et al.
2003; Meyer et al. 2006) reinforces our concerns about possible
misrepresentations of averaged true optical depths.  If nearly all of the C~I
absorptions that we observed consisted of clusters of needle-like features that
had velocity dispersions $b\lesssim 0.4\,{\rm km~s}^{-1}$ created by low
temperatures and low turbulent (or shear) velocities, and they are
separated\footnote{Narrow features that have small separations and thus are
partly blended are of no concern, since they behave much like features with much
larger $b$ values.} by more than about $2.5b$, we could have in principle
erroneously introduced systematic errors that could explain the observed
differences between our $f$-values and the published ones, even though
the lines were observed at a wavelength resolving power $R=200,000$ (see
Appendix~A of JT01).  This phenomenon would have had to operate in a
consistent fashion from one case to the next, since we almost always
found reconstructed profiles (using our $f$-values) to show good
agreement with the observations of both strong and weak lines.  The
uniform persistence of this misrepresentation effect seems rather
unlikely, but we have no grounds for saying that it is completely
impossible.

We now present a single example that
disproves the proposition that our results were incorrect because we
were continually misled by narrow substructures in the profiles.  The
substance of our argument is that for one special case we can compare
the observed and predicted absorption features from a velocity complex
for which the values of $b$ caused by thermal Doppler broadening for any
subcomponents must not be very much less than the resolution of the STIS
spectrograph.

\begin{figure}[b!]
\epsscale{0.8}
\plotone{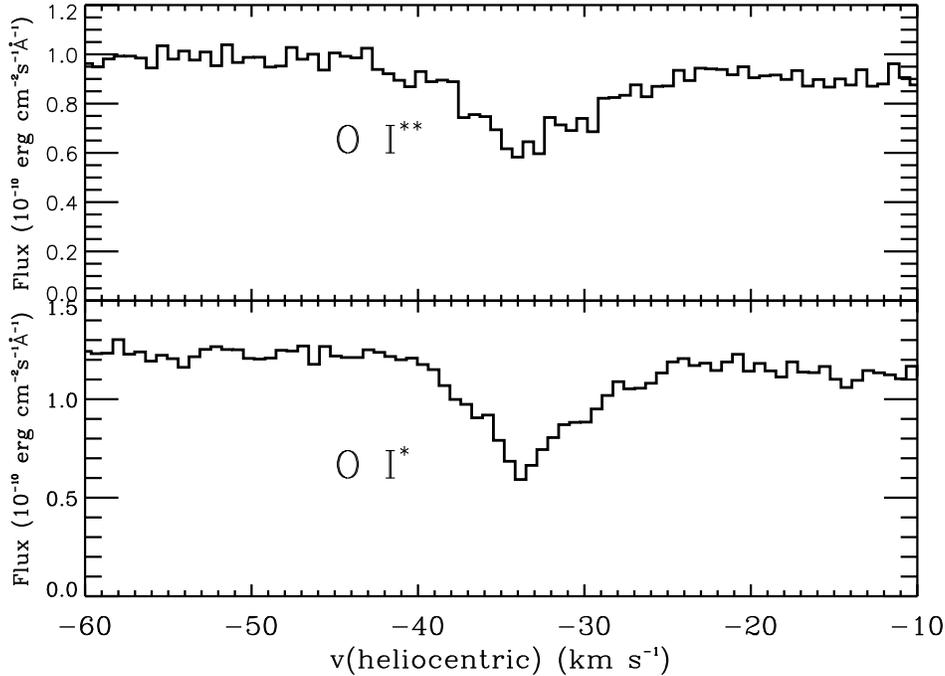}
\caption{Absorption profiles (unnormalized) of O~I$^{**}~\lambda 1306$
({\it top panel\/}) and O~I$^*~\lambda 1304$ ({\it bottom panel\/})in
the spectrum of HD$\,$210839.\label{hd210839_oi}}
\end{figure}

\begin{figure}[t!]
\plotone{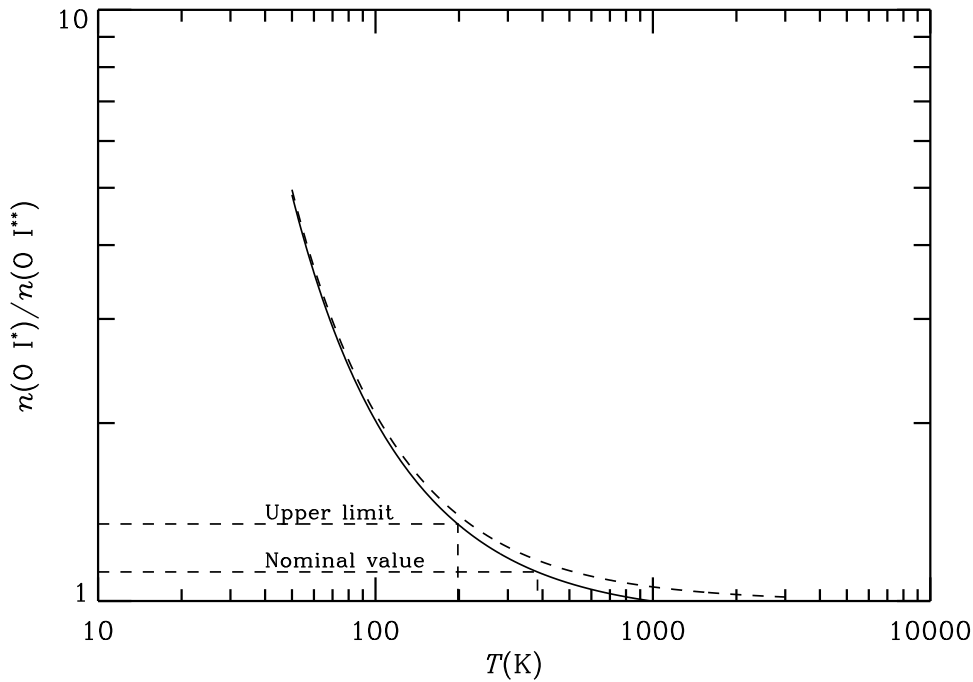}
\caption{The expected ratio of $n({\rm O~I}^*)/n({\rm O~I}^{**})$ for
collisions in an environment where $n({\rm H~I})\leq 100\,{\rm cm}^{-3}$
({\it solid curve\/}) and $n({\rm H~I})=1000\,{\rm cm}^{-3}$ ({\it
dashed curve\/}).  For the velocity component at $-34\,{\rm km~s}^{-1}$ in front 
of $\lambda$~Cep, our nominal value and upper limit for $N({\rm
O~I}^*)/N({\rm O~I}^{**})$ are indicated by the horizontal dashed lines,
and the temperatures $T=200$ and $390\,$K that apply to these values
are shown by the vertical dashed lines.\label{oifsl}}
\end{figure}

A cluster of absorption profiles with a peak at about $-34\,{\rm
km~s}^{-1}$ toward HD$\,$210839 ($\lambda$~Cep) is unusual because it 
shows
exceptionally strong features from O~I$^*$ and O~I$^{**}$, as shown in
Fig.~\ref{hd210839_oi}, and the C~I fine-structure excitation indicates
that it is at an unusually high pressure.  The ratio $n({\rm
O~I}^*)/n({\rm O~I}^{**})$ is a good indicator of the local kinetic
temperature of the gas.  Figure~\ref{oifsl} shows how this ratio should
vary with temperature.  This relationship follows from solutions for the
equilibrium equation based on the H~I collision rate constants as a
function of temperature given by Abrahamsson et al. 
 (2007)\footnote{Abrahamsson et al. did not 
list rates for $T>1000\,$K.  The collision rate constants that we needed for $T$
somewhat above 1000$\,$K are simply extrapolations from their values
just below 1000$\,$K.} and the radiative decay rates of Galav\'is et al. 
 (1997).  We find from the absorptions produced by 
the 1304.86$\,$\AA\ and 1306.03$\,$\AA\ transitions that over the velocity
interval $-40 < v < -20\,{\rm km~s}^{-1}$, $\log N({\rm O~I}^*)=13.410\pm
0.018$ and $\log N({\rm O~I}^{**})=13.359\pm 0.023$.  The nominal value
of 1.12 for the ratio of the two is shown by the lower of the two
horizontal dashed lines in Fig.~\ref{oifsl}, while the other dashed line
just above it indicates our estimate of 1.35 for an upper limit based on
the $+2\sigma$ error for $N({\rm O~I}^*)$ and $-2\sigma$ limit for
$N({\rm O~I}^{**})$.  It follows that our favored value for $T$ is
390$\,$K, and our most conservative (i.e., low) value is 200$\,$K, both
of which are substantially lower than an upper limit $T<660\,$K based on
a fit that gave $b=0.53\,{\rm km~s}^{-1}$ for the strongest component of
the K~I line in this same velocity complex observed at a resolving power
of $0.56\,{\rm km~s}^{-1}$ (FWHM) by Welty \& Hobbs  (2001).  (The
weaker K~I component at a more positive velocity yielded a $b$ value
that indicated that $T<1550\,$K.)  Since the thermal Doppler broadening
for $T=390\,$K is expected to yield a value for $b$ of only $0.41\,{\rm
km~s}^{-1}$, there is either some extra broadening due to turbulence or
the profiles might be split into even smaller separate components that
were not resolved by Welty \& Hobbs.

\placefigure{hd210839_oi}
\placefigure{oifsl}


One might argue that the near agreement of the O~I$^*$ and O~I$^{**}$
column densities could arise from these features themselves being
strongly influenced by unresolved, saturated absorption spikes within
the velocity complex.  Extreme saturations of this sort would tend to
deceive us into deriving nearly the same column densities for the two
species.  However, we can test for this by examining weaker features
from the 1040$\,$\AA\ multiplet covered by the {\it Far Ultraviolet
Spectroscopic Explorer\/} ({\it FUSE\/}).  One such spectrum of
HD$\,$210839 is available in the MAST archive.  No absorption by the
O~I$^*$ 1040.94$\,$\AA\ transition is readily apparent in the
spectrum; a formal measurement of the equivalent width over the
appropriate velocity range yields $0.8\pm 1.3\,{\rm m\AA}$.  For the
column density of O~I$^*$ that we derived, one would have expected to
find $W_\lambda (1040.94\,{\rm \AA})=2.2\pm 0.1\,{\rm m\AA}$.  If we had
underestimated $N({\rm O~I}^*)$ because the 1304.86$\,$\AA\ feature had
hidden saturation, the equivalent width of the weaker feature in the {\it FUSE\/} 
spectrum would have
been larger than this expectation.  We are unable to perform the same test for 
O~I$^{**}$
because the 1041.69$\,$\AA\ feature suffers from interference from the
Lyman 6$-$0~R(6) transition of H$_2$ at nearly the same wavelength.

A separate argument that disfavors a strong internal saturation of the
1304.86$\,$\AA\ feature and a less strong effect with the
1306.03$\,$\AA\ feature is that the shapes of the two absorptions do not
differ from each other appreciably.  If anything, the O~I$^*$ absorption
seems more strongly peaked in the center, an effect opposite to an
expectation that saturation could be occurring in the strongest part of
the profile.  This effect might indicate the presence of slightly cooler
gas at velocities very near the central velocity of $-34\,{\rm
km~s}^{-1}$ (the ratio of apparent optical depths at this central point
is 1.42, which would indicate $T\approx 120\,$K if we were to use the
curve shown in Fig.~\ref{oifsl}).

Returning to the topic of subcomponents within the C~I, C~I$^*$ and
C~I$^{**}$ profiles, for our conservative lower limit $T=200\,$K we
expect the Doppler broadening to produce a $b$ value for carbon atoms of
at least $0.53\,{\rm km~s}^{-1}$, which is somewhat less than the STIS
instrumental profile function with $b=0.90\,{\rm km~s}^{-1}$ (if its
shape is Gaussian).  We can build a model of this complex composed of
separate components, each with $b=0.53\,{\rm km~s}^{-1}$, on top of a
low amplitude, broad shoulder.  This model has been tailored such that,
after it has been smoothed by the STIS instrumental spread function, it
duplicates exactly the shape of the C~I$^*$ profile that we derived.  It
also creates structures that are consistent with the 3 K~I component
parameters derived by Welty \& Hobbs (2001), 
after we acknowledge the fact that
the thermal broadening of the K~I should be less than that for C~I.

If we could have resolved the partly blended complex of profiles
perfectly, we would have found that the apparent optical depth
$\tau_{\rm a} (v)$ [in this case equal to the true optical depth $\tau
(v)$] divided by $f\lambda$ of the transition should remain constant for different
transitions at all velocities $v$, regardless of line strength.  However, after the
profile function has been smoothed by the instrumental profile, there
can in principle be some violation of this equality.  Our model was
constructed in a manner to create the worst conceivable situation that
would aggravate this violation, i.e., we adopted the lower limit for $T$
instead of the nominal one, and we assumed the measurable broadening of
profiles seen in the C~I and K~I spectra, beyond the instrumental
smoothing, was created by clusters of narrow profiles rather than a
smooth turbulent broadening.  This conservative model indicates that as
the value of $f\lambda$ increases by one order of magnitude from 5 to
50, which corresponds to peak values of $\tau (v)$ equal to 0.24 and
2.4, respectively, the average discrepancy in $\tau_{\rm a}/f\lambda$
over the strongest portions of the profile ($-36 < v < -28\,{\rm
km~s}^{-1}$) is only 0.041~dex.  This is only about 13\% of the observed
rate of divergence between our $f$-values and those that have been
published elsewhere.

\begin{figure}
\epsscale{0.75}
\plotone{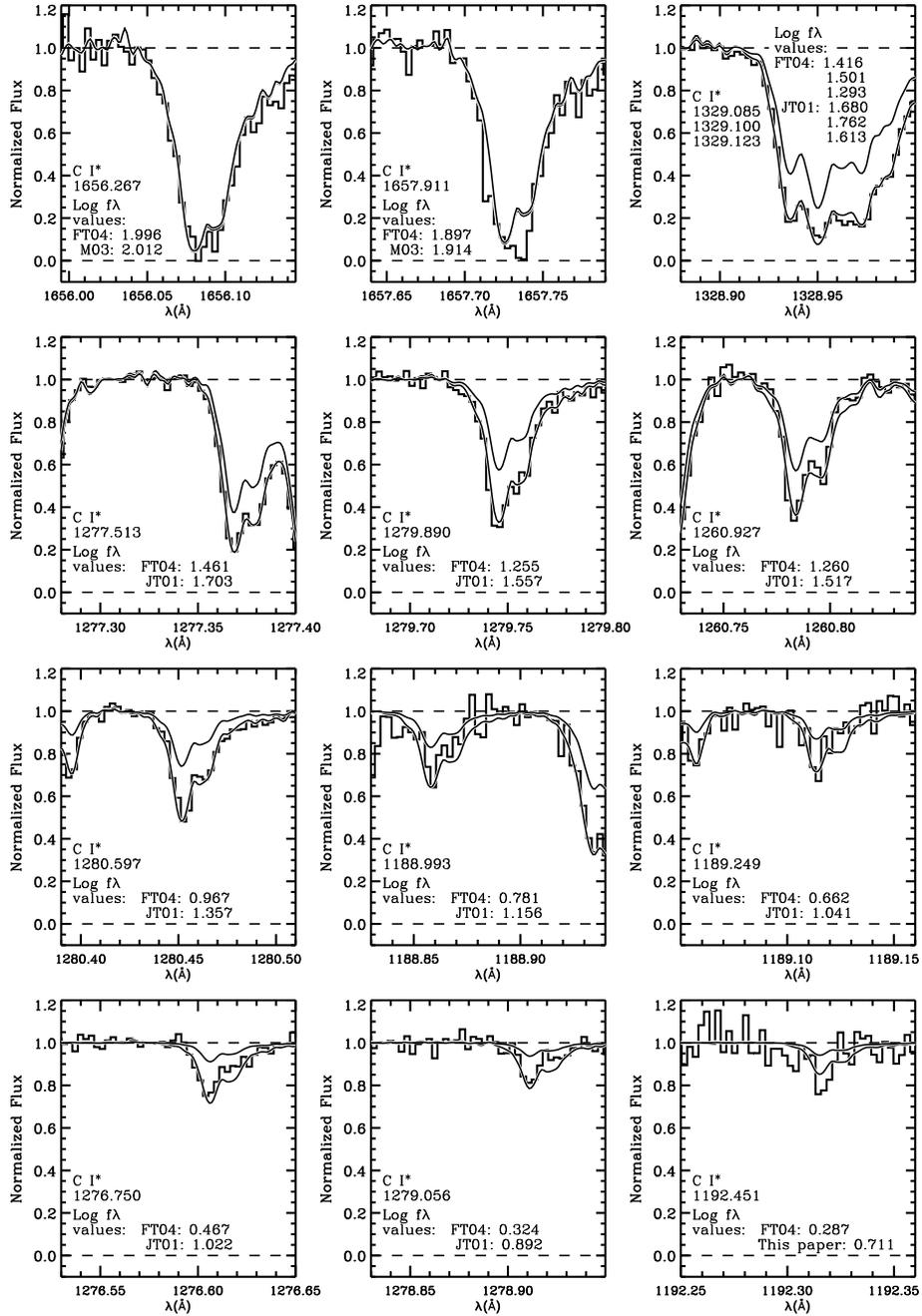}
\caption{Selected absorptions from the $-34\,{\rm km~s}^{-1}$ C~I$^*$
component in the spectrum of  HD$\,$210839 ($\lambda$~Cep) that are not
contaminated by overlapping features created by the components between
$-20$ and $0\,{\rm km~s}^{-1}$.  The observed fluxes ({\it
histogram-style traces\/}) have been normalized to our best estimate for
the continuum level in each case.  The different panels are ordered in a
sequence of decreasing values of $\log(f\lambda)$ determined by JT01. 
The upper smooth curve in each panel represents a reconstruction of the
profile using the published f-value, while the lower curve arises from
the f-value derived by JT01.  [Key to published $f$-value sources:
M03~=~Morton  (2003), FT04~=~ Froese Fischer \& 
Tachiev (2004).]\label{hd210839_display}}
\end{figure}

We have demonstrated the expected insignificance of
misrepresentations reflected by $\tau_{\rm a}$ when progressing from
very weak absorptions to absorptions of moderate strength (10 times stronger), 
where we consider $\tau_{\rm a}$ to be simply an instrumentally smoothed 
version the real optical depth $\tau$.  Moving on, we
are now in a position to interpret comparisons between the observed
absorption features and ones that are reconstructed from our derived
column densities vs. velocity, on the assumption that the computed
values of $\tau_{\rm a}$ give the correct residual intensities (to
within about 0.041~dex).  The different displays in
Fig.~\ref{hd210839_display} show this comparison for the complex at
$-34\,{\rm km~s}^{-1}$ for both the $f$-values derived by JT01 and those
listed by Froese Fischer \& Tachiev (2004)  (FT04).  We have restricted
the choices in this display to transitions of C~I$^*$, and we have
rejected cases where there was interference from other transitions that
had absorptions from other, less negative velocity components that were
overlapping.  It is clear that as the lines become weaker, the
disparity between the observations and the predictions using the FT04
$f$-values increases, while those using the JT01 $f$-values match the
observations to within the noise levels in each case.

\placefigure{hd210839_display}

It is important to emphasize that the investigation discussed here addresses the 
accuracy of the relative $f$-values from one multiplet to the next and not the 
correctness of all of the multiplets taken together.  Had we adopted a multiplet 
other than the strongest one at 1657$\,$\AA\ as a standard, all of our derived 
$f$-values would have been lower by some constant factor.

\end{document}